\documentclass{cernyrep}
\usepackage{texnames}
\usepackage[T1]{fontenc}
\usepackage[bookmarks, colorlinks=true, linktoc=page, pdftex, linkcolor=black, citecolor=black, urlcolor=blue]{hyperref}
\newcommand{\picturefolder}{.}
\newcommand{\tune}{{Q}}
\newcommand{\inttune}{{N}}
\newcommand {\vp}      {\Delta \vec{u}}
\newcommand {\vd}      {\vec{d}}
\newcommand {\vdisp}    {\vec{D}}
\newcommand {\vk}      {\Delta \vec{\theta}}
\newcommand {\nump}    {N}
\newcommand {\numk}    {M}
\newcommand {\matA}    {{\bf A}}
\newcommand {\matZ}    {{\bf Z}}
\newcommand {\matV}    {{\bf V}}

\newcommand {\matW}    {{\bf W}}
\newcommand {\matR}    {{\bf R}}
\newcommand {\matS}    {{\bf S}}
\newcommand {\matG}    {{\bf G}}

 \newcommand{\rd}{{\rm d}}

\usepackage{fancyhdr}
\fancyhfoffset{4 mm}
\fancypagestyle{ARTTITLE}{%
\fancyhf{} 
\lhead{\small{Proceedings of the 2018 CERN--Accelerator--School course on \it{Beam Instrumentation}, Tuusula, (Finland)}} 
\lfoot{Available online at \url{https://cas.web.cern.ch/previous-schools}}
\rfoot{\thepage\hspace*{3mm}}
 
}

\begin{document}

\title{Linear Imperfections}

\author {J\"{o}rg Wenninger}

\institute{CERN, Geneva, Switzerland}

\begin{abstract}
This lecture gives an overview of the impacts on linear machine optics of machine imperfections due to incorrect field settings and misalignments. The effects of imperfections in dipole, quadrupole, and sextupole magnets are presented, along with beam observables and  correction techniques that may be used to restore the nominal machine parameters. The main concepts of orbit correction are discussed in detail, because the principles underlying those techniques can be used for other corrections.
\end{abstract}


\maketitle 

\thispagestyle{ARTTITLE}

\section{Introduction}

In this first section we present a brief summary of the main fields in an accelerator and the associated concepts. For details on linear accelerator physics we refer the reader to the introduction by M.~Sands \cite{SANDS} or other books on accelerator physics, such as \Bref{WIEDEMANN}.

\begin{figure}[btp]
  \begin{center}
\includegraphics[width=0.8\linewidth]{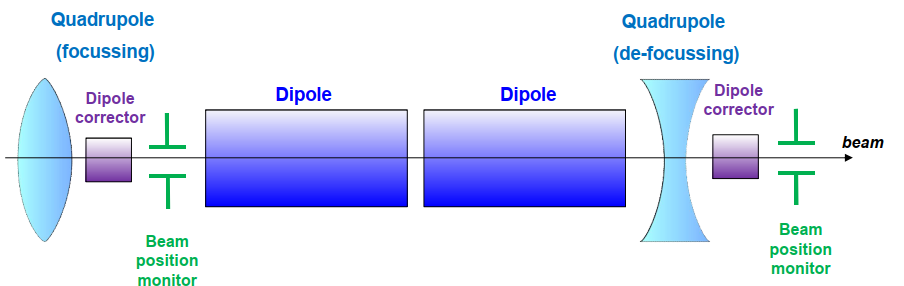}
   \end{center}
\caption{Schematic layout and main components of a basic accelerator cell, in this case a FODO cell}
\label{fig:acc-cell}
\end{figure}

\begin{figure}[tbp]
  \begin{center}
\includegraphics[width=0.4\linewidth]{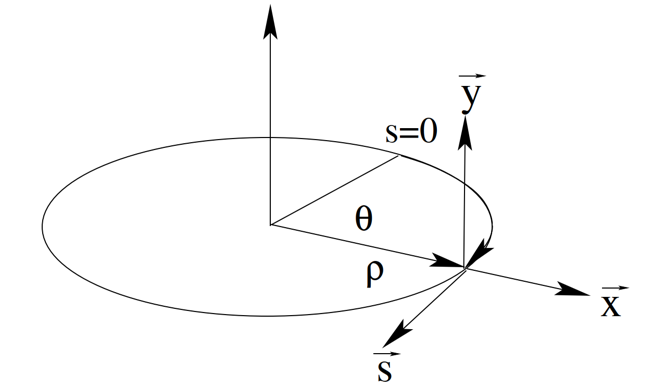}
   \end{center}
\caption{The local right-handed coordinate system $(x,y,s)$ used to describe the particle position in an accelerator}
\label{fig:coordinates}
\end{figure}

An accelerator is usually composed of a number of basic `cells'. The cell layouts of accelerators come in many variants (see, for example, Fig.~\ref{fig:acc-cell}); a cell usually contains some of the following magnetic elements: dipole magnets to bend the beams, quadrupole magnets to focus the beams, beam position monitors  to measure beam position, small dipole corrector magnets for beam steering, and sextupole magnets to control off-energy focusing. Magnets with higher field order are used in some machines to mitigate against collective effects (e.g.\ octupoles for Landau damping), to control the non-linear optics, or to compensate for higher-order field errors generated by imperfections in some of the main magnets. The coordinate system used for accelerators is shown in Fig.~\ref{fig:coordinates}.

The Lorentz force $\overrightarrow{F_L}$ due to a magnetic field $\overrightarrow{B}$ that acts on a particle with charge $q$ and speed $\overrightarrow{v}$ is given by
\begin{equation}
  \overrightarrow{F_L} = q  \overrightarrow{v} \times \overrightarrow{B}  \: .
  \label{eq:lorentz}
\end{equation}
The Lorentz force is always orthogonal to the direction of particle motion.

A dipole magnet is the simplest magnet with two magnetic poles that generate, ideally, a perfectly homogeneous magnetic field $B_0$. Dipole magnets are used to bend the beam on its reference path.  A~dipole corrector magnet, often also called an orbit corrector or steerer, is a small version of dipole magnet dedicated to steering the beam trajectory or closed orbit.

\begin{figure}[tbp]
  \begin{center}
\includegraphics[width=0.35\linewidth]{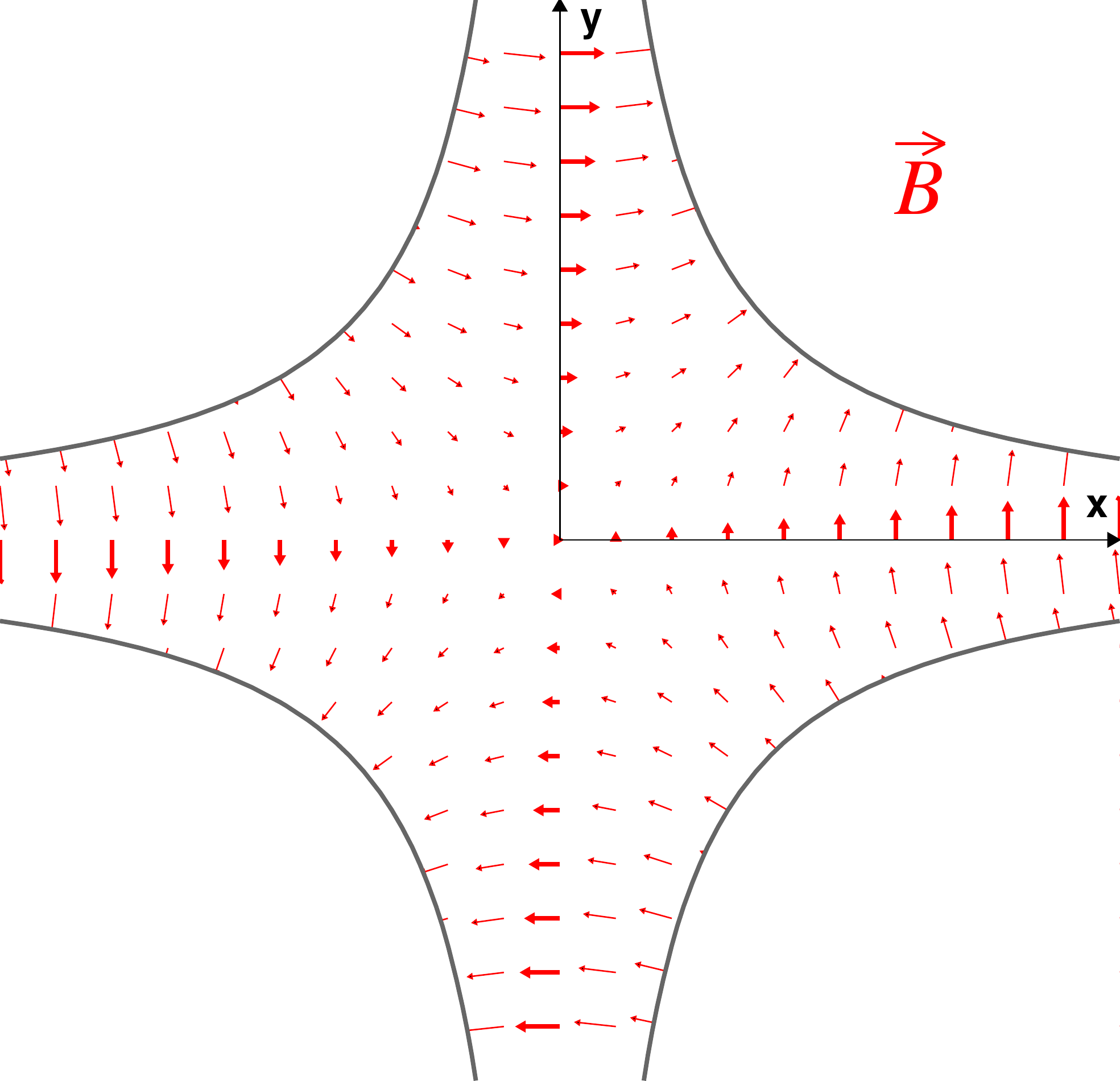} \hspace{10mm}
\includegraphics[width=0.35\linewidth]{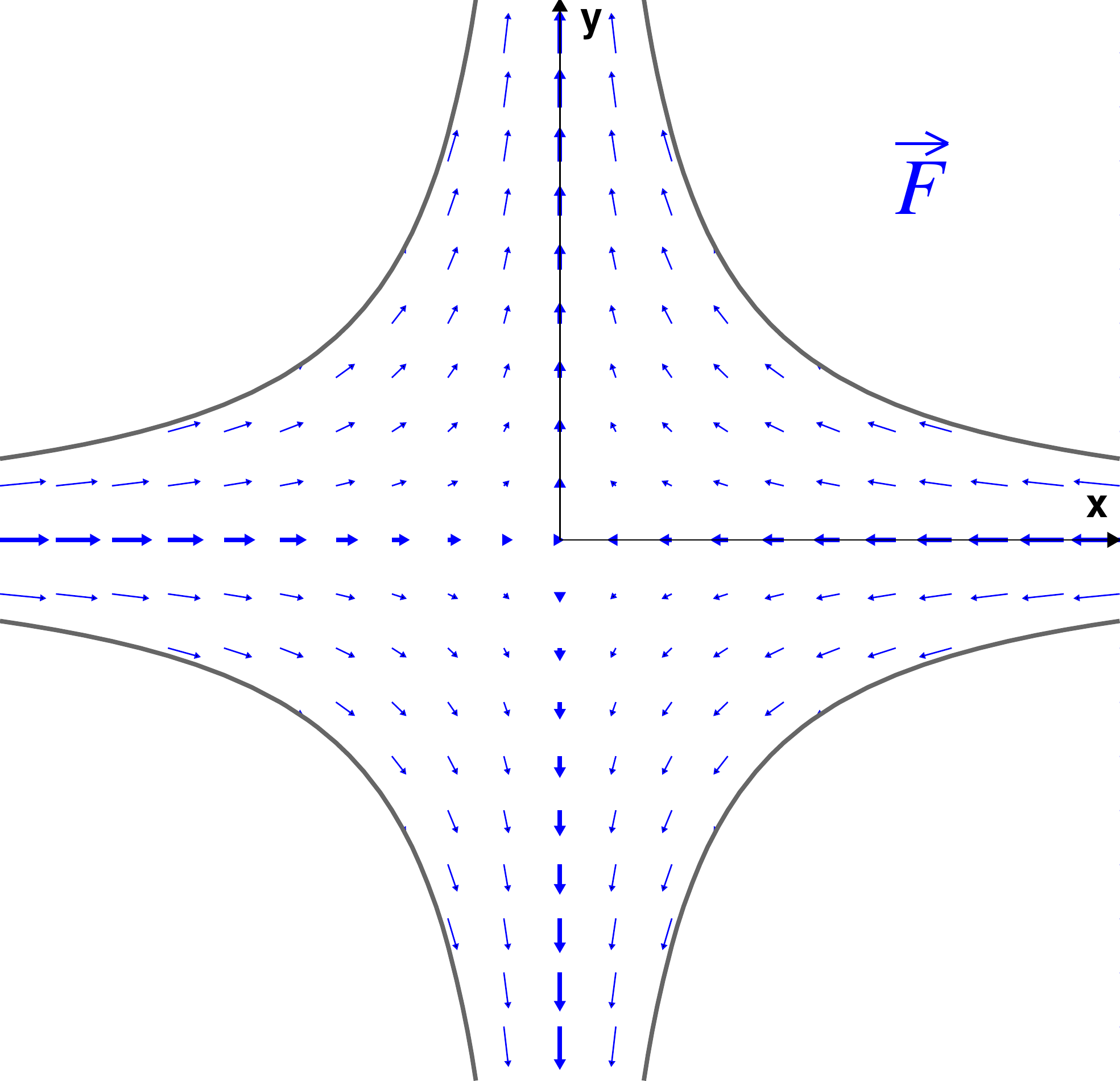}
   \end{center}
\caption{Schematic of a quadrupole magnet: magnetic field $\vec{B}$ (left) and Lorentz force $\vec{F}$  (right) on a particle with positive charge coming out of the plane of this page. Both the magnetic field and the force grow linearly with the distance from the quadrupole axis. On the axis there is no field (force). For this polarity the field is focusing in the horizontal plane (force direction points towards the axis) and defocusing in the vertical plane (force direction points away from the axis).}
\label{fig:quad-b-f}
\end{figure}

A quadrupole magnet has four magnetic poles of parabolic shape, as shown in Fig.~\ref{fig:quad-b-f}. It provides a field (force) that increases linearly with the distance from the quadrupole centre:
\begin{equation} \label{eq:quad}
  B_x = g y\:,   \qquad  B_y = g x \:,
\end{equation}
where $B_x$ and $B_y$ are the horizontal and vertical components of the magnetic field, respectively,  and $g=\partial B/\partial x$ is the quadrupole field gradient in units of tesla per metre [T$/$m]. The linear force gradient of a quadrupole provides focusing of the beam in one plane and defocusing in the other plane. The effect of a quadrupole is similar to that of an optical lens, except that optical lenses in general  (de)focus on both planes. The sign of the gradient $g$ defines the plane for which the quadrupole is focusing or defocusing; for a horizontally focusing quadrupole, by convention $g > 0$ in the coordinate system shown in Fig.~\ref{fig:coordinates}.

\begin{figure}[tbp]
  \begin{center}
\includegraphics[width=0.35\linewidth]{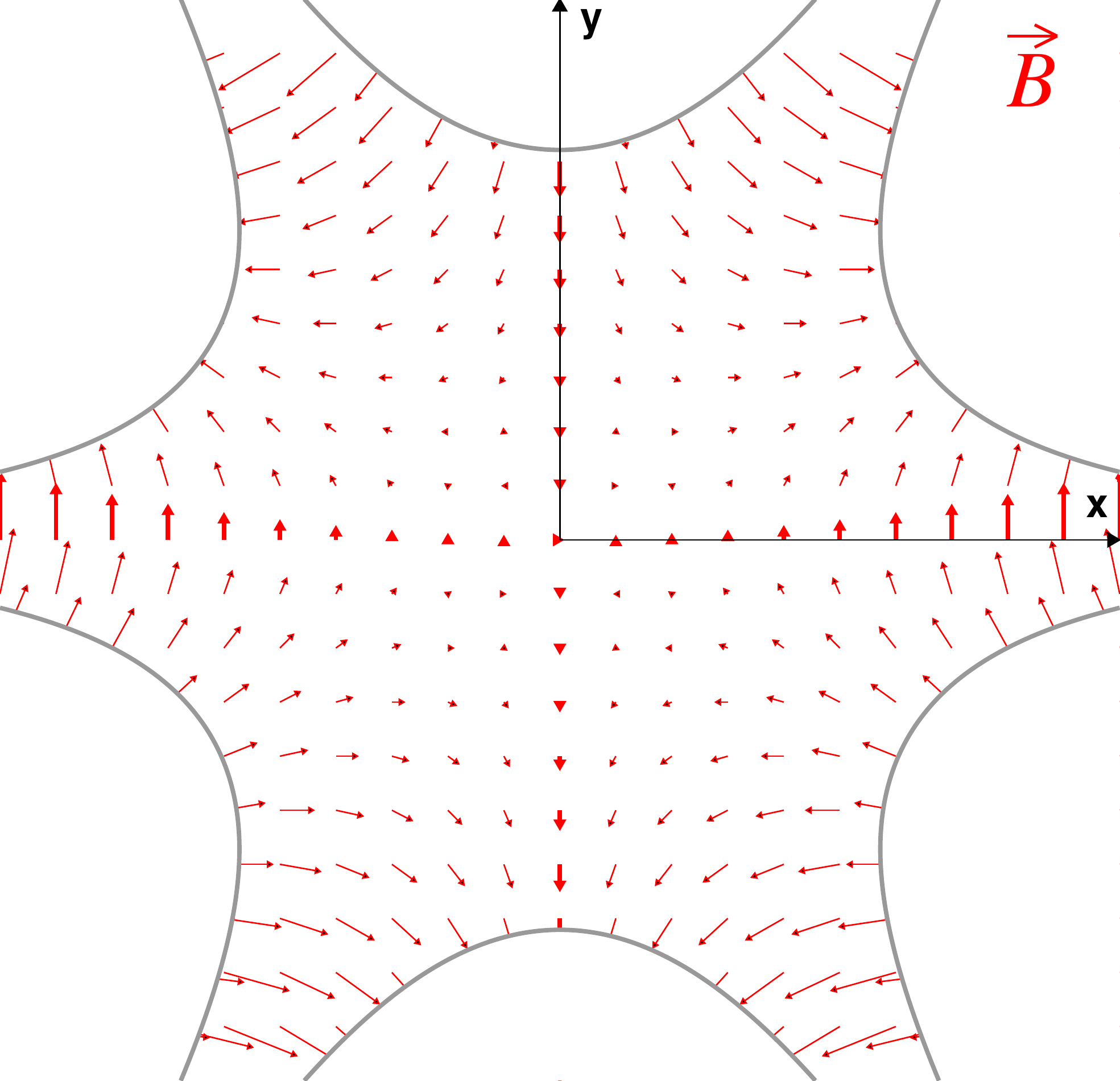} \hspace{10mm}
\includegraphics[width=0.35\linewidth]{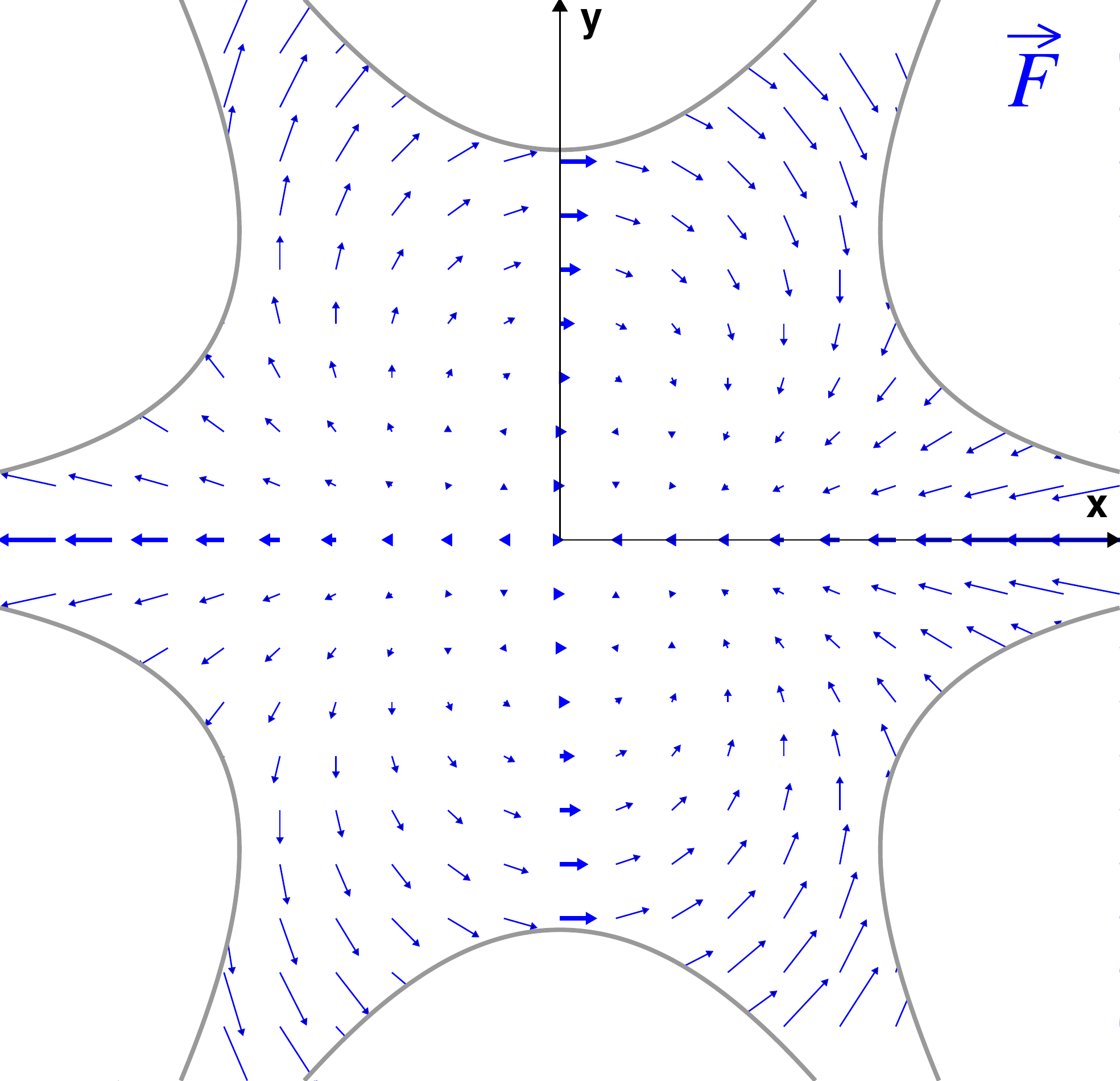}
   \end{center}
\caption{Schematic of a sextupole magnet: magnetic field $\vec{B}$ (left) and Lorentz force $\vec{F}$  (right) for a particle with positive charge coming out of the plane of this page.}
\label{fig:sext-b-f}
\end{figure}

Figure~\ref{fig:sext-b-f} depicts a sextupole magnet with a non-linear field given by
\begin{equation} \label{eq:sext}
  B_x = g' x y\: ,   \qquad B_y = \tfrac{1}{2}\, g' (x^2 - y^2)\:,
\end{equation}
where $g'=\partial^2 B/\partial x^2$ is the sextupolar gradient in units of tesla per metre squared [T$/$m$^2$].  Sextupoles are used to correct errors in the linear optics, for example the chromaticity (tune change with momentum) to be discussed later. Sextupoles generate linear optics errors through misalignments. A sextupole can be viewed as a quadrupole with a gradient that increases with distance from the axis.

Accelerator lattice designers and modelling programs usually parametrize quadrupole and sextupole strengths by normalized strengths $K_1$ and $K_2$. For a particle with elementary charge $e$ and momentum $p$, the normalized quadrupole gradient $K_1$ (in units of [1$/$m$^2$]) is defined as
\begin{equation} \label{eq:k1}
  K_1 = \frac{e g}{p} = 0.2998 \,\frac{g \,\mathrm{[T/m]}}{p \,\mathrm{[GeV/\mathit{c}]}} \;.
\end{equation}
The focal length $f$ of a quadrupole is given by $1/f = K_1 l$ where $l$ is the quadrupole length.
Similarly, the normalized sextupole gradient $K_2$ (in units of [1$/$m$^3$]) is defined  as
\begin{equation} \label{eq:k2}
  K_2 = \frac{e g'}{p} = 0.2998 \,\frac{g' \,\mathrm{[T/m^2]}}{p \,\mathrm{[GeV/\mathit{c}]}} \;.
\end{equation}

\begin{figure}[b]
  \begin{center}
\includegraphics[width=0.8\linewidth]{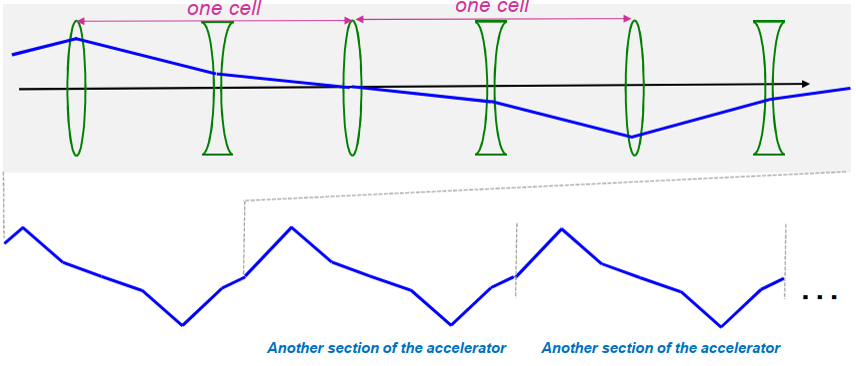}
   \end{center}
\caption{Particle motion in an accelerator lattice composed of alternating focusing and defocusing quadrupoles}
\label{fig:particle-in-lattice}
\end{figure}

\begin{figure}[tbp]
  \begin{center}
\includegraphics[width=0.8\linewidth]{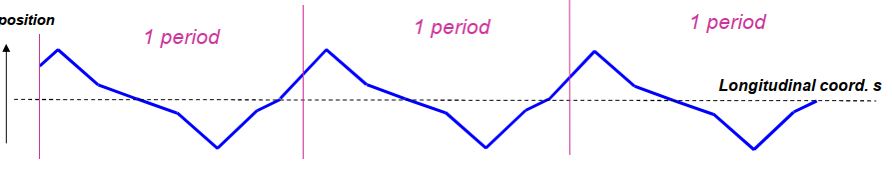} \\
\includegraphics[width=0.8\linewidth]{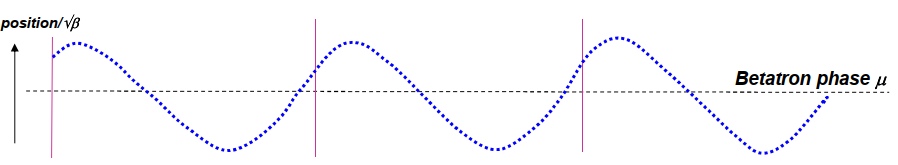}
   \end{center}
\caption{Particle motion in an accelerator lattice in physical units $x(s)$ and $s$ (top), and the same motion in normalized coordinates $x(s)/\sqrt{\beta(s)}$ and $\mu(s)$ (bottom); the coordinate transformation generates a periodic sinusoidal oscillation.}
\label{fig:particle-in-lattice-norm}
\end{figure}

Figure~\ref{fig:particle-in-lattice} represents schematically the movement of a particle in the accelerator lattice. The lattice quadrupoles  act as focusing and defocusing lenses and make the particle bounce back and forth in the transverse directions. In this example the lattice is perfectly periodic and composed of three identical periods. The motion is represented in physical units corresponding to the transverse position $x(s)$ (or~$y(s)$) and the longitudinal coordinate $s$ in Fig.~\ref{fig:particle-in-lattice}. The number of oscillation periods for one turn of the machine is called the machine tune ($Q$) or betatron tune; in this example $Q$ is around 2.75, i.e.\ two periods plus three-quarters of a period. It is possible to change the coordinates by replacing the longitudinal position $s$ (in metres) with the betatron phase advance $\mu(s)$ (in degrees per radian) and normalizing the transverse position by the betatron function $\beta(s)$, $x(s) \Rightarrow x(s)/\sqrt{\beta(s)}$. In this normalized coordinate system the oscillation is transformed into a sinusoidal motion with constant amplitude, the betatron oscillation
\begin{equation} \label{eq:norm-beta}
  \frac{x(s)}{\beta(s)} = A \sin(\mu(s) + \mu_0) \:,
\end{equation}
which is often more convenient (and simpler) for analysing the beam motion in those coordinates (see Fig.~\ref{fig:particle-in-lattice-norm}); here  $A$ is the invariant amplitude of the motion.  The betatron function $\beta(s)$ defines the beam envelope, while the betatron phase defines the phase of the betatron oscillation.

\section{From model to reality}

\begin{figure}[hbt]
  \begin{center}
\includegraphics[width=0.8\linewidth]{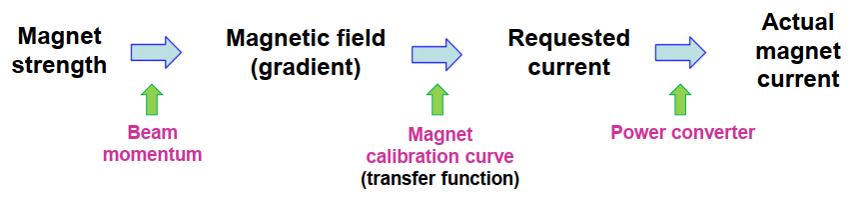}
   \end{center}
\caption{From model to reality in the case of a magnetic strength, for example $K_1$ of Eq.~\eqref{eq:quad}}
\label{fig:model-I}
\end{figure}

The machine model defined by the accelerator designer must be converted into electromagnetic fields and eventually into currents for the power converters that feed the magnet circuits. The model positions must be translated into physical positions in the accelerator tunnel.
Imperfections (i.e.\ errors) are introduced when the model is transferred to the real machine and can arise from mechanical tolerances, measurement uncertainties, etc. For magnetic fields the errors may be  due to imprecise knowledge of the beam momentum, magnet measurement uncertainties, and  power converter regulation, as shown in Fig.~\ref{fig:model-I}. Alignment errors of accelerator components are another common source of imperfections. To ensure that the accelerator elements are in the correct positions,  their alignment in the tunnel must be very accurate, at sub-micrometre level  for certain linear collider variants such as CLIC \cite{CLIC-DR}, to microns or tens of microns in synchrotron light sources, or to roughly 100 microns in hadron accelerators.
The alignment process for a magnet involves first the precise determination of the magnetic axis in the laboratory with reference to the element alignment markers used by survey teams, and then the precise in-situ alignment (position and angle) of the element in the tunnel.

As a consequence of imperfections, the actual accelerator may differ from the model to an extent that the accelerator may not function well or even at all. For example, the beam may not circulate due to misalignments, or the optics may be incorrect due to field and alignment errors. Over the past few decades, instruments and tools have been developed to measure and correct accelerator parameters in control rooms and to restore design models or update the actual machine model. In many cases the tools are applied iteratively when an accelerator is bootstrapped and commissioned.
Here we give an overview of imperfections that affect the linear machine optics and discuss how to measure and correct them.

\begin{figure}[tbp]
  \begin{center}
\includegraphics[width=0.6\linewidth]{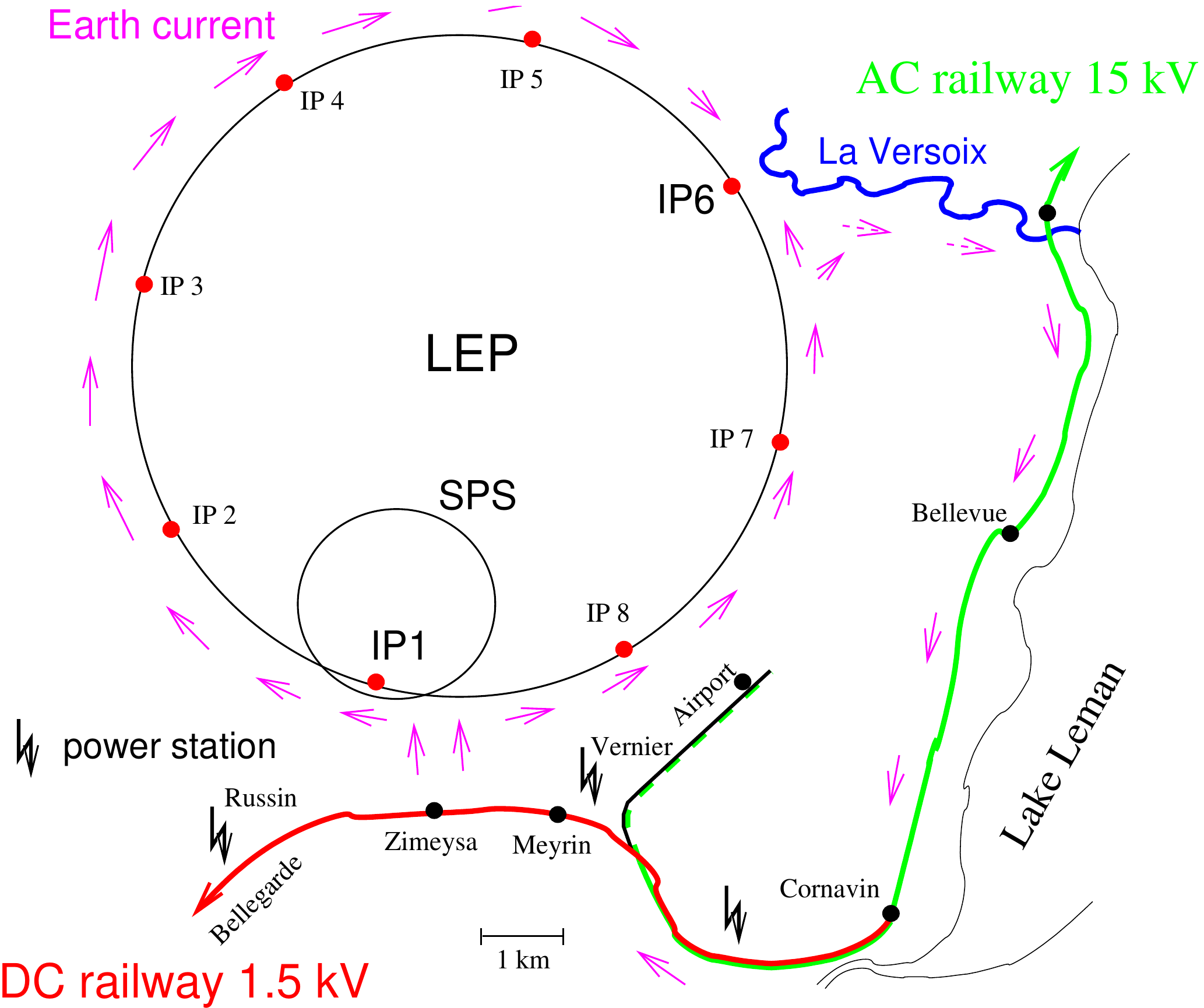}  \\
\vspace{8mm}
\includegraphics[width=0.7\linewidth]{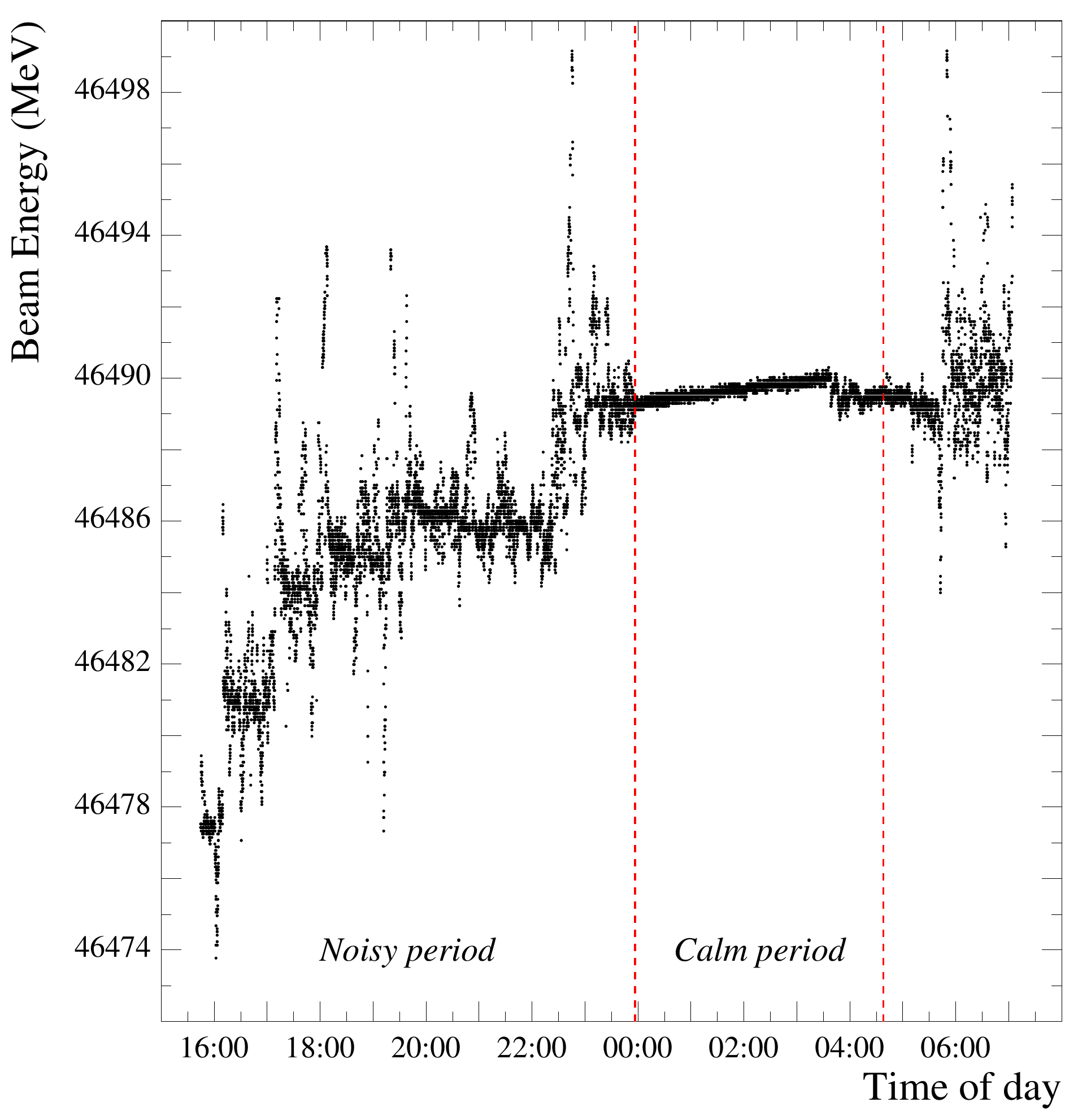}
   \end{center}
\caption{The path of earth currents flowing over the LEP vacuum chamber that were generated by the DC railway line near Geneva (top). The currents generated by the railway line flowed back over the LEP vacuum chamber, where they generated a slow magnetic field rise that was monitored by NMR probes installed in a sample of dipole magnets (bottom). The quiet period corresponds to nighttime, when there are few trains circulating over the European railway network.}
\label{fig:lep-energy}
\end{figure}

\section{Beam momentum}

The momentum $p$ of a particle of charge $q$ in a storage ring is defined by the integral of the bending field along the beam orbit~\cite{LEP1-ENERGY, LHC-ENERGY}:
\begin{equation}\label{eq:energy}
  p = \frac{q}{2 \pi} \oint B(s)\,\rd s = Z \times 47.7 \,[\mathrm{MeV/\mathit{c
}/T\,m}] \; \oint B(s)\,\rd s  \: ,
\end{equation}
with $q = Ze$, where $e$ is the elementary charge of the electron.

Energy errors may arise from incorrect dipole field settings, for example due to a calibration error of the main dipole field. External sources may also influence the momentum, as was the case for the Large Electron--Position Collider (LEP), where the magnetic bending was creeping up by $\Delta p/p \simeq 10^{-4}$ due to earth currents generated by a DC railway line~\cite{LEP1-ENERGY}; see Fig.~\ref{fig:lep-energy}. Earth currents of 1--2\,A flowed over the LEP vacuum chamber, where they induced a slow rise in the  dipole field triggered by the current spikes visible on NMR field probes. Many of the spikes were induced by trains running on French railways far from the Geneva area. For LEP the impact was significant mainly because the dipole field was very low, around 50--100\:mT, compared to most accelerators, which operate in the range of teslas. In addition the beam energy had to be known to the level of approximately  $10^{-5}$ for the experimental physics programme.

\begin{figure}[tbp]
  \begin{center}
\includegraphics[width=0.7\linewidth]{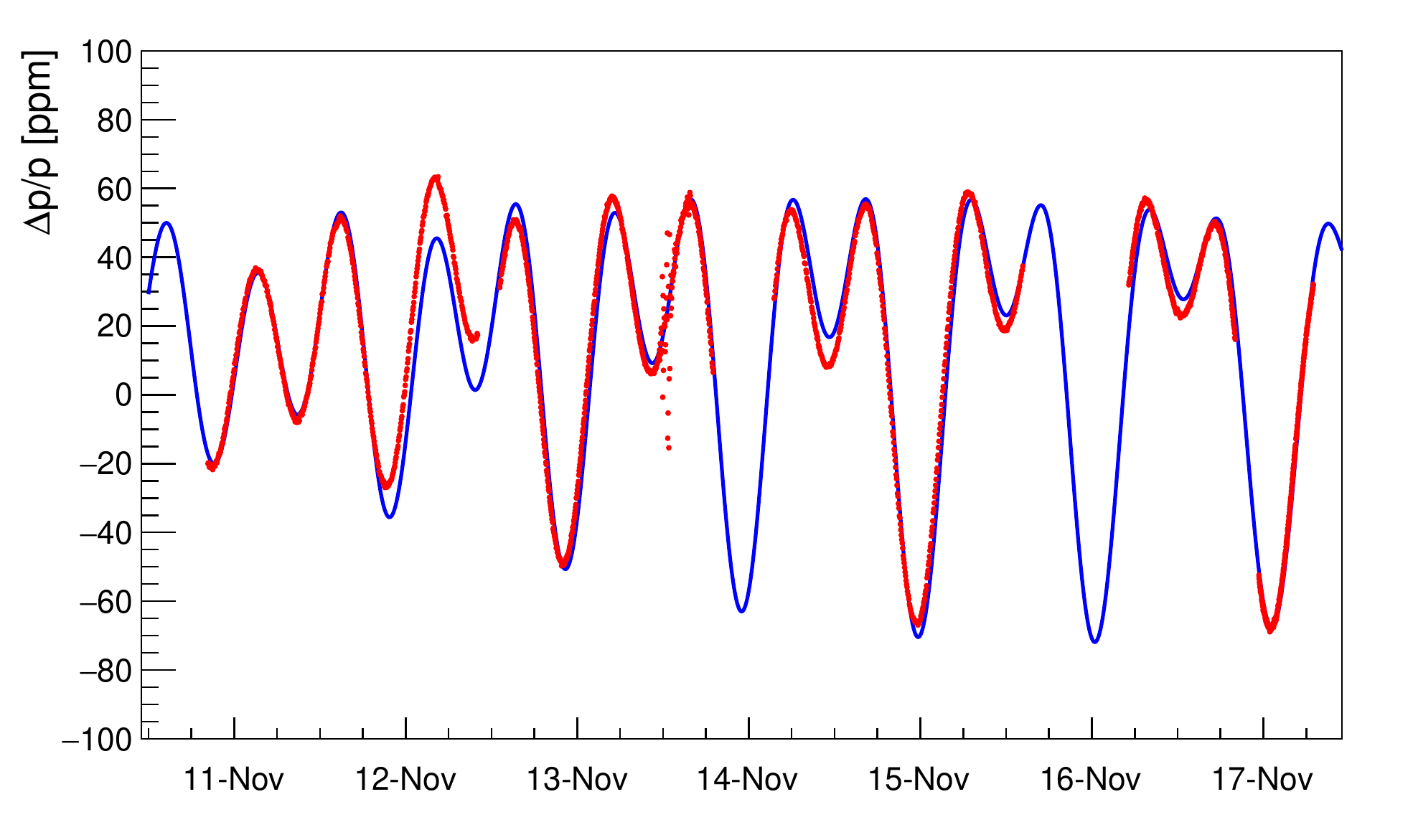}
   \end{center}
\caption{Relative beam momentum variation at the LHC (expressed in parts per million [ppm]) over a period of almost one week in November 2016 due to the influence of terrestrial tides \cite{LHC-ENERGY}. The tides modulate the circumference of the LHC by around 1\,mm peak-to-peak for a circumference of 26.7\,km. The circumference modulation results in a modulation of the momentum. The blue line is a prediction, and  the red points are measurements.}
\label{fig:tides}
\end{figure}

Alternatively, the field integral in Eq.~\eqref{eq:energy} may also change if the orbit length is not correct and the beam is not centred on average in the quadrupoles. The energy error introduced by an orbit length error $\Delta L$ (or radio frequency error $\Delta f_{\rm RF}$) is given by
\begin{equation}\label{eq:energy-redial}
  \frac{\Delta p}{p} = \frac{1}{\alpha} \frac{\Delta L}{L} = - \frac{1}{\alpha - 1/\gamma^2} \frac{\Delta f_{\rm RF}}{f_{\rm RF}}\:,
\end{equation}
where $\alpha$ is the momentum compaction factor; for large machines $\alpha$ is in the range of $10^{-3}$ to $10^{-5}$. Orbit length errors arise from incorrect setting of the radio frequency or from  geological changes in the machine circumference. Circumference changes due to terrestrial tides were observed and monitored at the LEP \cite{LEP1-ENERGY}, where their effect on the beam energy was at the level of a few multiples of $10^{-4}$. Figure~\ref{fig:tides} displays an observation of tides  inducing momentum variations at the Large Hadron Collider (LHC), which is now installed in the former 26.7\,km  LEP tunnel~\cite{LHC-ENERGY}.

\section{Orbit and dispersion}

The next category of imperfections is defined by the presence of an unintended deflection along the path of the beam. Such errors are in general the first ones encountered when a beam is initially injected into an accelerator. Dipole orbit correctors are added to the accelerator lattice to compensate for the effects of unintended deflections. An orbit corrector is used to generate a deflection of opposite sign and amplitude to compensate as locally as possible for an unintended deflection.

\begin{figure}[tbp]
  \begin{center}
\includegraphics[width=0.7\linewidth]{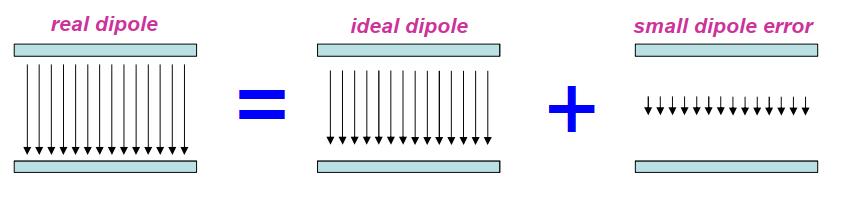} \\
\vspace{5mm}
\includegraphics[width=0.7\linewidth]{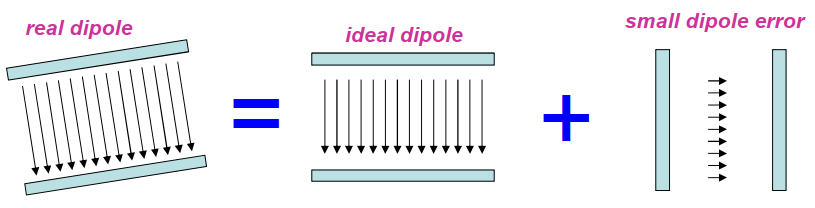} \\
\vspace{5mm}
\includegraphics[width=0.7\linewidth]{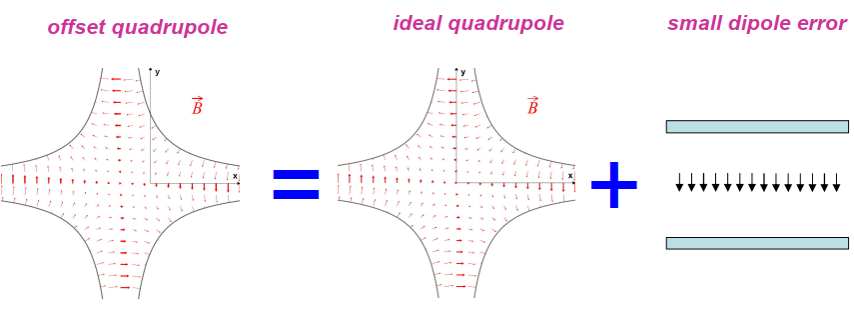}
   \end{center}
\caption{Sources of undesired deflections: dipole magnet calibration errors (top), dipole magnet rotation (middle), and misaligned quadrupoles (bottom).}
\label{fig:b1-errors}
\end{figure}

The first source of undesired deflections is a field error (deflection error) of a dipole magnet. This can be due to an error in the magnet current or in the calibration table that defines the current required for a given field. The latter kind of error may be introduced through a limited field measurement or limited magnetic model accuracy. The imperfect dipole can be represented as a perfect dipole plus a small dipole error. A small rotation of a dipole magnet has the same effect, but the field error appears in the orthogonal plane as shown in Fig.~\ref{fig:b1-errors}. The second source of undesired deflections is misalignment of a quadrupole magnet. The magnetic field of a quadrupole shifted by $x_0$ in the horizontal plane becomes
\begin{equation} \label{eq:quad-offset}
  B_x = g y \:,  \qquad B_y = g (x - x_0) = g x - g x_0\:,
\end{equation}
with an additional constant term $B_y = - g x_0$. This term corresponds to a constant vertical field, which generates a horizontal deflection of all particles of the beam. The misaligned quadrupole can therefore be represented as a perfectly aligned quadrupole plus a small undesired dipole.

\subsection{Closed orbit}

\begin{figure}[tbp]
  \begin{center}
\includegraphics[width=0.3\linewidth]{\picturefolder/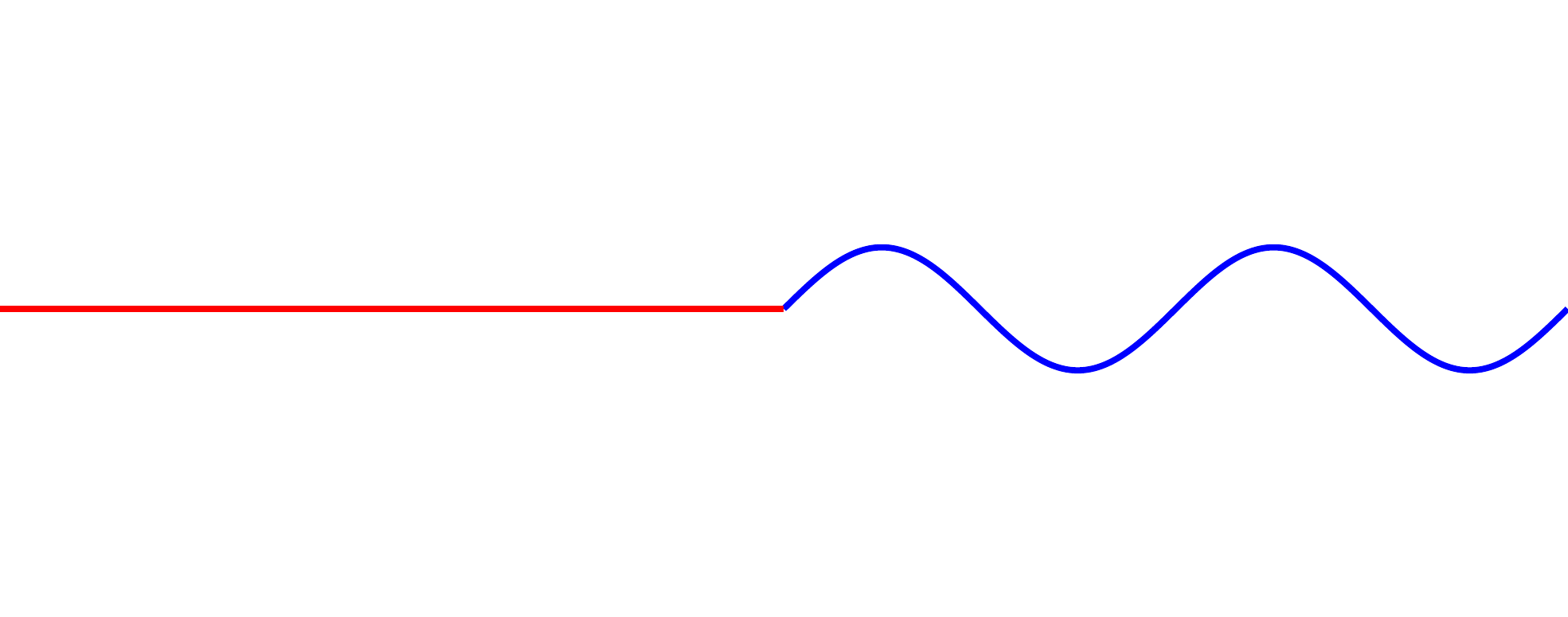} \hspace{2cm} \includegraphics[width=0.3\linewidth]{\picturefolder/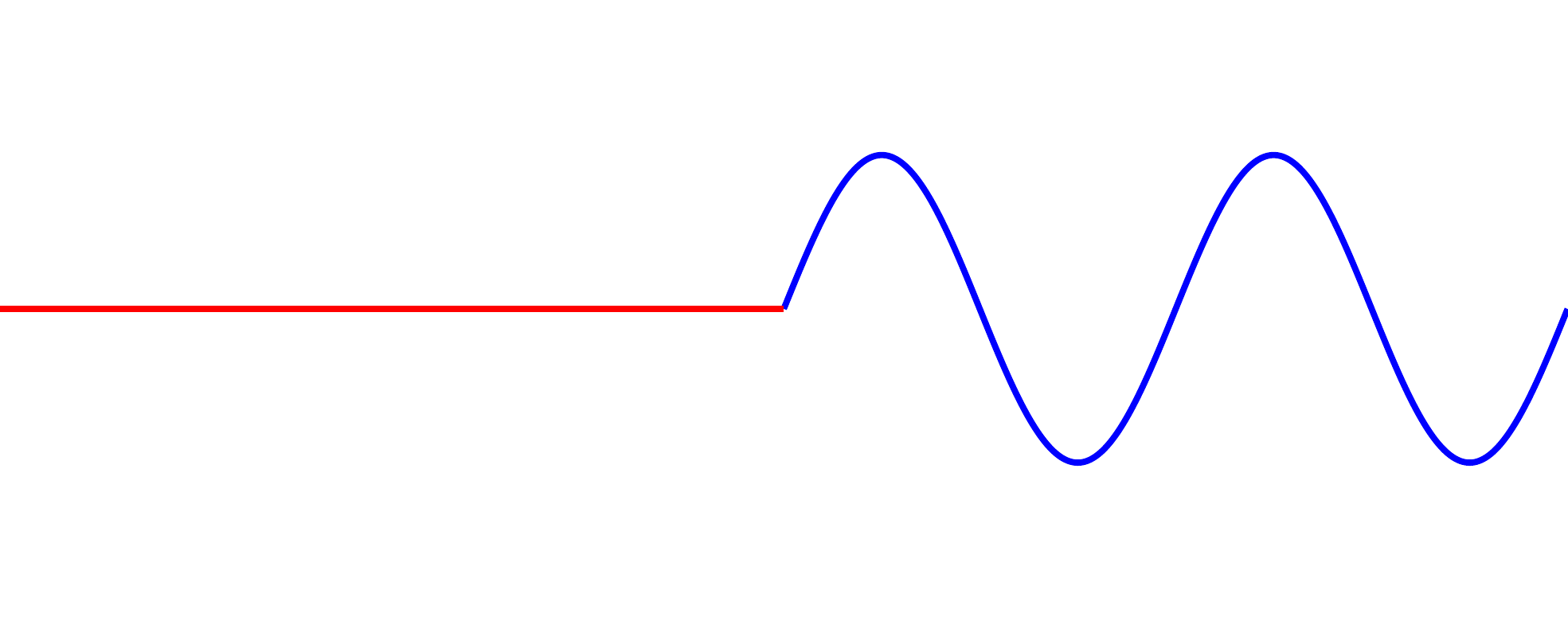} \\
\includegraphics[width=0.3\linewidth]{\picturefolder/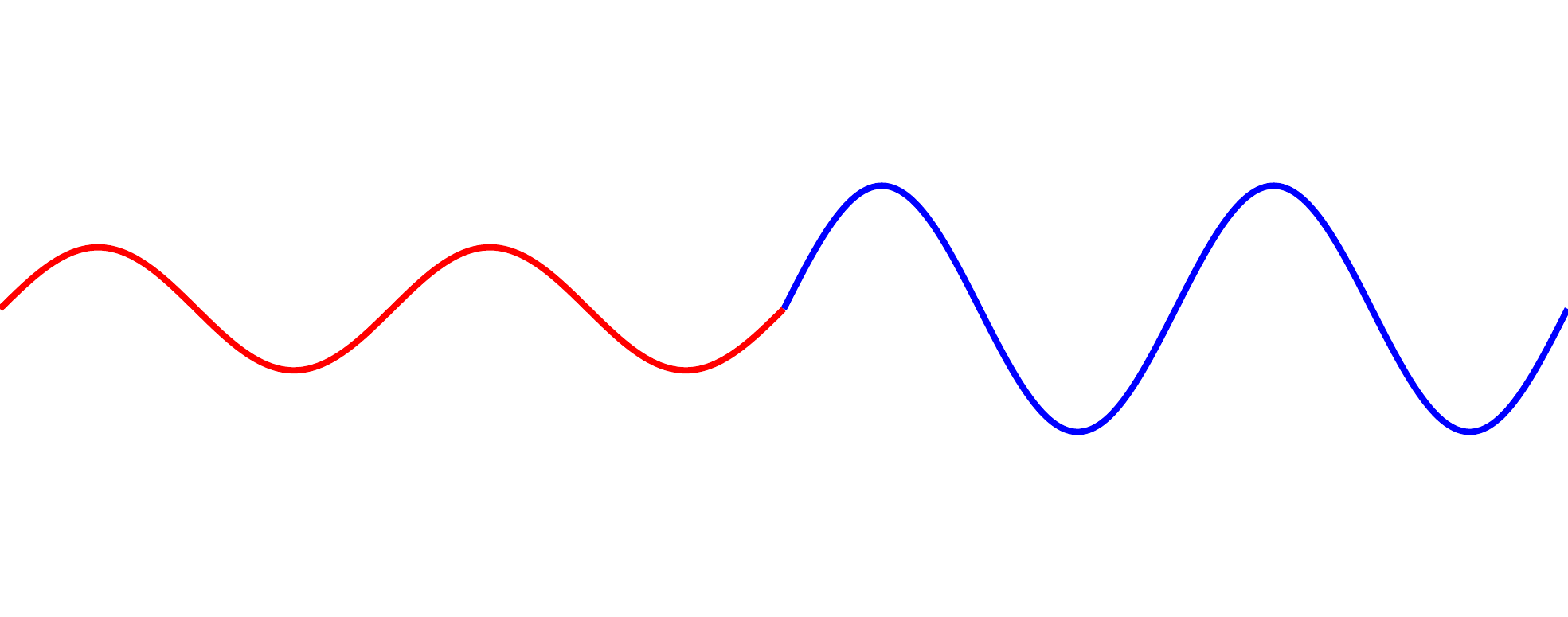} \hspace{2cm}
\includegraphics[width=0.3\linewidth]{\picturefolder/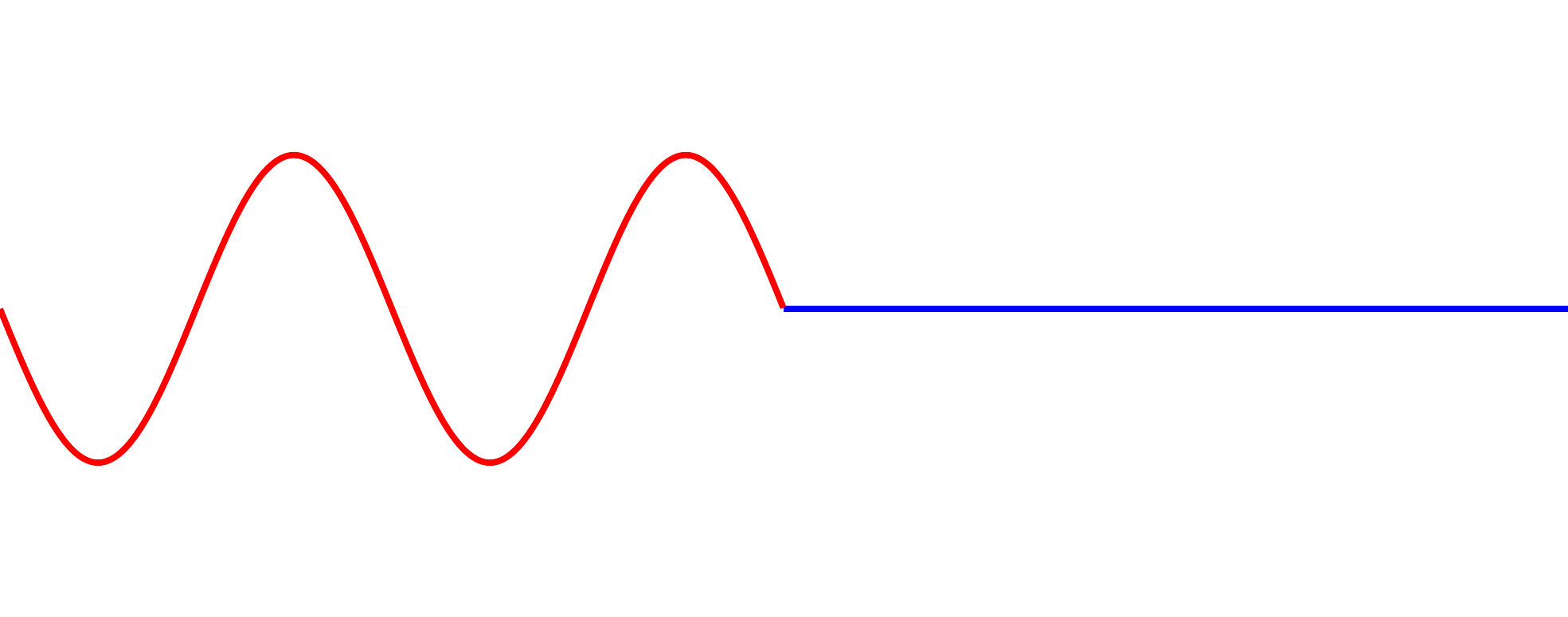} \\
\includegraphics[width=0.3\linewidth]{\picturefolder/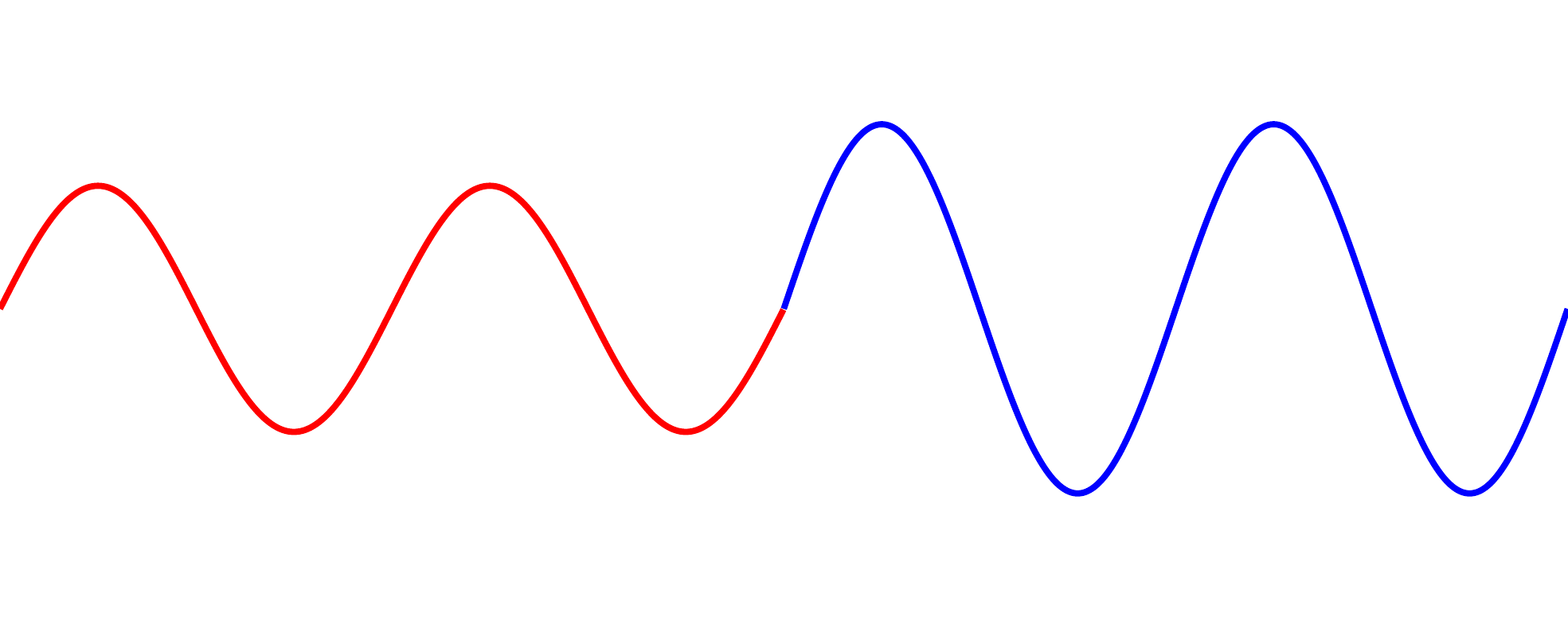} \hspace{2cm}
\includegraphics[width=0.3\linewidth]{\picturefolder/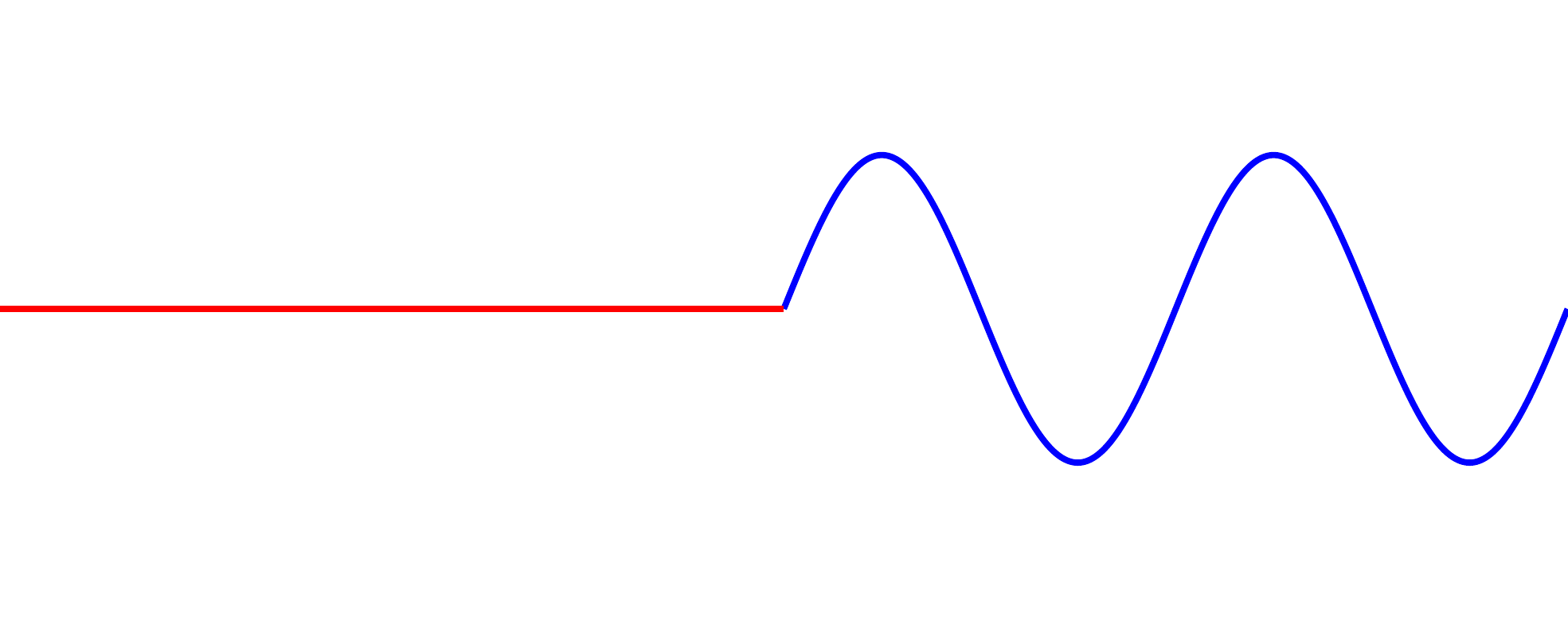} \\
\includegraphics[width=0.3\linewidth]{\picturefolder/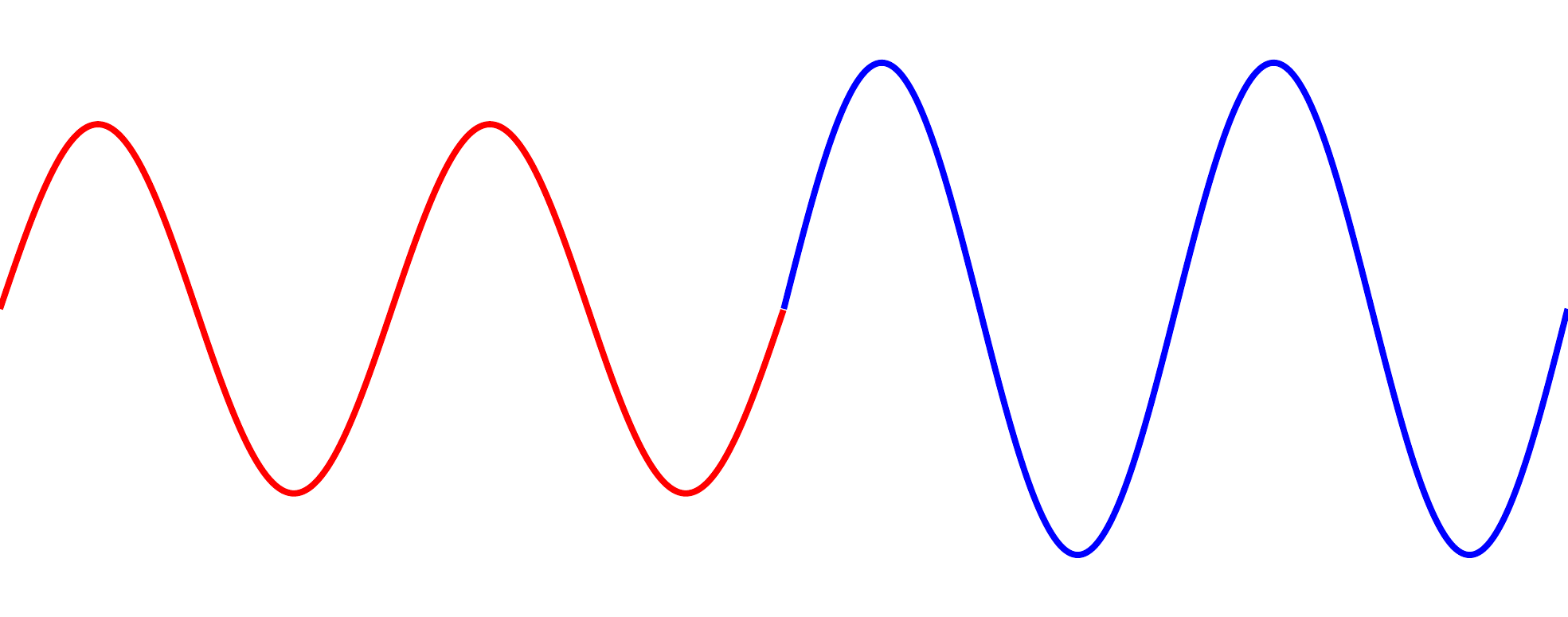} \hspace{2cm}
\includegraphics[width=0.3\linewidth]{\picturefolder/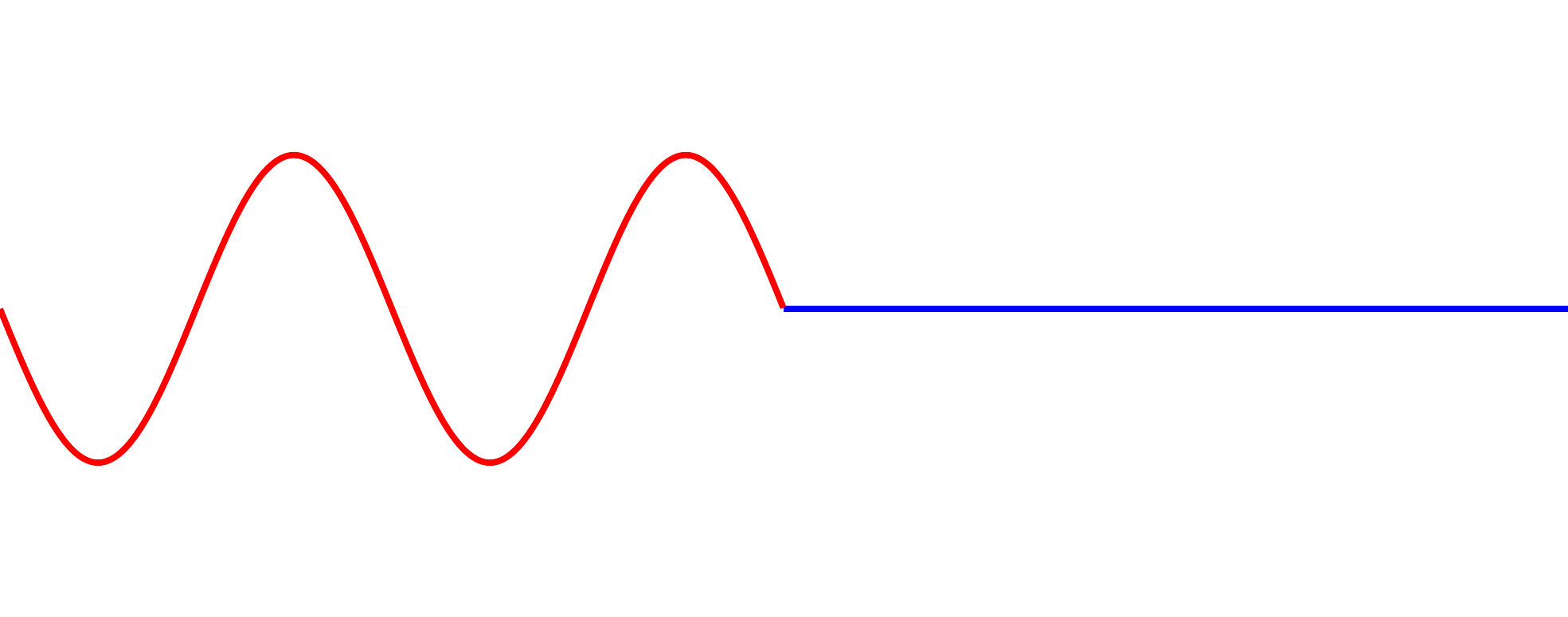}
   \end{center}
\caption{Beam trajectory in a circular accelerator on four consecutive turns for an integer tune $\tune = \inttune$ (left column) and for a half-integer tune $\tune = 0.5 + \inttune$  (right column), where $\inttune \in \mathbb{N}$. The beam comes in from the left and encounters the deflection at the point where the line colour changes from red to blue. After the kick, the beam propagates and comes back on the next turn from the left on the next line. For an integer tune, the deflections add up coherently on each turn, so the amplitude grows without limit. For a half-integer tune, the trajectory changes from turn to turn but  does not diverge.}
\label{fig:co-tuneN}
\end{figure}

To illustrate in simple terms the impact of an undesired deflection on the beam trajectory in a circular machine, a simple model is presented in Fig.~\ref{fig:co-tuneN}. For a machine with an integer tune value, $Q = N$ where $N \in \mathbb{N}$, the deflections add up on every turn. Thus the amplitude diverges and the particles do not stay within the accelerator vacuum chamber. The divergence is due to a resonance, the integer resonance, which occurs when $Q = N$ where $N \in \mathbb{N}$. A circular machine cannot be operated with such a tune value. On the other hand, the trajectory remains bounded for a half-integer tune, $Q = 0.5 + N$ where $N \in \mathbb{N}$, because the deflections are compensated for on every other turn.

\begin{figure}[tbp]
  \begin{center}
\includegraphics[width=0.8\linewidth]{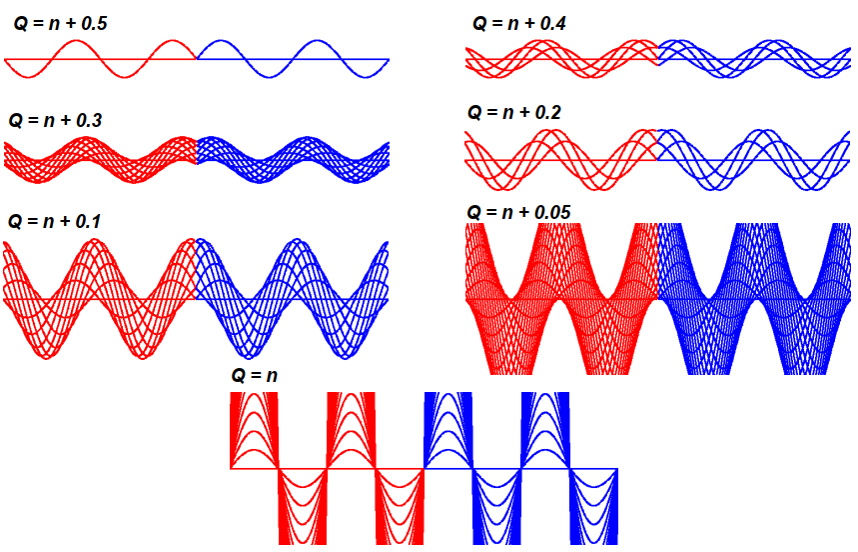}
   \end{center}
\caption{Superposition of 50 consecutive turns (see also Fig.~\ref{fig:co-tuneN}) for fractional tune values $q= 0.5$, 0.4, 0.3, 0.2, 0.1, 0.05, and 0. The vertical scale is the same for all values of~$q$.}
\label{fig:traj-tune}
\end{figure}

To elucidate the influence of the tune on the beam orbit, the first 50 turns are superposed in Fig.~\ref{fig:traj-tune} as a function of  the fractional tune value, $q$, such that $Q = q + N$ where $N \in \mathbb{N}$. Figure~\ref{fig:traj-tune} highlights the fact that the particle oscillates around a mean value which depends on $q$, and the amplitudes diverge as $q$ approaches 0 (i.e.\ the integer resonance).
The stable mean value around which the particles oscillate, visible in Fig.~\ref{fig:traj-tune}, is called the closed orbit.
Every particle in the beam performs betatron oscillations around the closed orbit. The general expression for the closed orbit $x(s)$ in the presence of a deflection $\theta$ is
\begin{equation}
  x(s) = \frac{ \sqrt{\beta(s) \beta_{\theta}} \:\cos(| \mu(s) - \mu_{\theta} | - \pi Q)\, \theta}{2 \sin(\pi Q)}  \;.
\end{equation}
In this expression $\beta(s)$ and $\mu(s)$ are the betatron function and phase advance at the position $s$, $Q$ is the tune, and  $\beta_{\theta}$ and $\mu_{\theta}$ are the betatron function and phase advance at the location of the kick~$\theta$.

\begin{figure}[tbp]
  \begin{center}
\includegraphics[width=0.28\linewidth]{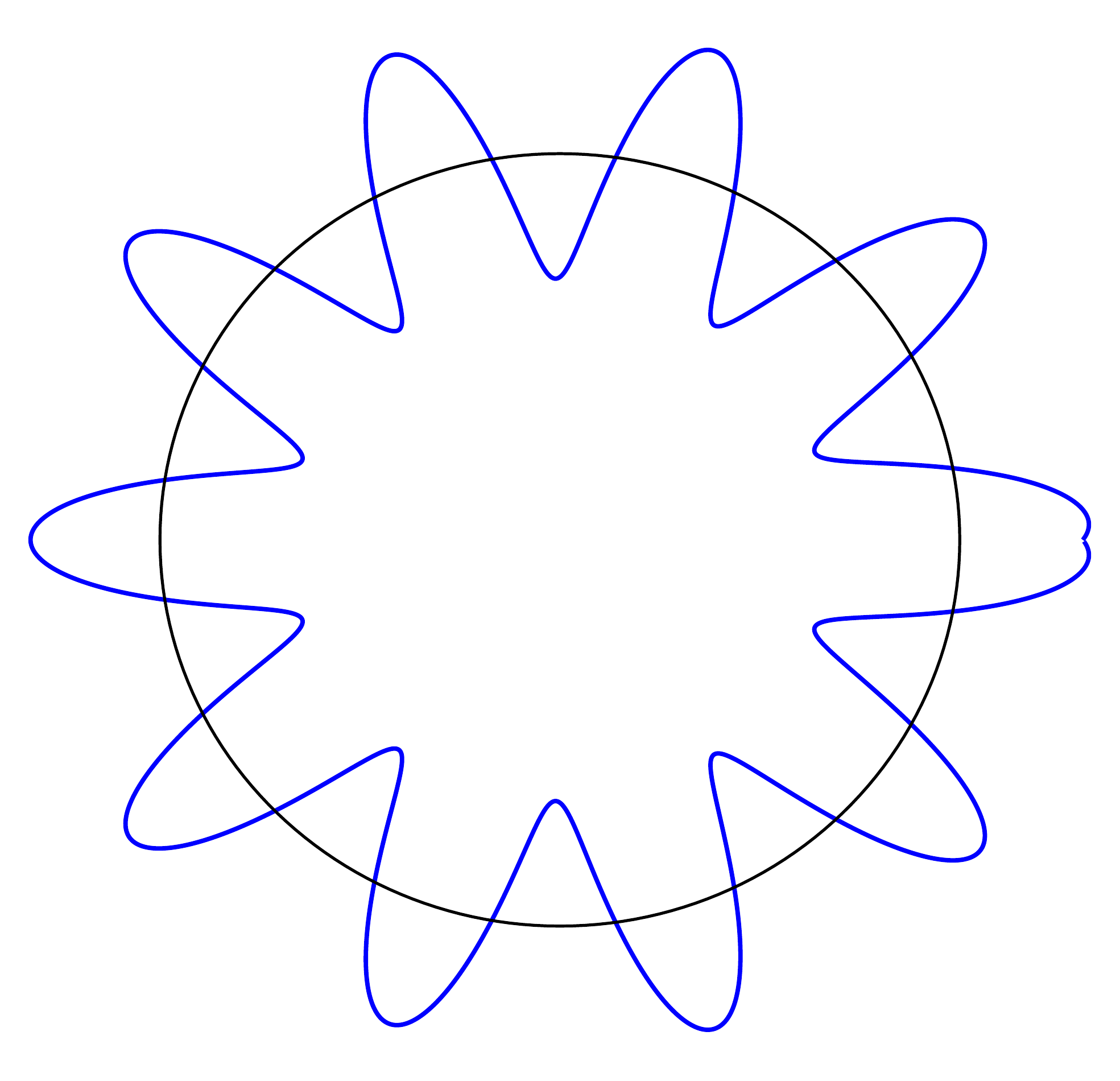}
\includegraphics[width=0.28\linewidth]{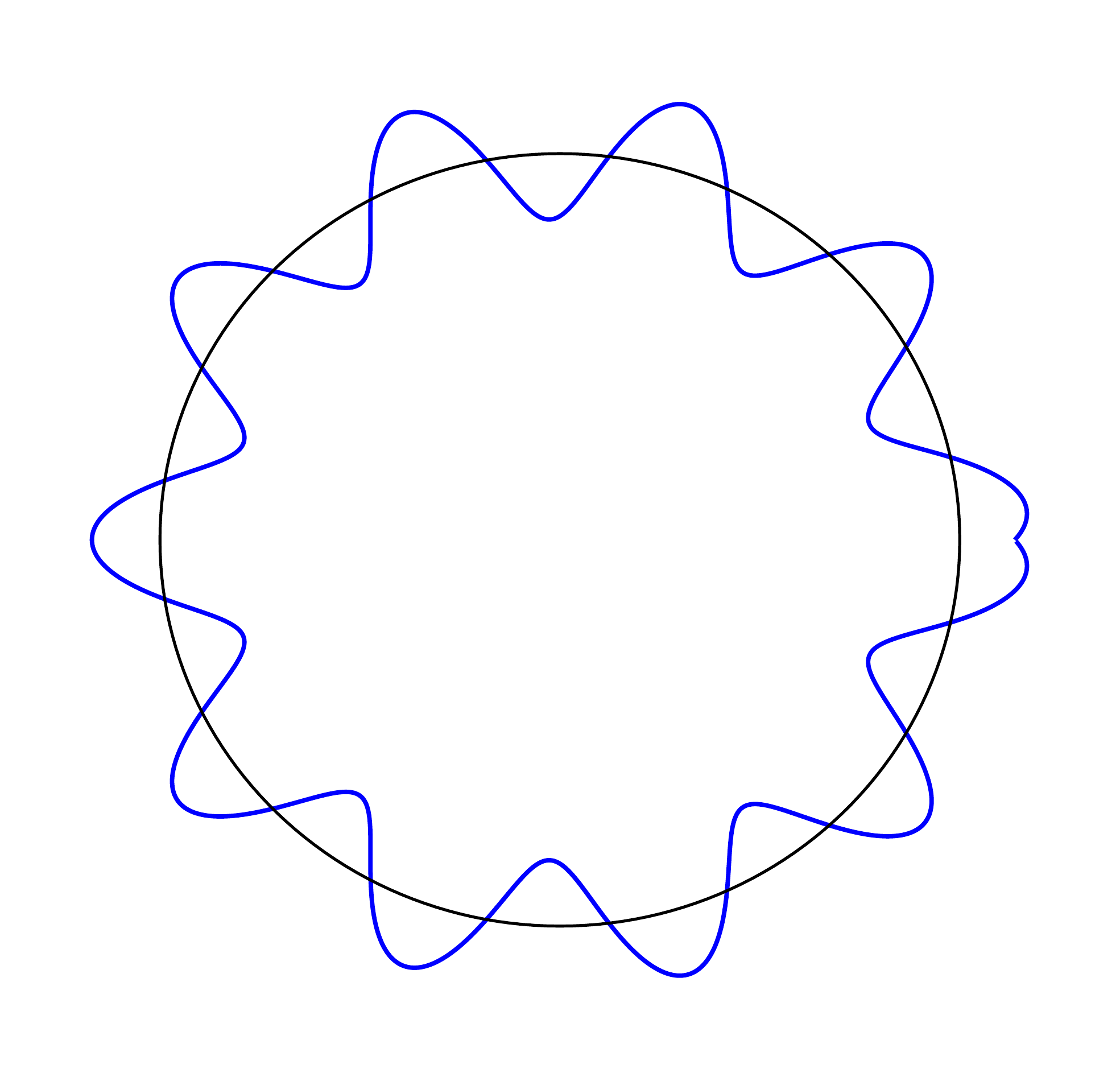}
\includegraphics[width=0.28\linewidth]{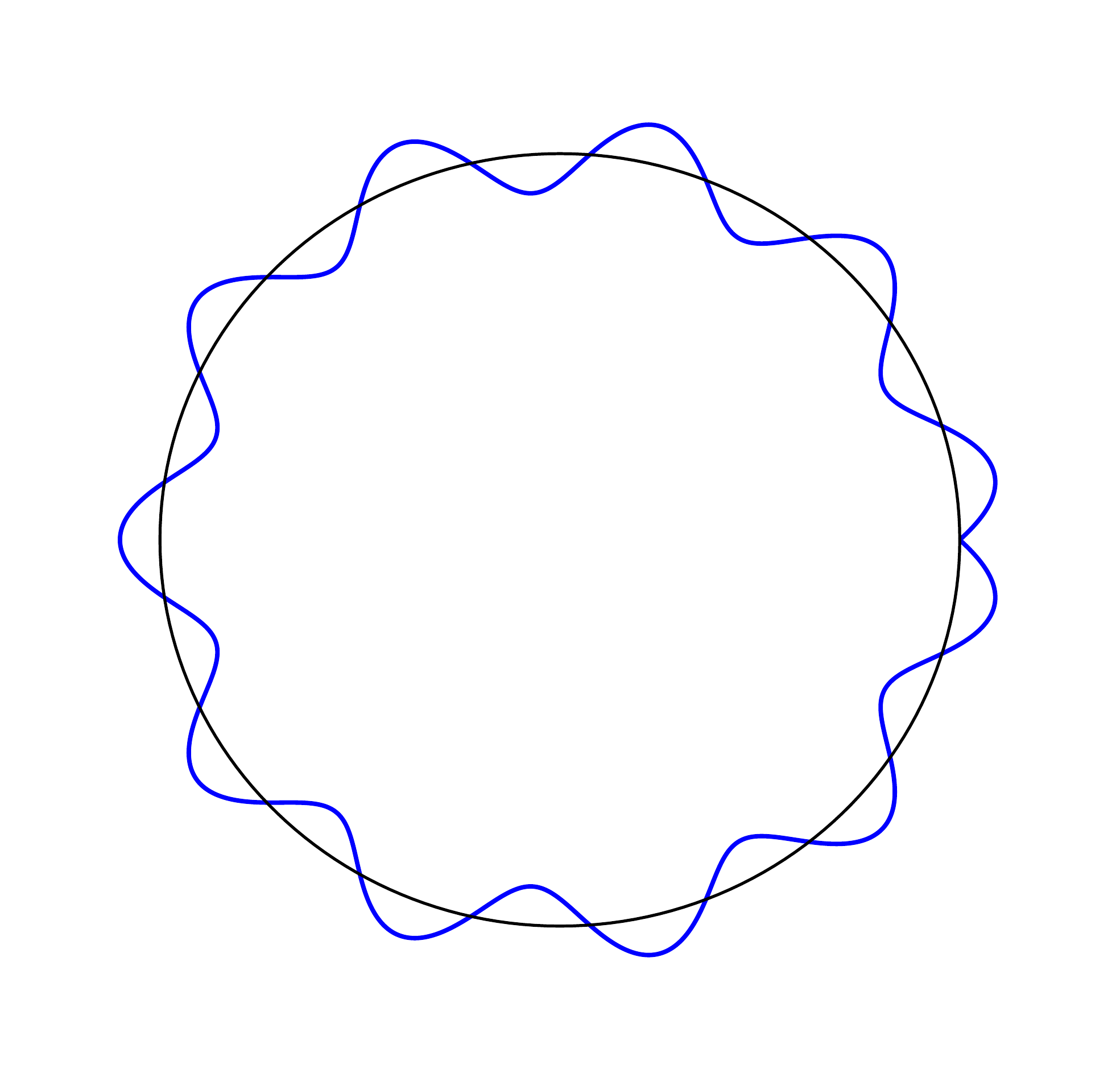} \\
\includegraphics[width=0.28\linewidth]{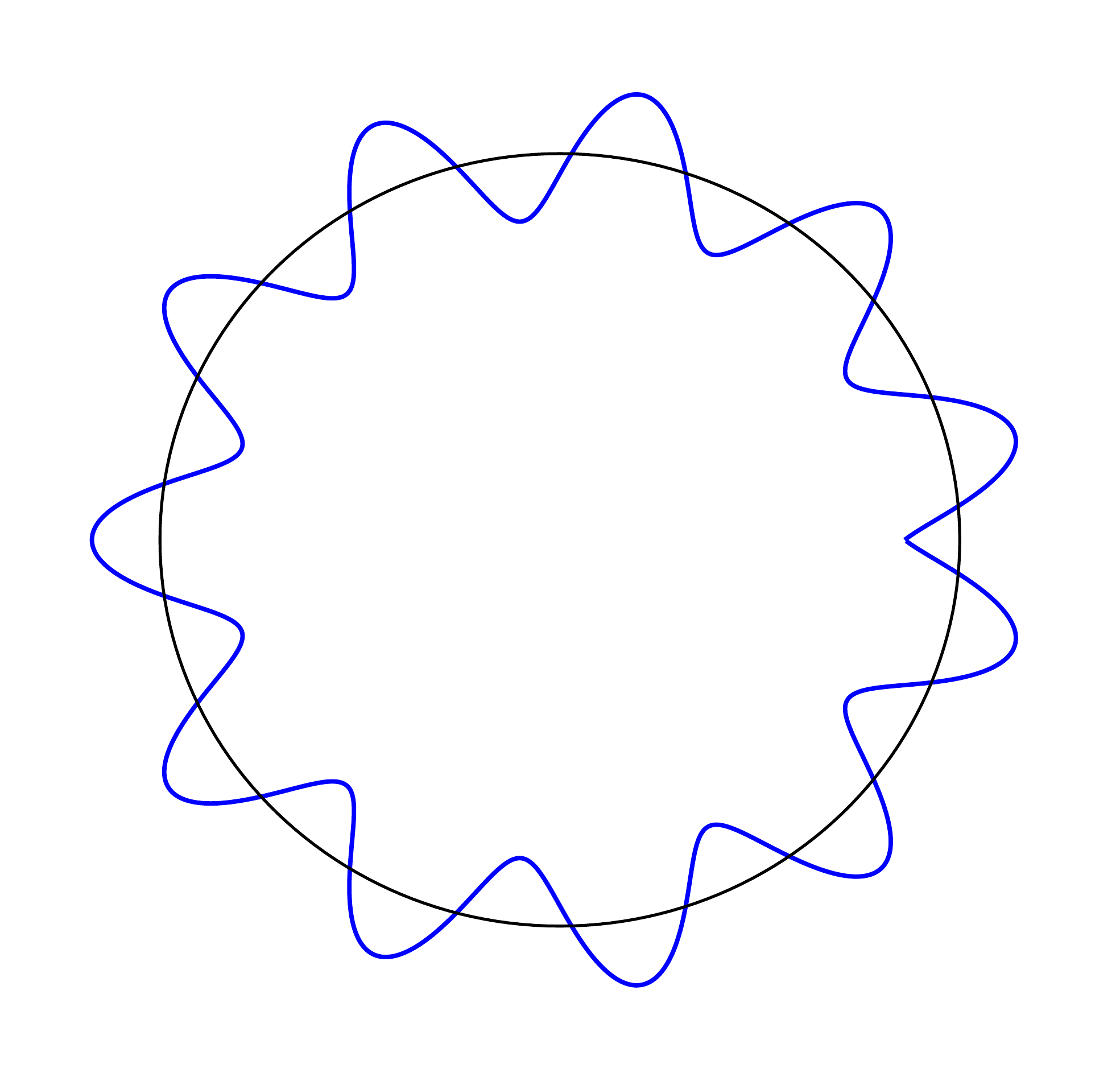}
\includegraphics[width=0.28\linewidth]{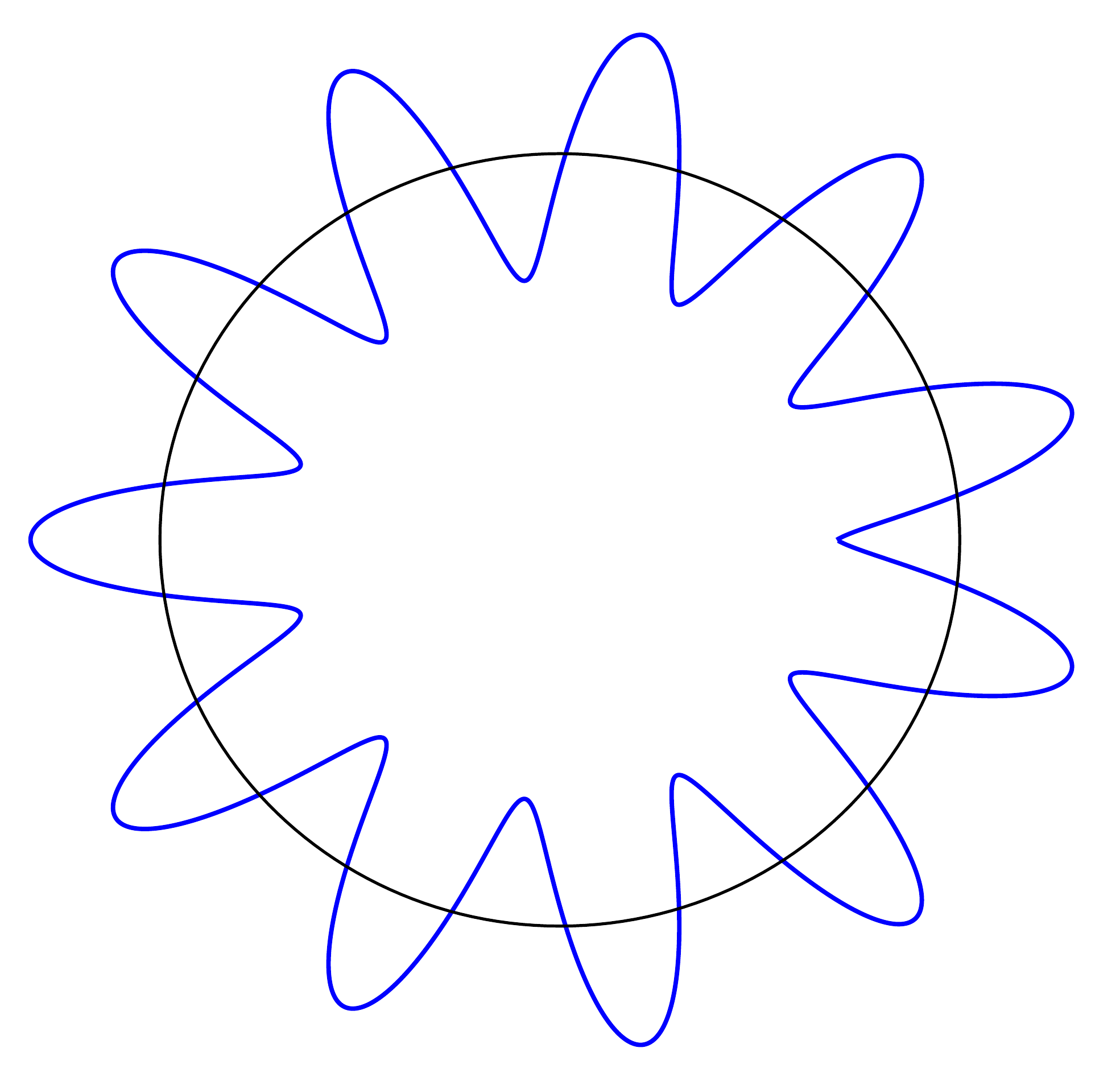}
   \end{center}
\caption{Closed orbit for $\tune$ values of 10.1, 10.2, 10.5, 10.8, and 10.9 (from top left to bottom right). The deflection (kick) is in the horizontal direction on the right for each value of $q$. Note the change of amplitude due to the tune value.}
\label{fig:co-tune}
\end{figure}

Figure~\ref{fig:co-tune} displays sketches of horizontal orbit deviations along a ring due to a single kick for a tune of $Q = 10 + q$ for selected values of $q$. The kink at the location of the kick $\theta$ is clearly visible, and it can be used to localize the location of the deflection from an orbit measurement. The angle change at the kink corresponds exactly to~$\theta$.

In a real machine the beam position is sampled by $\nump$ beam position monitors (BPMs) distributed along the machine.
The position response (change) $\Delta u_i$ $(u = x,y)$ of the beam at BPM $i$ due to a deflection $\Delta \theta_j$ from a source $j$ is given in linear approximation by
\begin{equation} \label{eq:rij-co}
  \Delta u_i = R_{ij}  \Delta \theta_j = \frac{ \sqrt{\beta_i \beta_j}\: \cos(| \mu_i - \mu_j | - \pi Q) }{2 \sin(\pi Q)} \,\Delta \theta_j
\end{equation}
for a circular accelerator and by
\begin{equation} \label{eq:rij-st}
\Delta u_i = \begin{cases}
R_{ij}  \Delta \theta_j = \sqrt{\beta_i \beta_j} \sin(\mu_i - \mu_j) \Delta \theta_j & \quad \text{if } \mu_i > \mu_j  \, ,\\
 0 & \quad \text{if } \mu_i \leq \mu_j
\end{cases}
\end{equation}
for a linear accelerator. The beam position at all BPMs can be represented by a vector
\begin{equation}
\vp = \left( \begin{array}{c} \Delta u_1 \\ \Delta u_2 \\ \vdots \\ \Delta u_{\nump} \end{array} \right) \, ,
\end{equation}
and the corrector strengths (kicks) can be represented by a vector
\begin{equation}
\vk = \left( \begin{array}{c} \Delta \theta_1 \\ \Delta \theta_2 \\ \vdots
    \\ \Delta \theta_{\numk} \end{array} \right) \, .
\end{equation}
The relationship between the positions and deflections can be expressed in terms of a matrix ${\bf R}$ called the response matrix:
\begin{equation}
\label{eq:orbit-response} \vp = {\bf R} \vk\: .
\end{equation}
The elements $R_{ij}$ of the matrix ${\bf R}$ are those given in Eqs.~\eqref{eq:rij-co} and \eqref{eq:rij-st}. The response matrix ${\bf R}$ obviously contains a lot of information on the machine optics; tools that take advantage of this fact to determine and correct the lattice functions will be presented in a later section.

\subsubsection{Orbit correction}

Given a measured orbit $\vd$, the goal of orbit correction is to find a set of corrector deflections
$\vk$ that satisfy the relation
\begin{equation} \label{eq:orbit-corr}
 \vd + {\bf R} \vk = 0 \; .
\end{equation}
In general the number of BPMs ($N$) and the number of correctors ($M$) are not identical, and Eq.~\eqref{eq:orbit-corr} is either over-constrained ($N
> M$) or under-constrained ($N < M$). In the former and more frequent case, Eq.~\eqref{eq:orbit-corr} cannot be solved exactly.
Instead, an approximate solution must be found, and commonly used least-squares algorithms minimize the quadratic residual
\begin{equation}
\label{eq:orbit-corr-quad} S = \| \vd + {\bf R} \vk \|^2 \: .
\end{equation}
Algorithms for beam steering aim to minimize the least-squares error, and it is  assumed that the response matrix ${\bf R}$ is known well enough to obtain a convergent correction. Equation~\eqref{eq:orbit-corr} is quite generic; problems in optics and dispersion correction can be cast in a similar form, and the algorithms used for steering may also applied to other linear imperfection problems.

\subsubsection{MICADO algorithm}

The problem of correcting the orbit deterministically came up a long time ago in the first machines. B.\ Autin and Y.\ Marti of CERN published a note in 1973 describing an algorithm, which they named MICADO, that is still in use today in many machines \cite{MICADO}.
The intuitive principle of MICADO is rather simple. Each column of ${\bf R}$ corresponds to the response of all BPMs to one of the correctors.
MICADO compares the response of every corrector (i.e.\ each column in turn) with the orbit to be corrected, $\vd$, by calculating the scalar products
\begin{equation}
\sigma_j = \sum_{i=1}^{i=N} d_i R_{ji}     \quad \text{ and } \quad \rho_j = \sum_{i=0}^{i=N} R_{ji} R_{ji}
\end{equation}
for all correctors, i.e.\ for $j \in [1,M]$. MICADO selects the corrector that has the best match (correlation) with the orbit corresponding to the largest value $\sigma_j^2/\rho_j$. This corrector will yield the greatest reduction in the quadratic sum $S$ of Eq.~\eqref{eq:orbit-corr-quad} by setting the right kick value $\Delta \theta_j$.
This procedure can be iterated using the remaining correctors until the orbit is good enough (i.e.\ stop after $K$ steps using $K$ correctors) or as good as it can be by using all available correctors.

\begin{figure}[tbp]
  \begin{center}
\includegraphics[width=0.8\linewidth]{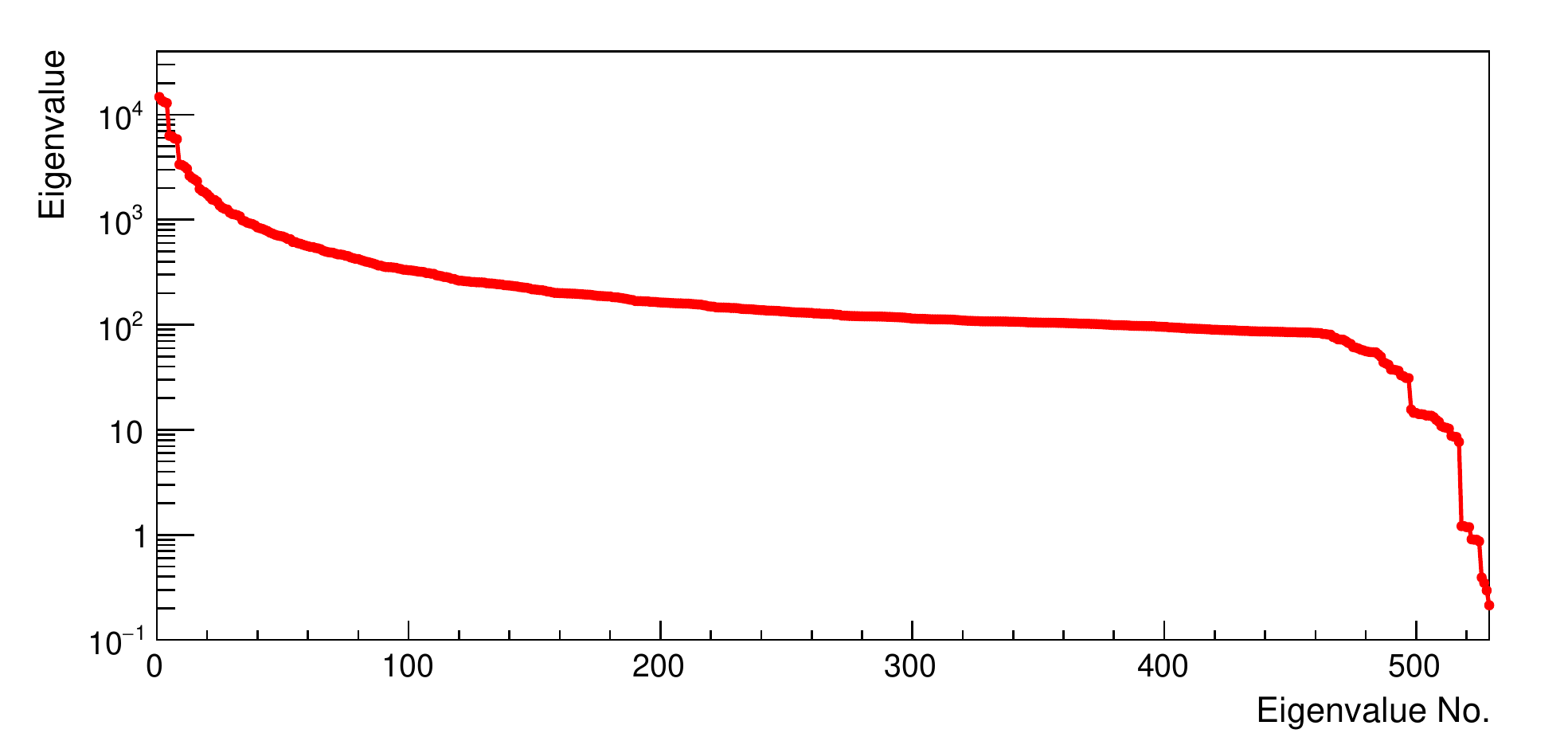}
\includegraphics[width=0.48\linewidth]{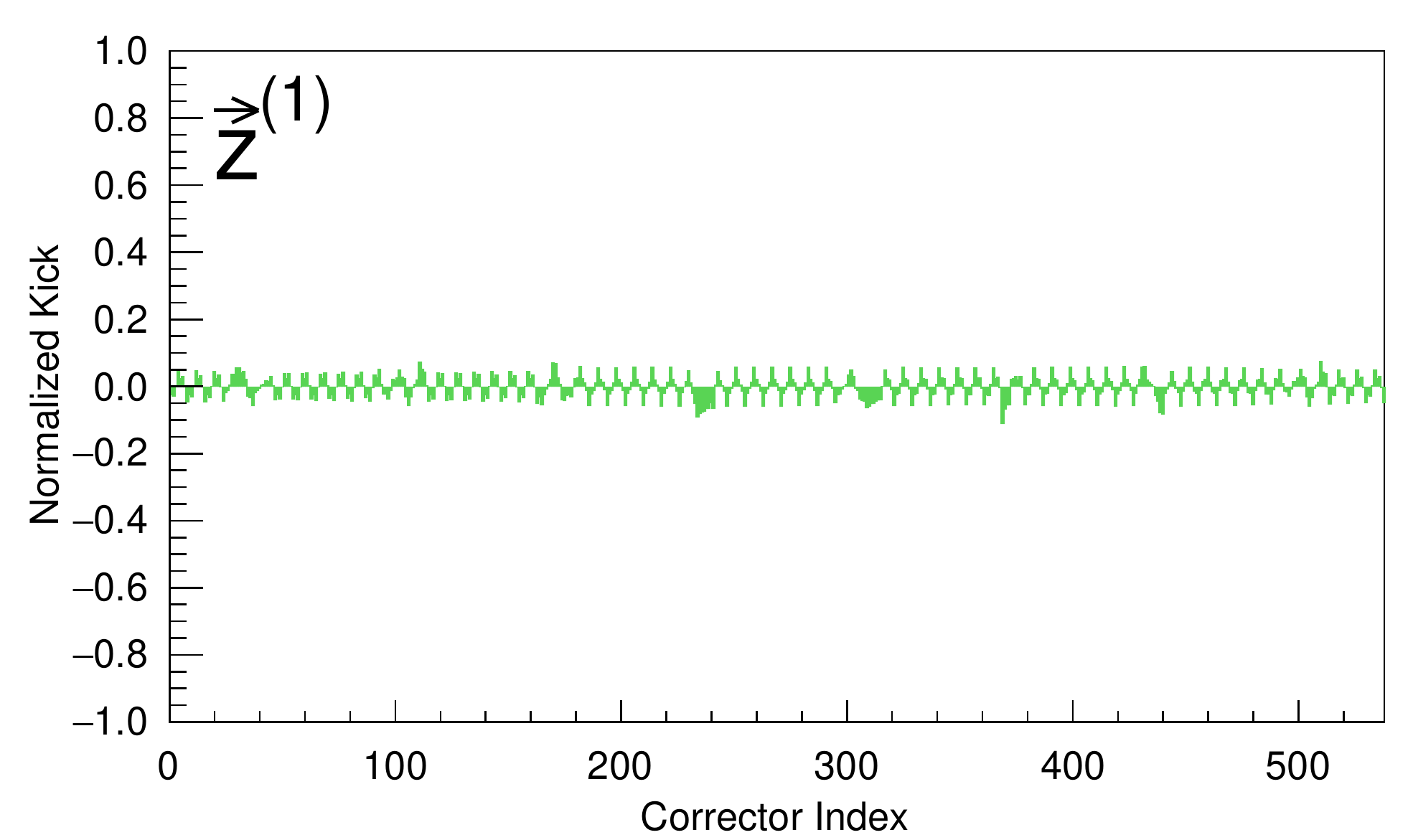}
\includegraphics[width=0.48\linewidth]{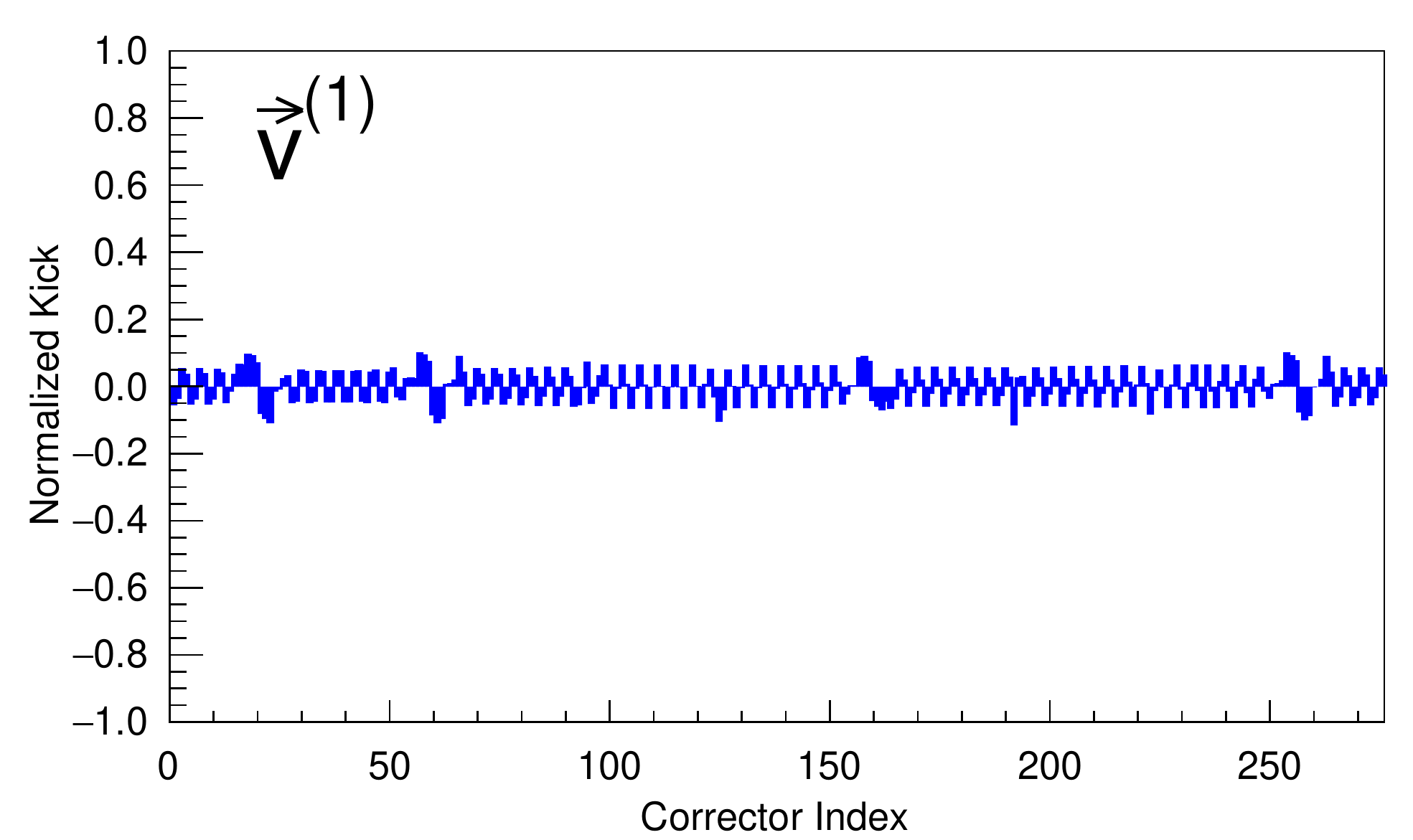}
\includegraphics[width=0.48\linewidth]{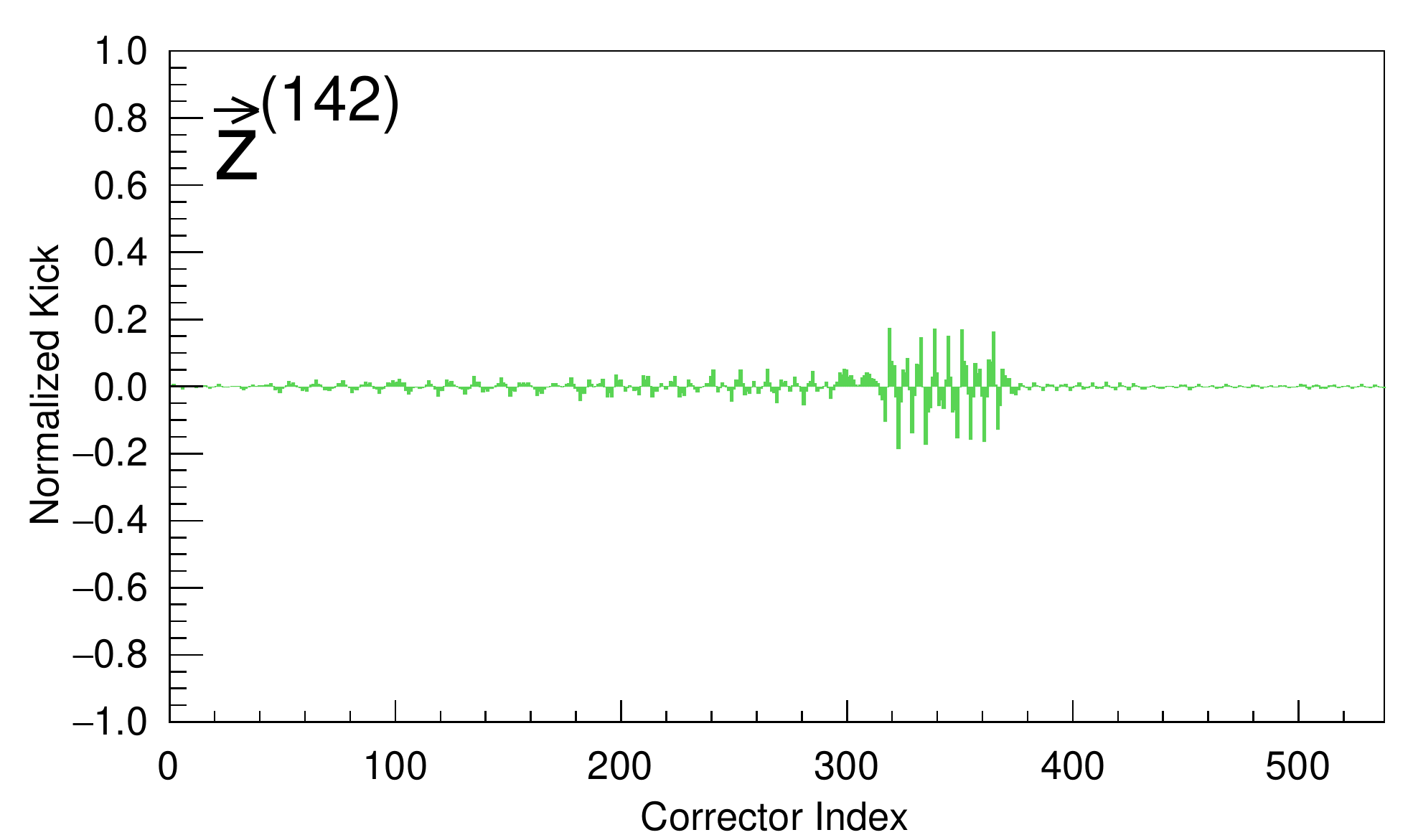}
\includegraphics[width=0.48\linewidth]{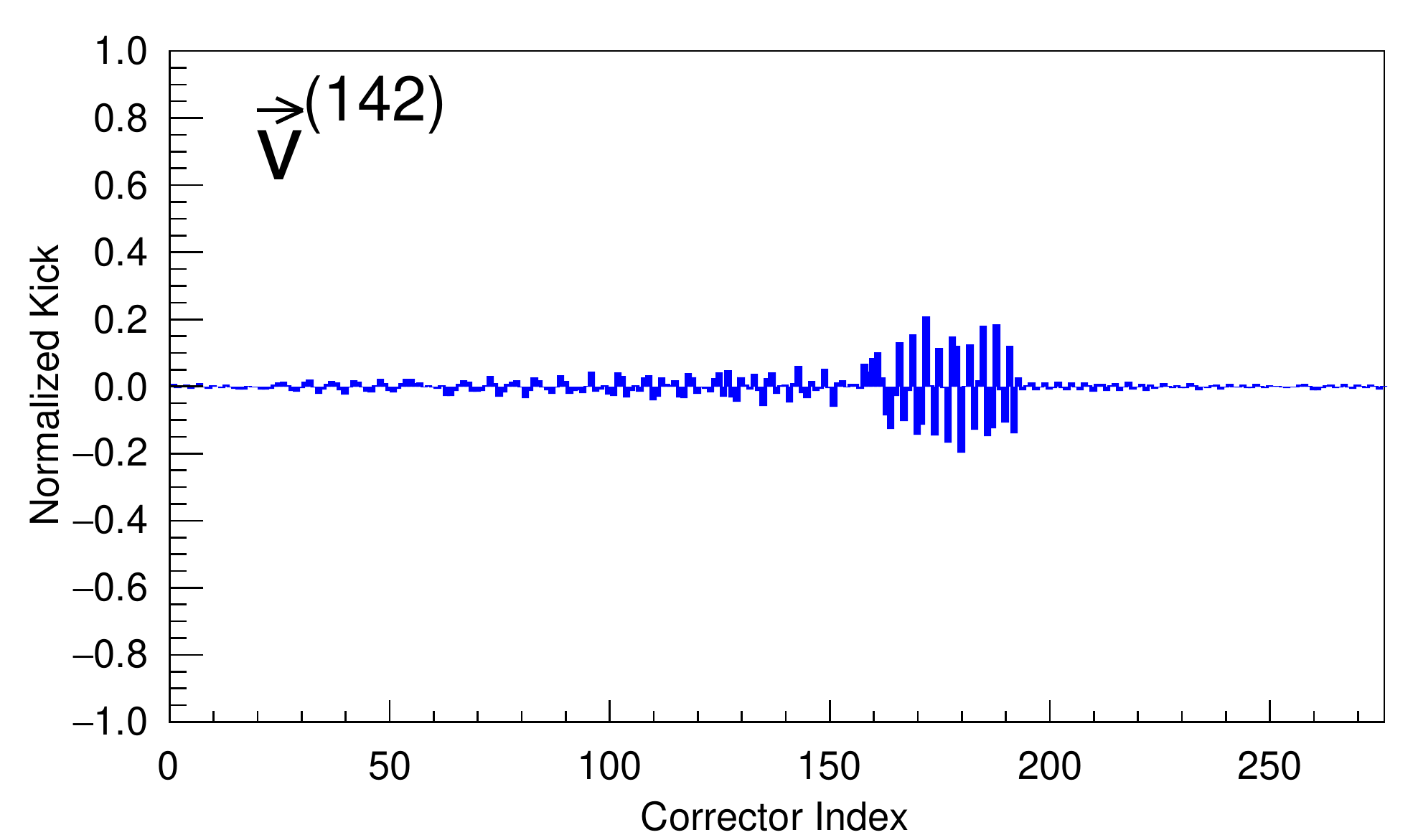}
\includegraphics[width=0.48\linewidth]{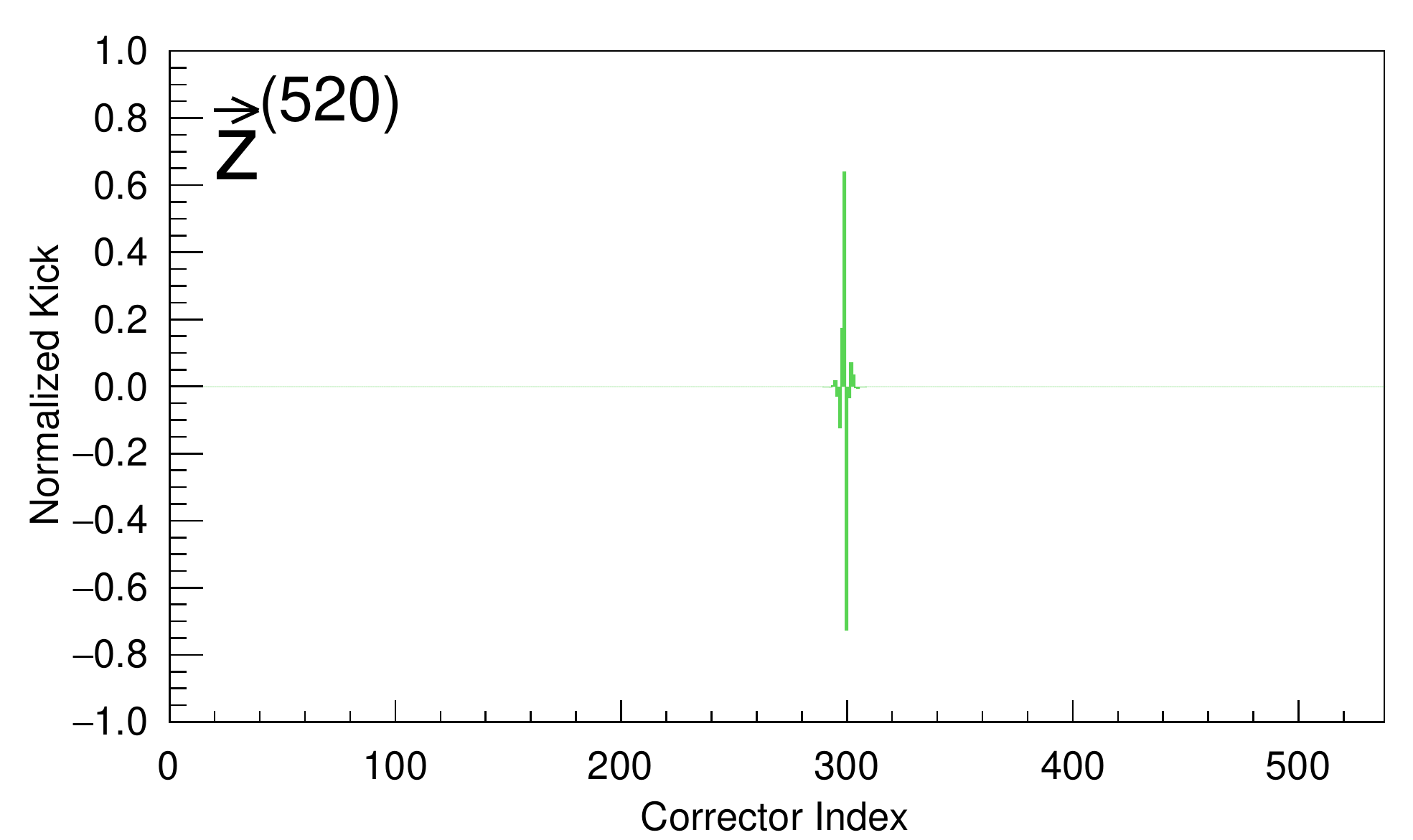}
\includegraphics[width=0.48\linewidth]{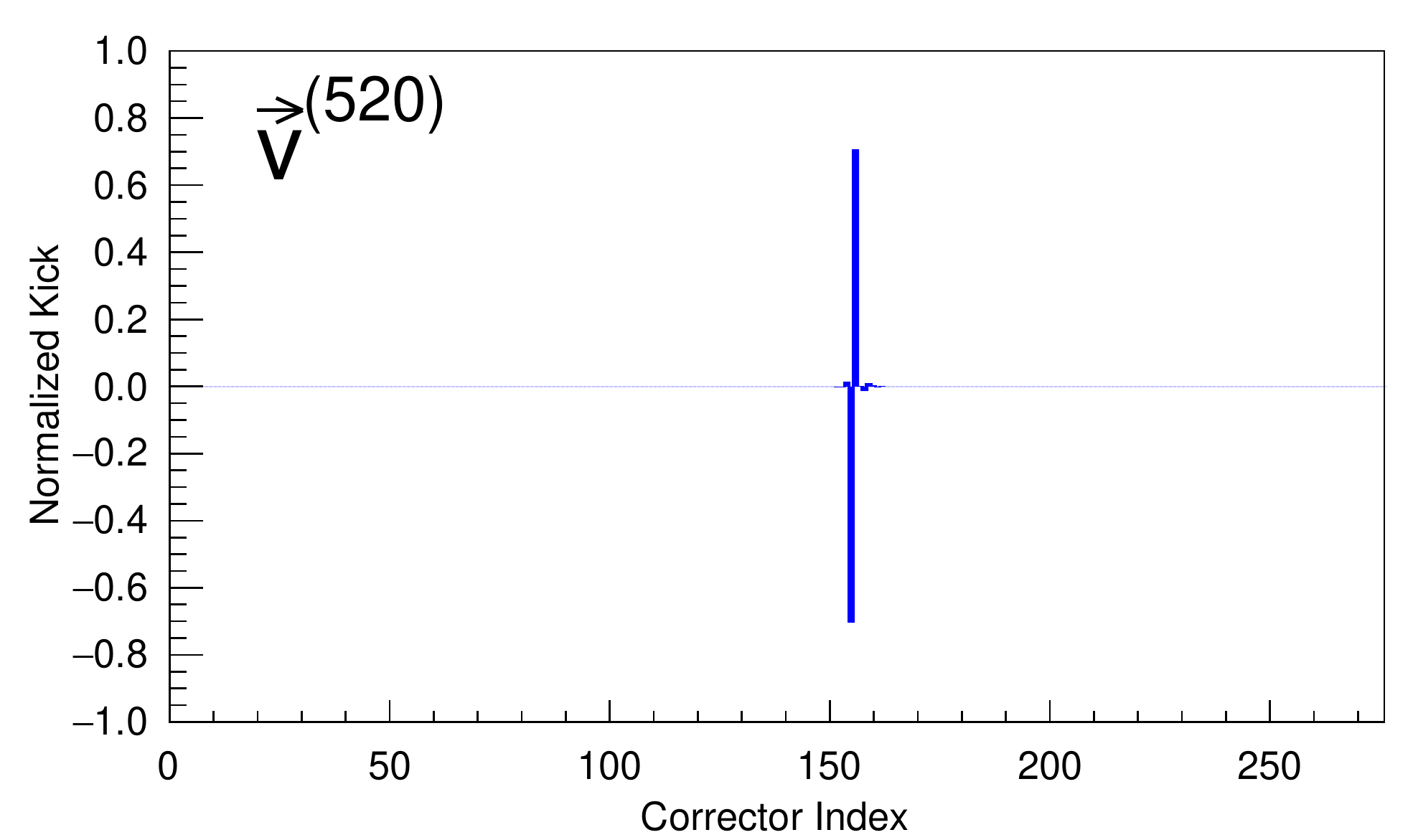}
   \end{center}
\caption{Example of the SVD eigenvalue spectrum (top) and selected eigenvectors $\vec{z}_i$ and $\vec{v}_i$ (lower three rows) for the horizontal plane of the LHC, illustrating how the solutions become more local as the eigenvalue decreases.}
\label{fig:svd-lhc-results}
\end{figure}

\subsubsection{SVD algorithm}

Singular value decomposition (SVD) is a generic operation applicable to any matrix $\matA$ of dimension $N \times M$; $\matA$ is decomposed into three matrices $\matZ$, $\matW$, and $\matV$, where $\matW$ is a diagonal `eigenvalue' matrix and $\matV$ is a square matrix that is also orthonormal. This algorithm has become the favourite tool for orbit correction, for reasons that will be explained below.
For $N \geq M$, the SVD of matrix ${\bf A}$ has the form~\cite{P3}
\begin{equation} \label{eq:svd-decomp}
\matA = {\bf ZWV}^{\rm t}\,,
\end{equation}
or
\begin{equation} \label{eq:svd}
\matA = \left[
\begin{array}{cccc}
z_{1}^{(1)} & z_{1}^{(2)} & \cdots & z_{1}^{(M)}   \\
z_{2}^{(1)} & z_{2}^{(2)} & \cdots & z_{2}^{(M)}  \\
 \vdots & \vdots & & \vdots  \\
z_{N}^{(1)} & z_{N}^{(2)} & \cdots & z _{N}^{(M)}
\end{array}
\right] \,
\begin{bmatrix}
w_{1} \quad \quad \quad\quad  \\
 w_{2} \quad \quad  \quad  \\
\quad   \ddots  \quad  \\
 \quad  \quad \quad w_{M}
\end{bmatrix}
 \left[
\begin{array}{cccc}
v_{1}^{(1)} & v_{2}^{(1)} & \cdots & v_{M}^{(1)}  \\
v_{1}^{(2)} & v_{2}^{(2)} & \cdots & v_{M}^{(2)}  \\
 \vdots & \vdots & &  \vdots  \\
v_{1}^{(M)} & v_{2}^{(M)} & \cdots & v_{M}^{(M)}
\end{array}
\right]\, .
\end{equation}
The superscript `t' denotes matrix transposition; $\matZ$ is an $N \times M$ matrix whose column vectors $\vec{z}^{(\alpha)}\ (\alpha = 1, \ldots , M)$ form an orthonormal set, with $\matZ^{\rm t}\matZ = {\bf I}$ where ${\bf I}$ is the identity matrix; $\matW$ is an $M \times M$ diagonal matrix with non-negative elements; and $\matV^{\rm t}$ is the transpose of the $M \times M$ matrix $\matV$ whose column vectors $\vec{v}^{(\alpha)}\ (\alpha = 1, \ldots , M)$ form an orthonormal set, with $\matV^{\rm t}\matV = \matV \matV^{\rm t} = {\bf I}$.
From Eq.~\eqref{eq:svd} it follows that for $\alpha = 1, \ldots , M$,
\begin{equation}
\matA \vec{v}^{(\alpha )} =w_{\alpha } \vec{z}^{(\alpha )}\: ,
\quad \matA^{\rm t} \vec{z}^{(\alpha )} = w_{\alpha } \vec{v}^{(\alpha )}
\end{equation}
and
\begin{equation}
\matA \matA^{\rm t} \vec{z} ^{(\alpha )} =w_{\alpha }^{2} \vec{z} ^{(\alpha
)}\: , \quad \matA^{\rm t} \matA \vec{v} ^{(\alpha )} = w_{\alpha
}^{2} \vec{v} ^{(\alpha )}\:,
\end{equation}
where $\vec{z} ^{(\alpha )}$ is an eigenvector of matrix $\matA \matA^{\rm t}$ and $\vec{v} ^{(\alpha )}$  is an eigenvector of matrix $\matA^{\rm t} \matA$.

When none of the diagonal elements $w_{\alpha }$ vanish, the solution to Eq.~(\ref{eq:orbit-corr}) is obtained by inverting Eq.~\eqref{eq:svd-decomp} and is given by
$\vec{\theta} = -{\bf VW}^{-1}{\bf Z}^{\rm t} \vd$.

Figure~\ref{fig:svd-lhc-results} presents the horizontal plane spectrum of 532 eigenvalues and a few selected eigenvectors $\vec{z} ^{(\alpha )}$  and $\vec{v} ^{(\alpha )}$ for the  LHC. The eigenvalues are sorted from largest to smallest, and as the eigenvalues decrease, the associated eigenvectors correspond to increasingly local `structures' on the orbit ($\vec{z} ^{(\alpha )}$).

\begin{figure}[tbp]
  \begin{center}
\includegraphics[width=0.53\linewidth]{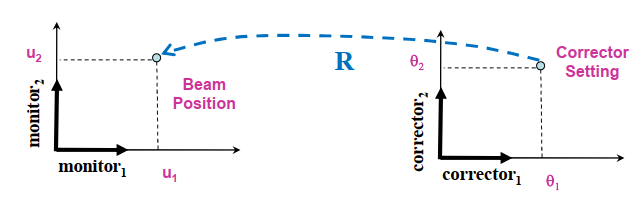} \hspace{5mm}
\includegraphics[width=0.42\linewidth]{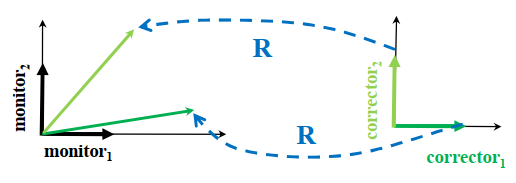} \\ \vspace{5mm}
\includegraphics[width=0.7\linewidth]{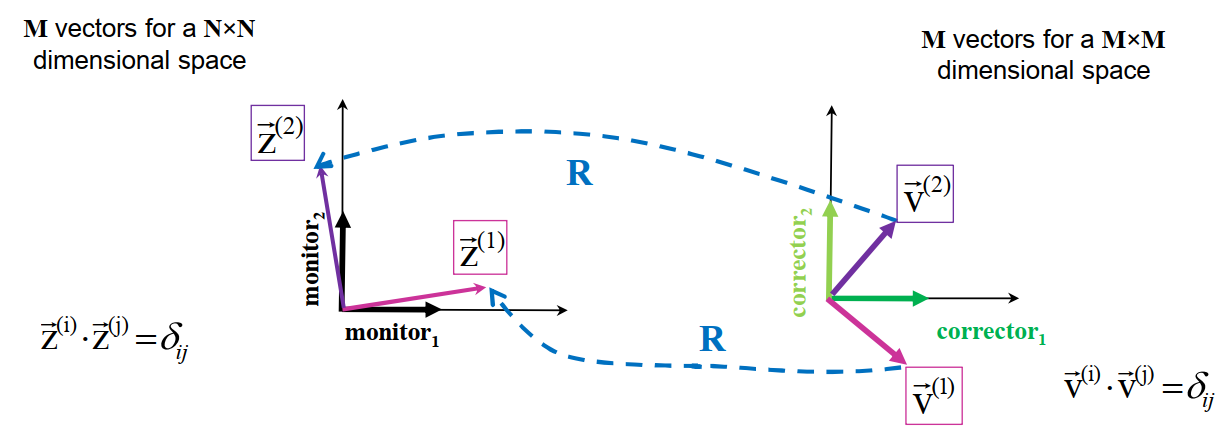}
   \end{center}
\caption{The response matrix $\matR$ maps points in the corrector space to points in beam position space (top left). Two orthogonal directions in corrector space do not  in general  map to orthogonal directions in position space (top right), preventing a straightforward inverse mapping.  SVD  identifies eigenvectors $\vec{v}$ that are orthogonal in corrector space and whose response in beam position space remains orthogonal (bottom).}
\label{fig:rmatrix-mapping}
\end{figure}

An explanation of how SVD helps to solve the correction problem is presented in Fig.~\ref{fig:rmatrix-mapping}.
The response matrix $\matR$ maps points in the `corrector space' to points in `beam position space'.
The natural basis vectors of those two spaces are the physical monitors and the orbit correctors. Unfortunately, $\matR$ maps orthogonal vectors (in the form of correctors) to non-orthogonal responses in monitor space. This makes the inverse process of obtaining the corrector patterns from the position space difficult.  SVD identifies an orthonormal basis of the corrector space, $\vec{v}^{(\alpha )}$, such that the position responses of the new basis vectors $\vec{z}^{(\alpha)}$ are also orthogonal in position space. It is now possible to express any beam position in terms of the $\vec{z}^{(\alpha )}$ vectors (up to an uncorrectable remainder) and to directly obtain the corresponding corrector pattern in terms of the $\vec{v}^{(\alpha)}$. In addition, every corrector setting can be decomposed uniquely into the $\vec{v}^{(\alpha)}$ basis, and every orbit can be decomposed into the $\vec{z}$ vectors plus a residual uncorrectable remainder.

The solution $\vec{\theta} = -{\bf VW}^{-1}{\bf Z}^{\rm t} \vd$ represented in terms of matrix multiplications can be re-expressed equivalently as a decomposition of the measured orbit into the orbit eigenvectors $\vec{z}^{(\alpha)}$ followed by correcting the effect of the $k$ largest eigenvectors (since $\vec{z}^{(\alpha)}$  is associated with $\vec{v}^{(\alpha)}$):
\begin{alignat}{4}
  \matZ^{\rm t} & \;\Delta\vec{d}  &      \qquad  \longrightarrow \qquad &   \vec{z}^{(i)} \cdot \Delta\vec{d} = c_i \; \forall i \leq k \: , \label{eq:svdsol1} \\
  (\matW^{-1} \matZ^{\rm t}) & \;\Delta\vec{d} &      \qquad  \longrightarrow \qquad &  c_i/w_i \; \forall i \leq k   \: ,\label{eq:svdsol2} \\
  - (\matV \matW^{-1} \matZ^{\rm t}) & \;\Delta\vec{d} &      \qquad  \longrightarrow \qquad &   - \sum_i^k  (c_i/w_i) \vec{v}^{(i)} \: . \label{eq:svdsol3}
\end{alignat}
Having one or more vanishing $w_{\alpha }$ indicates that the matrix is singular, and for orbit correction one discards the corresponding terms from Eqs.~\eqref{eq:svdsol2} and \eqref{eq:svdsol3}.

In practice it is often desirable to limit the number of eigenvalues used for the correction to control the root-mean-square (r.m.s.) strength of the orbit
correctors or to avoid small eigenvalues that are very sensitive to the accuracy of the model. The SVD algorithm is ideally suited to feedback application since the correction can be cast in the simple form of a matrix multiplication once the decomposition has been performed. This
provides a fast and reliable correction procedure for real-time feedback.

In summary, SVD can be used to solve the problem of determining a correction using $k$ out of $M$ eigenvalues.
When the eigenvalues $w_j$ are sorted in descending order, $w_{j+1} \leq w_j$, this operation corresponds intuitively to decomposition of the measured orbit into the orbit eigenvectors $\vec{z}^{(i)}$---which is unique---and then correcting the effect of the $k$ largest eigenvectors.

The two algorithms, MICADO and SVD, each have some advantages and drawbacks.
\begin{itemize}
  \item MICADO picks out individual correctors; for a perfect match of model and machine it will help to localize local sources.
  \item MICADO is well suited to identifying a single or a dominant source of  orbit perturbation for clean measurements.
  \item MICADO can, however, get into trouble if the matrix $\matR$ has singularities associated with poor BPM or corrector layout.
  \item SVD will always use all correctors.
  \item With few SVD eigenvalues for the correction, even a local perturbation will be corrected with many elements, which can be an advantage if the strength of the correctors is limited.
  \item The number of SVD eigenvalues controls the locality and quality of the correction; with more eigenvalues local structures will be corrected better. Limiting the number of eigenvalues provides a means of  avoiding corrections on noise, in particular with eigenvectors that provide large strength and  little position change.
    \item Since the SVD correction can be cast as a simple matrix operation, it is very suitable (and always used) for real-time orbit feedback.
\end{itemize}

\subsection{First turns}

The first problem encountered during machine commissioning is bringing the beam to the end of the linac, or circulating it in the storage ring. The difficulty of this task depends on the alignment errors and the length of the machine. For small accelerators it is usually not a serious issue, as the number of undesired deflections encountered over the length of the accelerator will not be too large; but for  a machine  many kilometres long, this is far from guaranteed, and trajectory excursions will build up randomly along the path of the beam $s$. For random errors the trajectory amplitudes will increase roughly in proportion to $\sqrt{s}$, as this is equivalent to a random walk process. For a typical r.m.s.\  alignment error $\sigma_{\rm a}$, it is possible to estimate the resulting r.m.s.\ orbit error $\sigma_{\rm orb}$ for a machine with a homogeneous lattice consisting of $N_{\rm c}$ FODO cells~\cite{OFB-LHC}:
\begin{equation}\label{eq:orbit-random-walk}
  \sigma_{\rm orb} \simeq \frac{|k| l_Q \beta_{\rm eff}} {4 \sin{\pi Q}} \sqrt{N_{\rm c}} \sigma_{\rm a} = \kappa \sigma_{\rm a}\:,
\end{equation}
where $k$ is the quadrupole gradient, $l_Q$ the quadrupole length, $Q$ the machine tune, and $\beta_{\rm eff}$ the effective betatron function over the cell. For the LHC, which has a length of 26.7\,km, $N_{\rm c} \approx 250$, and $\kappa \approx  20$--$30$ for the injection optics, with  $\sigma_{\rm a} \simeq 0.3$\,mm the expected orbit r.m.s.\ $\sigma_{\rm orb}$ is 6--9\,mm. This implies that beam excursions of $\pm 2 \sigma_{\rm orb}$ will exceed the mechanical aperture of the LHC vacuum chamber; it is therefore very unlikely that the beam will make a full turn without any correction.

\subsection{BPM measurement errors}

\begin{figure}[tbp]
  \begin{center}
\includegraphics[width=0.5\linewidth]{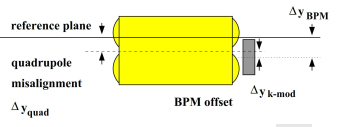}
   \end{center}
\caption{Illustration of BPM offset with respect to the nearest quadrupole axis}
\label{fig:bpm-offset}
\end{figure}

The quality of BPM measurements is of course crucial to obtaining a well-corrected orbit. BPMs are typically affected by the following errors:
\begin{itemize}
  \item measurement offsets that may be due to the electronics or  the mechanical alignment of the electrodes (see Fig.~\ref{fig:bpm-offset});
  \item scale errors and non-linearity of the position readings;
  \item intensity- and beam pattern-dependent effects leading to artificial measurement changes when the beam structure is changed.
\end{itemize}

The non-linearities and beam intensity systematics must be simulated or obtained from laboratory measurements. Non-linearities are due to electrode geometries and the electronics, and improper correction of such effects can bias beam-based measurements~\cite{BPM-NL}.

\begin{figure}[tbp]
  \begin{center}
\includegraphics[width=0.48\linewidth]{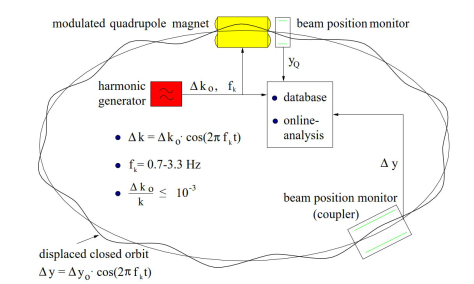}
\includegraphics[width=0.48\linewidth]{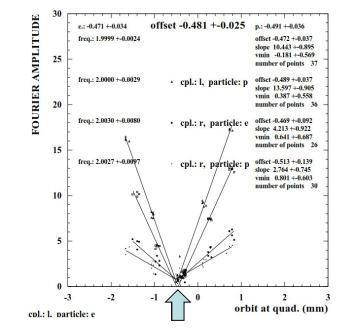}
   \end{center}
\caption{Illustration of the k-modulation technique for determining the offsets between the BPM readings and the magnetic axis of a quadrupole. The modulation of the quadrupole gradient leads to a time-varying orbit that propagates along the ring (left) or downstream in the case of a linac (right). When the amplitude of the oscillation is plotted as a function of the position of the beam in the quadrupole---here for the case of the LEP---the amplitude vanishes when the beam is aligned on the quadrupole axis~\cite{P1}.}
\label{fig:kmod-bpm}
\end{figure}

BPM offsets with respect to the adjacent quadrupole, a severe nuisance for orbit corrections, can be measured by a technique called k-modulation, whereby  the gradient of a selected quadrupole ($g$,  $K_1$) is modulated slightly at a frequency $f_{\rm mod}$ and the beam position is varied inside the quadrupole using orbit bumps. The beam orbit is then modulated at $f_{\rm mod}$, with a modulation amplitude that is proportional to the beam offset with respect to the centre of the quadrupole. The modulation amplitude vanishes when the beam is centred on the magnetic axis where the field of a quadrupole vanishes; see Eq.~\eqref{eq:quad}. Figure~\ref{fig:kmod-bpm} illustrates this technique for the LEP~\cite{P1}.

\subsection{Dispersion}

The bending of charged particles by magnetic fields depends on the momentum (Lorentz force); see Eq.~\eqref{eq:lorentz}.
For a beam subject to a magnetic deflection, the trajectories of the particles will be sorted by energy, which creates dispersion (in position). The dispersion $D(s)$ is closely related to the orbit (or trajectory), since it is the derivative of the beam position with respect to energy:
\begin{equation}\label{eq:dispersion}
  D(s) = \frac{\Delta u(s)}{\Delta p/p}\:,
\end{equation}
where $\Delta p/p$ is the relative energy offset.

\begin{figure}[tbp]
  \begin{center}
\includegraphics[width=0.37\linewidth]{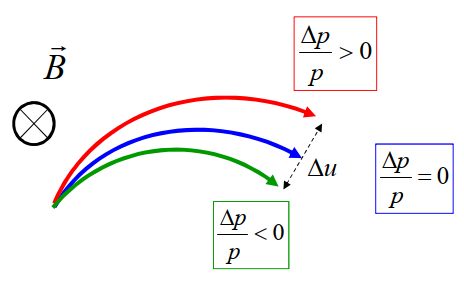}
   \end{center}
\caption{Definition of the dispersion due to a bending field that sorts the particles by their momentum}
\label{fig:disp-def}
\end{figure}

In a storage ring there is always non-zero horizontal dispersion due to the presence of bending magnets. For flat machines the vertical dispersion is usually zero by design. While for hadron machines the dispersion is in general not critical as long as it remains within reasonable boundaries, for e$^+$e$^-$ machines the vertical dispersion can lead to significant vertical emittance growth, as well as to serious luminosity or brightness performance loss. 

\noindent Dispersion errors may be driven by the following factors:
\begin{itemize}
  \item optics errors that distort the dispersion together with the general optics;
  \item coupling between the horizontal and vertical planes that transfers horizontal dispersion into the vertical plane;
  \item steering and alignment errors that generate dipolar deflection.
\end{itemize}
Optics and coupling corrections will be discussed later; for the moment we focus on the third point, namely dispersion driven by orbit deflections.

A technique that combines regular steering and dispersion correction, called dispersion-free steering (DFS)~\cite{P3} and used for the Stanford Linear Collider (SLC) and LEP e$^+$e$^-$ machines, will be outlined here. The principle of DFS is based on extension of the orbit response matrix to dispersion, including the dispersion response $\matS$ \cite{P3,DISP-RES} and a weight factor $\alpha$ between orbit and dispersion, $0 \leq \alpha \leq 1$. The dispersion response may be estimated analytically or from a simulation program such as MAD. Equation~\eqref{eq:orbit-corr} describing the orbit correction is expanded to include the dispersion $\vdisp$:
\begin{equation}
\label{eq:dfs}
 \begin{pmatrix} (1 - \alpha) \vd \\ \alpha \vdisp \end{pmatrix}
+ \begin{pmatrix} (1 - \alpha) \matR \\ \alpha \matS \end{pmatrix}  \vk = 0 \; .
\end{equation}
The combined orbit and dispersion correction system can be solved in exactly the same way as  the ‘normal’ steering.
This, however, does not apply to all sources of dispersion (e.g.\ dispersion due to coupling between planes).
DFS provides controlled correction of the orbit and the dispersion, with the dispersion dominated by deflection (errors). Details and an example of simultaneous correction of orbit and dispersion can be found in \Bref{P3}.

\section{Tune and coupling}

\begin{figure}[tbp]
  \begin{center}
\includegraphics[width=0.73\linewidth]{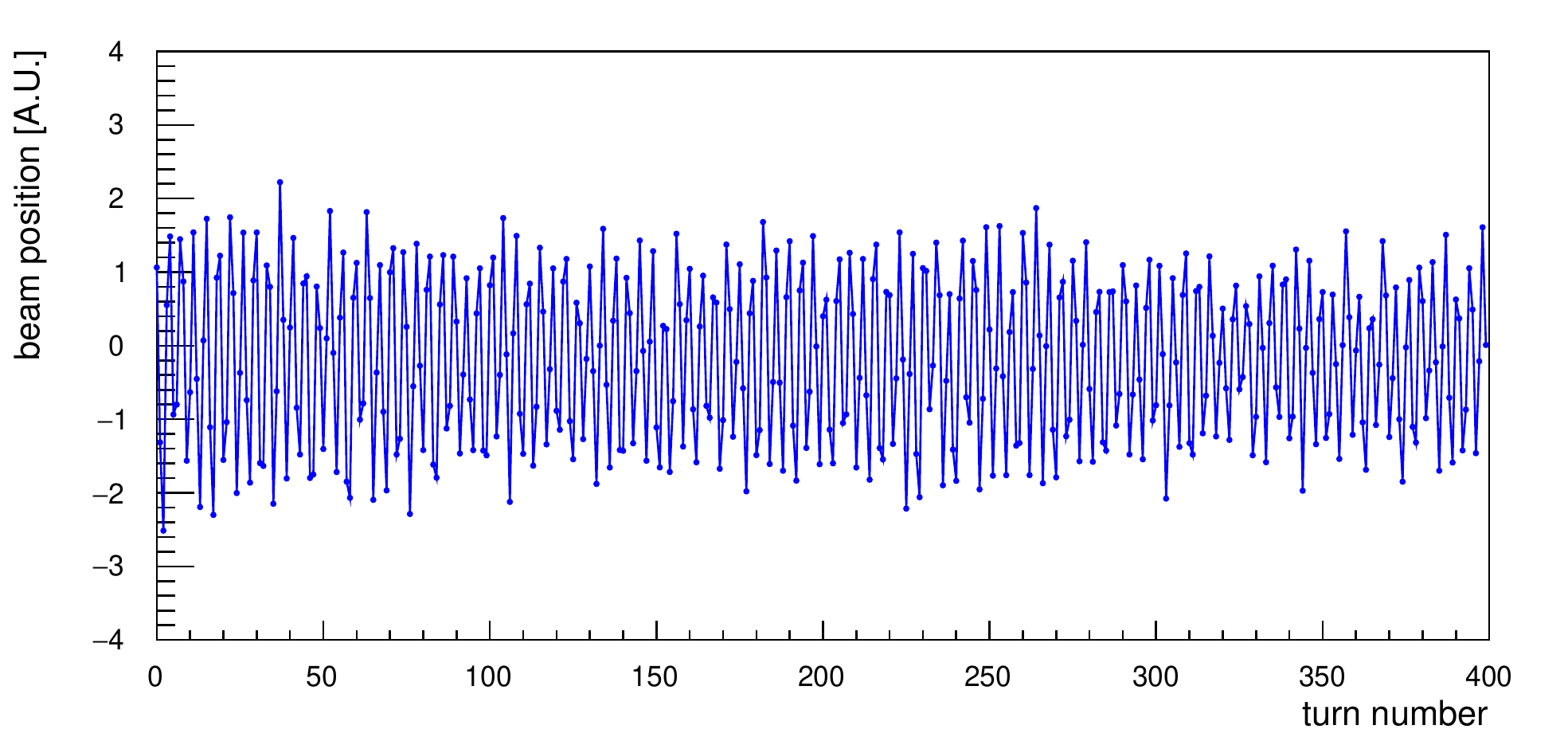}
\includegraphics[width=0.73\linewidth]{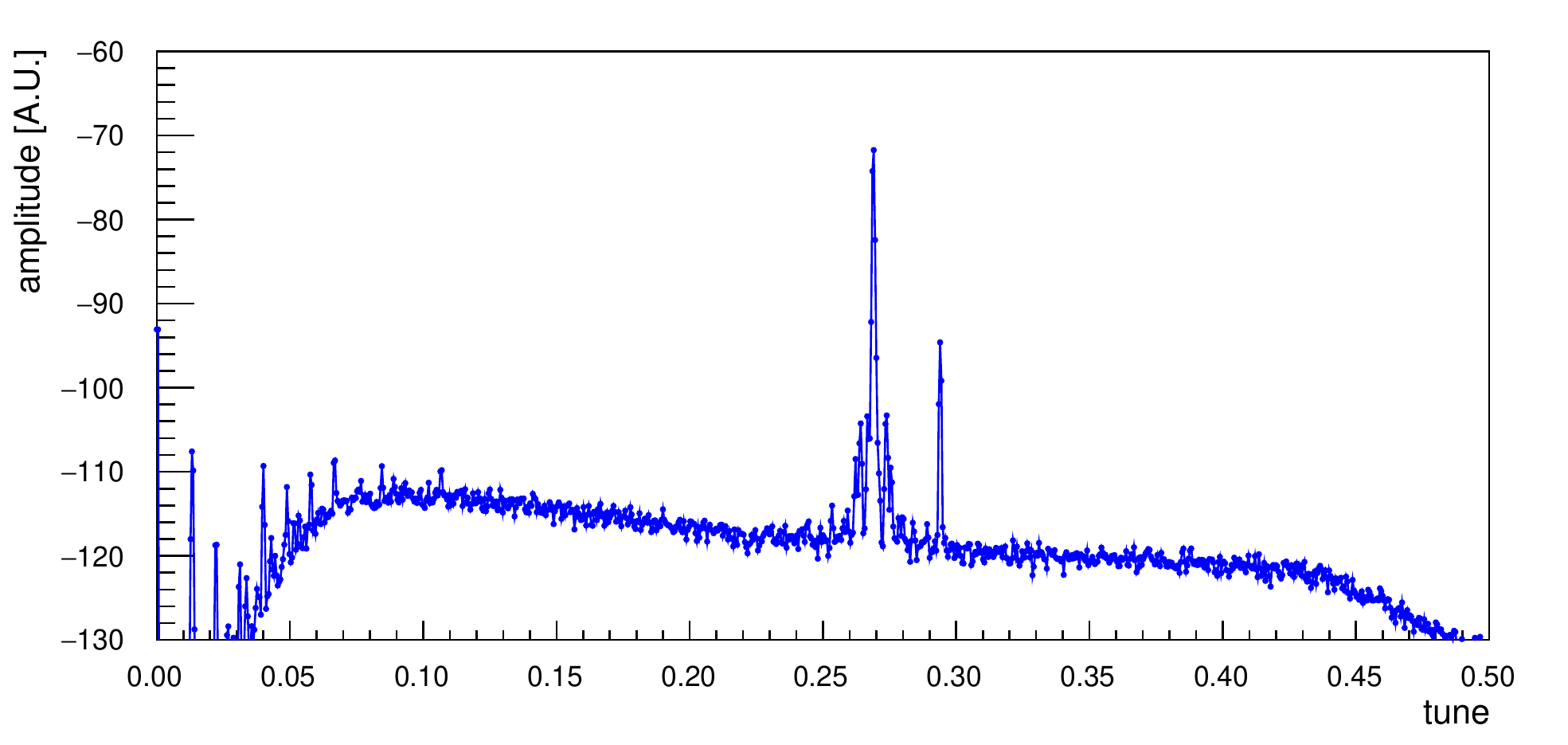}
   \end{center}
\caption{Raw beam position data at one BPM (top) and FFT of the same data revealing the machine tune (bottom). The spectrum exhibits a main line at $q = 0.265$ and a coupled tune just above $q = 0.290$.}
\label{fig:turns-tune}
\end{figure}

The machine tune, which corresponds to the number of betatron oscillations per turn, can be split into an integer part $N$ and a fractional part $q$, $Q = N + q$. The integer $N \in \mathbb{N}$ can be obtained by applying a kick to the orbit and then counting the number of oscillation periods. In large machines where $N$ reaches 50 to 400, the integer value may be wrong before correction if there are large uncorrected optics errors. An incorrect integer part can strongly perturb the orbit correction, leading to non-convergence of the MICADO or SVD algorithm due to the error in the response matrix $\matR$, even when the fractional part $q$ is correct; $q$ is  typically obtained from turn-by-turn data of a single BPM using either a kick to the beam or naturally occurring oscillations. Fast Fourier transform (FFT) of the oscillation data gives the tune frequency $q$, as shown in Fig.~\ref{fig:turns-tune}.

\begin{figure}[tbp]
  \begin{center}
\includegraphics[width=0.6\linewidth]{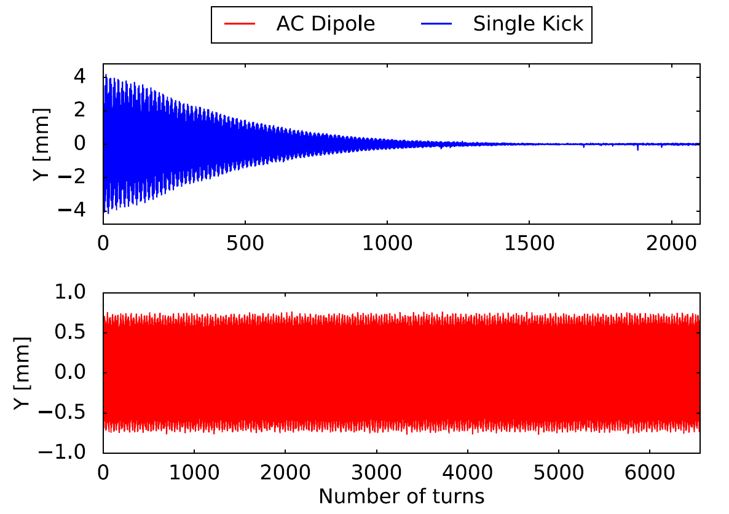}
   \end{center}
\caption{Single kick (top) and AC dipole (bottom)  excitation}
\label{fig:kicking-beam}
\end{figure}

\noindent The following are the two most common beam excitation methods:
\begin{itemize}
  \item a single kick  followed by free oscillation of the beam, damped by decoherence (tune spread between different particles due to magnetic non-linearities);
  \item an AC dipole-forced excitation at a fixed frequency, usually close to the tune.
\end{itemize}
Figure~\ref{fig:kicking-beam} shows examples of beam oscillations observed at a position monitor following a single and an AC dipole kick.
The advantage of the forced AC dipole oscillation is that a very long excitation time can be achieved, which provides better data quality. If the frequency is close to the tune but not inside the beam frequency spectrum, it yields emittance growth-free excitation for hadron beams. However, as this is a forced oscillation, it does not provide the tune directly, though it can be used for coupling and optics  measurements.

\begin{figure}[bht]
  \begin{center}
\includegraphics[width=0.65\linewidth]{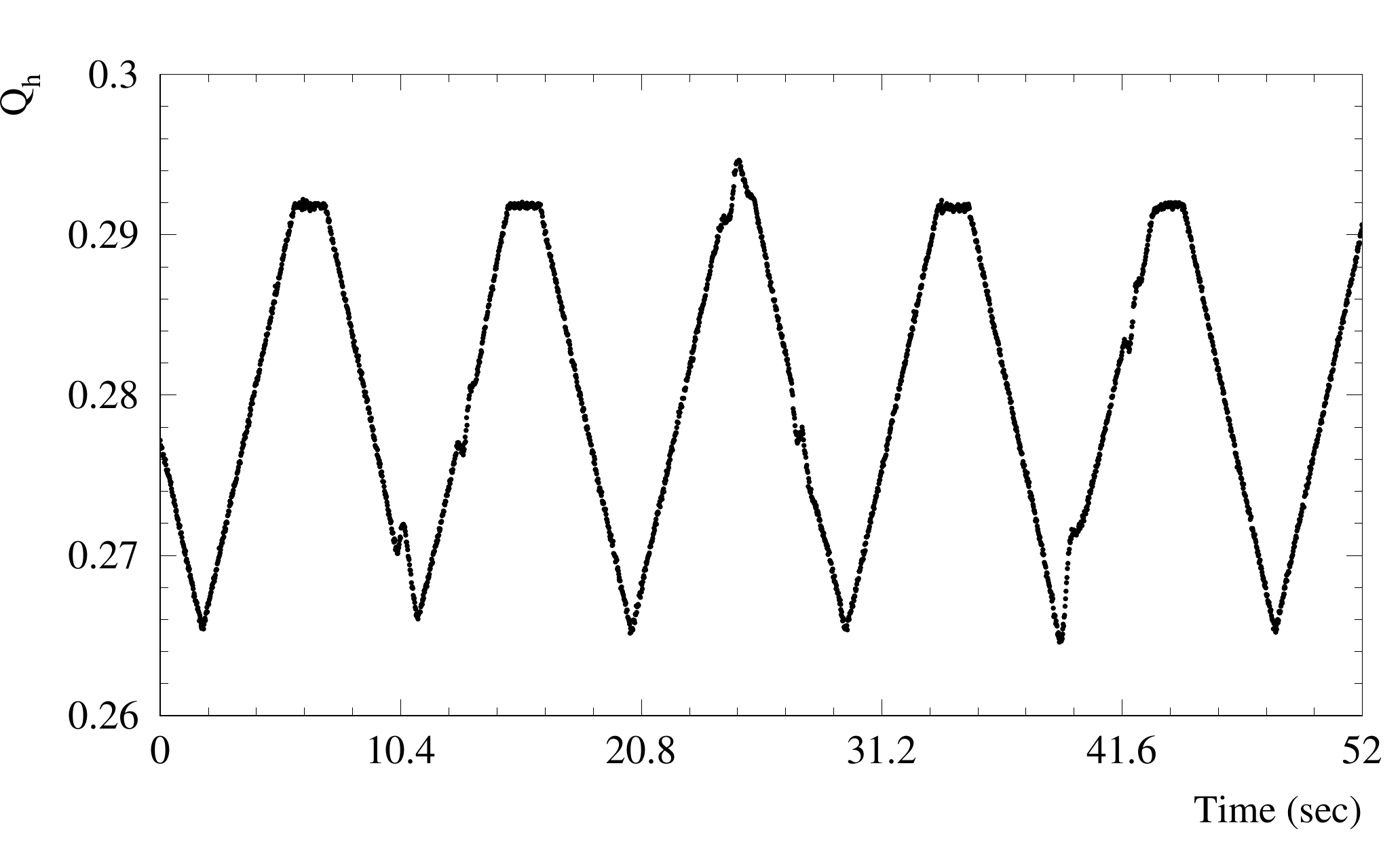}
   \end{center}
\caption{Example of tune tracking by a phase-locked loop (PLL) at the LEP; the distinct shape of the tune evolution in this example is due to a modulation of the beam energy through the radio frequency in view of a chromaticity measurement.}
\label{fig:tune-pll}
\end{figure}

A phase-locked loop (PLL) provides an alternative way to measure and continuously track tunes. An exciter shakes the beam very gently on the tune while remaining locked on the tune with the help of the beam response phase with respect to the excitation. This technique is ideal for e$^+$e$^-$ rings where damping will erase the effect of the excitation, but it is problematic for hadrons as it tends to produce emittance blow-up. A well-tuned PLL is able to track tunes with an accuracy of $\delta Q \sim 10^{-5}$ or better; an example is shown for the  LEP in Fig.~\ref{fig:tune-pll}. Such a system was used at the Relativistic Heavy Ion Collider (RHIC) to control tunes and coupling in ramp~\cite{Q-FB-RHIC}.

A direct and non-invasive technique for measuring tunes is based the Schottky spectrum of the beam. It  relies on noise in the beam, and no excitation is required, which makes it an ideal tool for hadron beams \cite{Q-SHOT-LHC}. The method is used extensively at low-energy hadron facilities. Due to the small signals, the detection electrodes must be close to the beam, or the beam must have a large charge $Ze$ (ions) because the Schottky signal amplitude is proportional to $Z^2$. For bunched beams the Schottky spectra can easily be dominated by large coherent beam oscillation signals, which dominate the response and make it challenging to extract the small incoherent signal. It should be noted that such a device is also able to provide chromaticity and emittance measurements~\cite{Q-SHOT-LHC}.

\begin{figure}[tbp]
  \begin{center}
\includegraphics[width=0.8\linewidth]{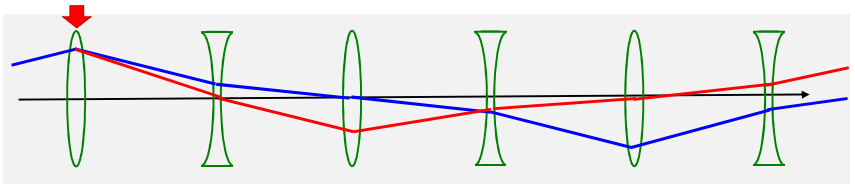}
   \end{center}
\caption{Sketch of the effect of a focusing error on the particle trajectory in the accelerator lattice. The focusing error induces a change in the phase of the oscillation and therefore also in the tune $Q$; in this case the tune increases.}
\label{fig:particle-in-lattice-deltak}
\end{figure}

A first source of tune error is the quadrupole gradient error; its effect on a particle oscillating in the lattice is shown schematically in Fig.~\ref{fig:particle-in-lattice-deltak}. The tune change $\Delta Q$ induced by a quadrupole strength change $\Delta k$ is given to first approximation by
\begin{equation}\label{eq:tune-deltak}
  \Delta Q \simeq \pm \frac{\beta_Q \Delta k l}{4 \pi}\:,
\end{equation}
where $l$ is the quadrupole length and $\beta_Q$ is the betatron function inside the quadrupole. The positive sign is for the horizontal plane and the negative sign for the vertical plane. More generally, such an error changes not only the tune but also the optics $(\beta(s), \mu(s))$ in the entire machine.

The magnetic field of a sextupole shifted by $x_0$ in the horizontal plane is a second source of tune (optics) errors (see Fig.~\ref{fig:coupling-error}), since the magnetic field  becomes
\begin{alignat}{4}
   B_x & = - 2 K_2 (x - x_0) y  \:, & \qquad  B_y & = - K_2 ((x - x_0)^2 - y^2) \: , \label{eq:sext-offset}\\
  B_x & = - 2 K_2 x y + 2 K_2 x_0 y  \:, & \qquad   B_y & = - K_2 (x^2 - y^2) + 2 K_2 x_0 x - K_2 x_0^2 \: ,  \label{eq:sext-offset-2}
\end{alignat}
with an additional quadrupolar field term $2 K_2 x_0$ and a small constant term corresponding to a deflection $- K_2 x_0^2$. A horizontally misaligned sextupole therefore generates an undesired quadrupole, and similarly a vertically shifted sextupole generates a skew quadrupole error. The strength of the undesired quadrupole is directly proportional to the shift~$x_0$.

While in the case of machine errors the tune change due to the gradient change is undesired, this effect can be used to adjust the tune over a small range during operation. Since in general the vertical and horizontal betatron functions differ at a quadrupole, the tune changes are different for the two planes (also in sign). By combining two (groups of) quadrupoles with different $\beta_x$ and $\beta_y$, it is possible to build combinations that affect the tune only in one plane, providing a set of two orthogonal knobs to trim the tunes independently in the two planes. For a set of two quadrupoles labelled 1 and 2, the tune changes in terms of gradient changes $\Delta k_1$ and $\Delta k_2$ are given by
\begin{equation}
\left( \begin{array}{c} \Delta Q_x \\ \Delta Q_y \end{array} \right)
\simeq \frac{1}{4 \pi} \left( \begin{array}{c} \beta_{x1} l_{Q1}  \\
- \beta_{y1} l_{Q1}  \end{array} \;
\begin{array}{c}   \beta_{x2} l_{Q2} \\
 - \beta_{y2} l_{Q2} \end{array} \right)
\left( \begin{array}{c} \Delta k_1 \\ \Delta k_2 \end{array} \right) \, .
\end{equation}
For properly selected quadrupoles this $2\times2$ matrix can be inverted to obtain $\Delta k_1$ and $\Delta k_2$ as functions of the desired tune changes $\Delta Q_x$ and $\Delta Q_y$. In large machines there are often distributed trim quadrupoles grouped in two (or more) families to spread out the correction and so reduce optics errors.

\subsection{Coupling}

Hadron machines usually operate very close to the diagonal in tune space; that is, the fractional tunes $q_x$ and $q_y$ are very close to each other, with $|q_x - q_y| \simeq 0.01$ typically. In such a configuration it is very important to decouple the horizontal and vertical planes. Large coupling can also be an issue for the vertical emittance in e$^+$e$^-$ machines.

\begin{figure}[tbp]
  \begin{center}
\includegraphics[width=0.35\linewidth]{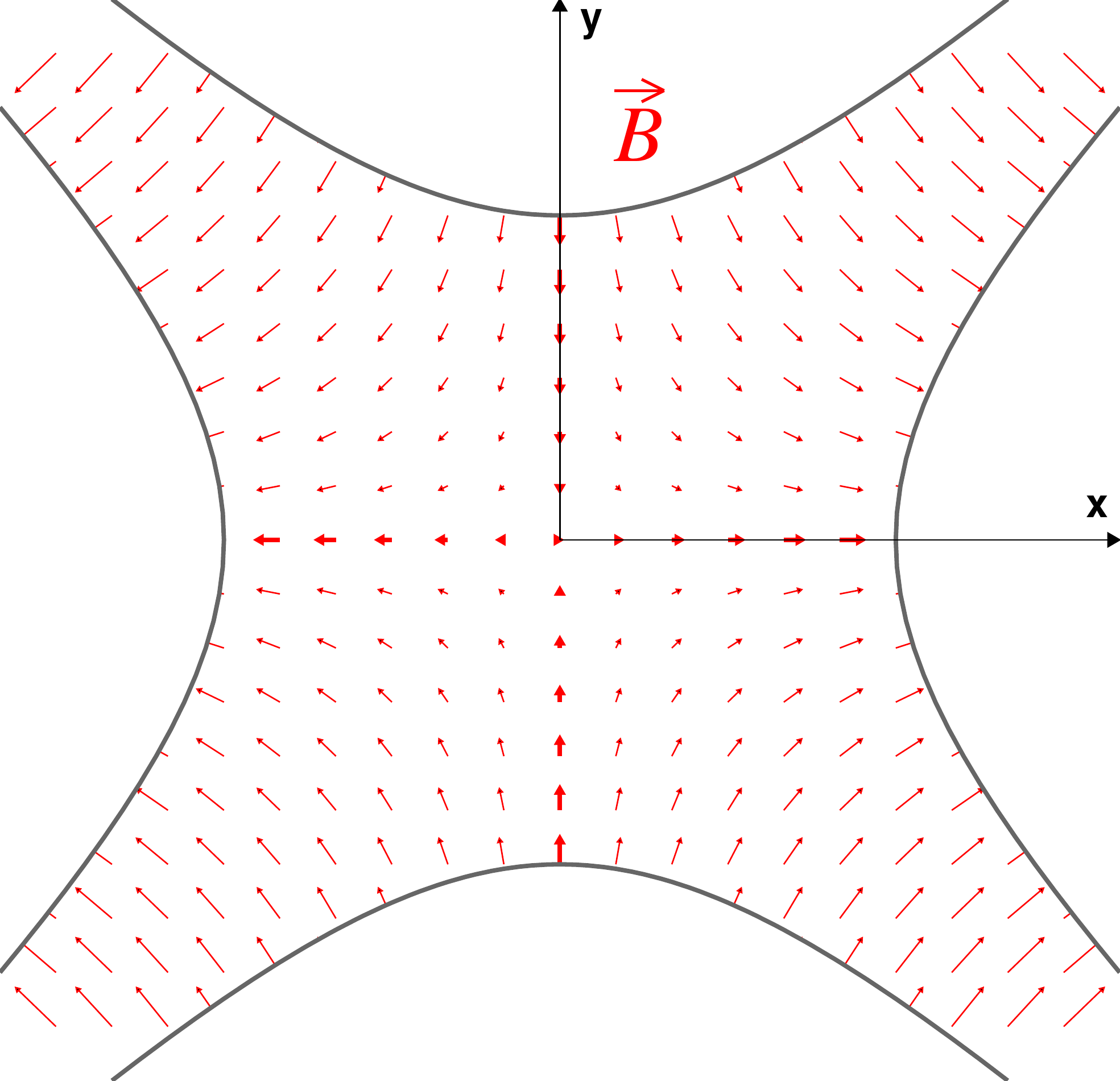} \hspace{10mm}
\includegraphics[width=0.35\linewidth]{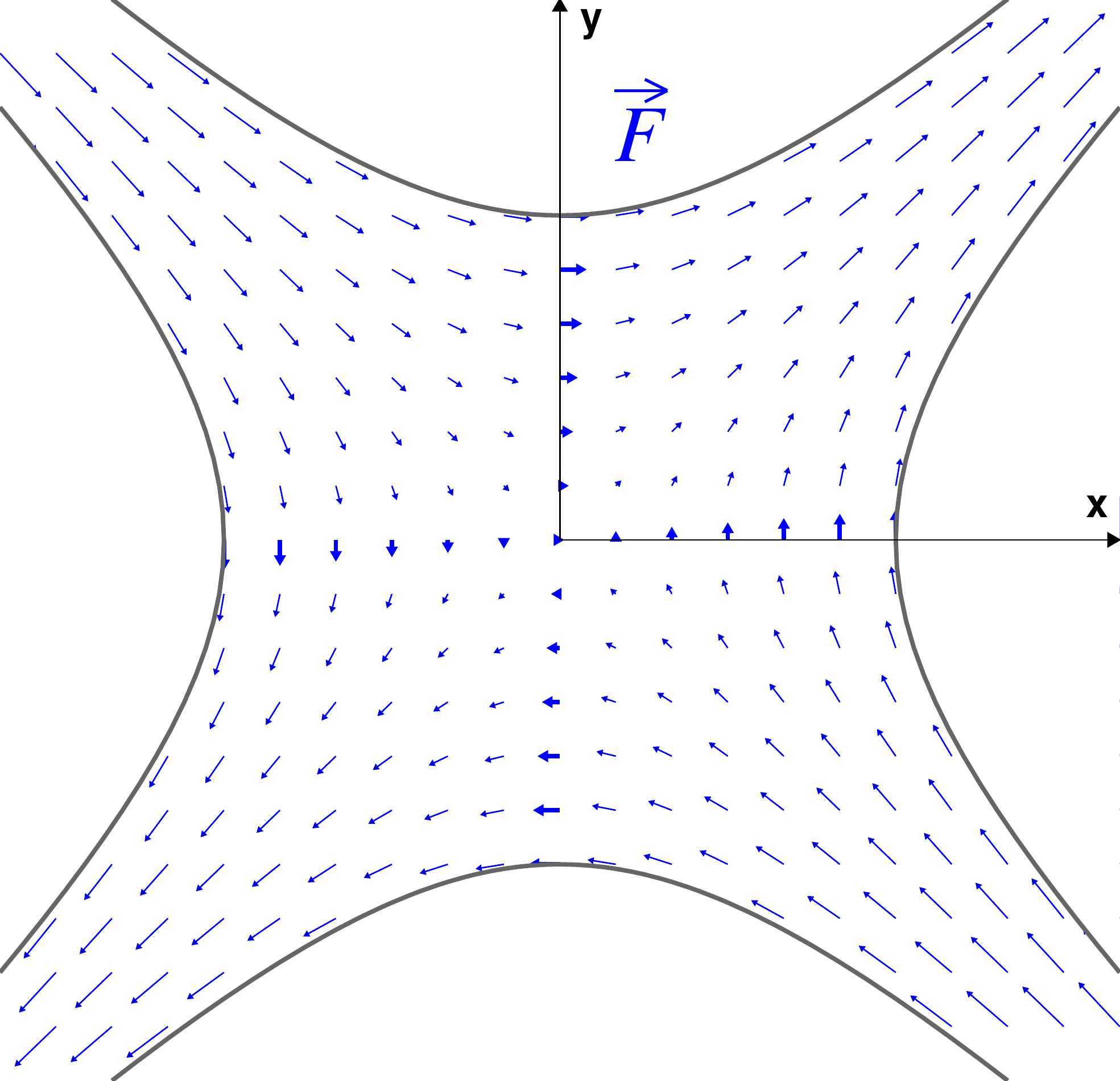}
   \end{center}
\caption{Schematic of a skew quadrupole magnet: magnetic field $\vec{B}$ (left) and Lorentz force $\vec{F}$  (right)  for a particle with positive charge coming out of the plane of this page. Both the magnetic field and the force grow linearly with distance from the quadrupole axis; on the axis there is no field (force). For the skew quadrupole, a horizontal offset leads to a vertical force and vice versa, which induces coupling between the two planes.}
\label{fig:quad-skew-b-f}
\end{figure}

\begin{figure}[tbp]
  \begin{center}
\includegraphics[width=0.8\linewidth]{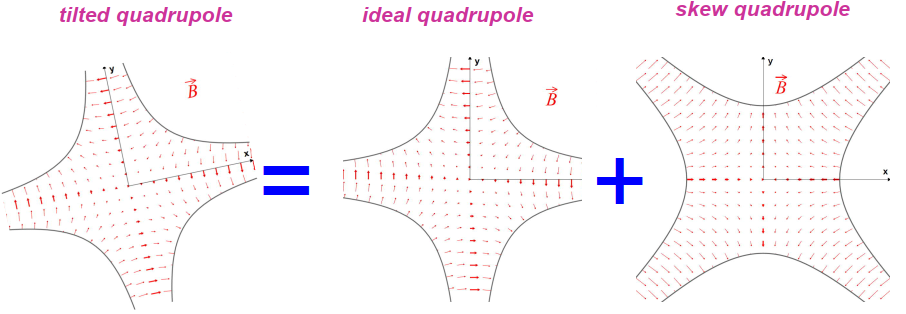} \\ \vspace{5mm}
\includegraphics[width=0.8\linewidth]{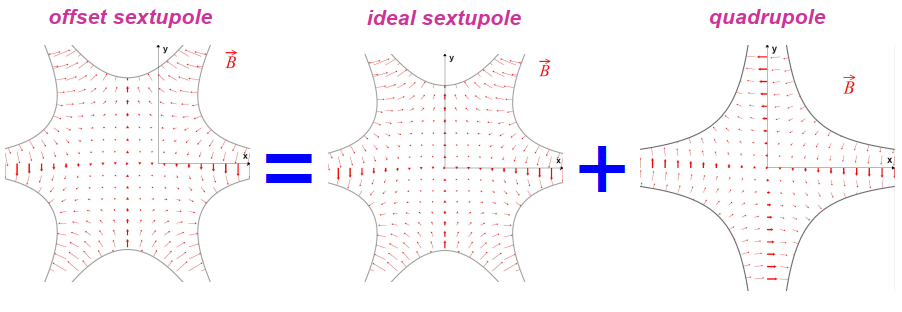}
   \end{center}
\caption{Sources of undesired coupling or quadrupole fields due to a rotated quadrupole (top) or a misaligned sextupole (bottom).}
\label{fig:coupling-error}
\end{figure}

If a quadrupole is rotated by 45$^{\circ}$, one obtains an element where the force (deflection) in $x$ depends on $y$ and vice versa: the horizontal and vertical planes are coupled. Such a quadrupole is called a skew quadrupole; Fig.~\ref{fig:quad-skew-b-f} illustrates the magnetic field and the Lorentz force for such a magnet. Small quadrupole rotations in the $x$--$y$ plane lead to coupling of the $x$ and $y$ planes because this introduces a small skew quadrupole as shown in Fig.~\ref{fig:coupling-error}. Conversely, coupling between the two transverse planes can be corrected by installing dedicated skew quadrupoles to compensate for alignment or skew quadrupolar field errors. Possible sources of undesired coupling in a machine include the following:
\begin{itemize}
  \item element misalignments (for example, roll angles of quadrupoles);
  \item skew quadrupolar field errors;
  \item solenoids (experiments, electron coolers, etc.);
  \item orbit offsets in sextupoles (or misaligned sextupoles) as shown in Fig.~\ref{fig:coupling-error}; a horizontal offset in sextupoles leads to a normal quadrupolar error and hence to a tune and optics change, while a vertical offset introduces spurious coupling.
\end{itemize}

\begin{figure}[tbp]
  \begin{center}
\includegraphics[width=0.3\linewidth]{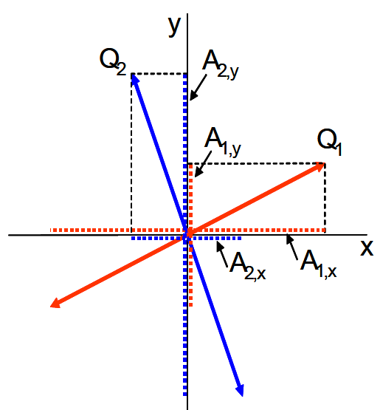}
   \end{center}
\caption{Tune eigenmodes $Q_1$ and $Q_2$ in the presence of coupling}
\label{fig:eigen-q}
\end{figure}

In the presence of coupling the two beam eigenmodes no longer coincide with $Q_x$ and $Q_y$. Instead, the eigenmodes are rotated in the $x$--$y$ plane as shown in Fig.~\ref{fig:eigen-q}. A simple way of characterizing the coupling coefficient $C-$ consists in measuring the crossed tune peak amplitudes, i.e.\ the vertical tune in the horizontal spectrum and vice versa. The coupling coefficient is then given by~\cite{Q-FB-RHIC}
\begin{equation}\label{eq:couple-tunes}
C- = \frac{2 \sqrt{r_1 r_2\, |Q_1 - Q_2|}}{1 + r_1 r_2}   \qquad \text{with } \;  r_1 = \frac{A_{1,y}}{A_{1,x}} \; \text{ and } \; r_2 = \frac{A_{2,x}}{A_{2,y}}\:,
\end{equation}
where $A_{i,u}$ is the amplitude of the peak of plane $u$ for mode $i$.  This is a simple measurement, but it does not provide any phase information. Only the local coupling is obtained, which can differ from the global coupling.

\begin{figure}[tbp]
  \begin{center}
\includegraphics[width=0.65\linewidth]{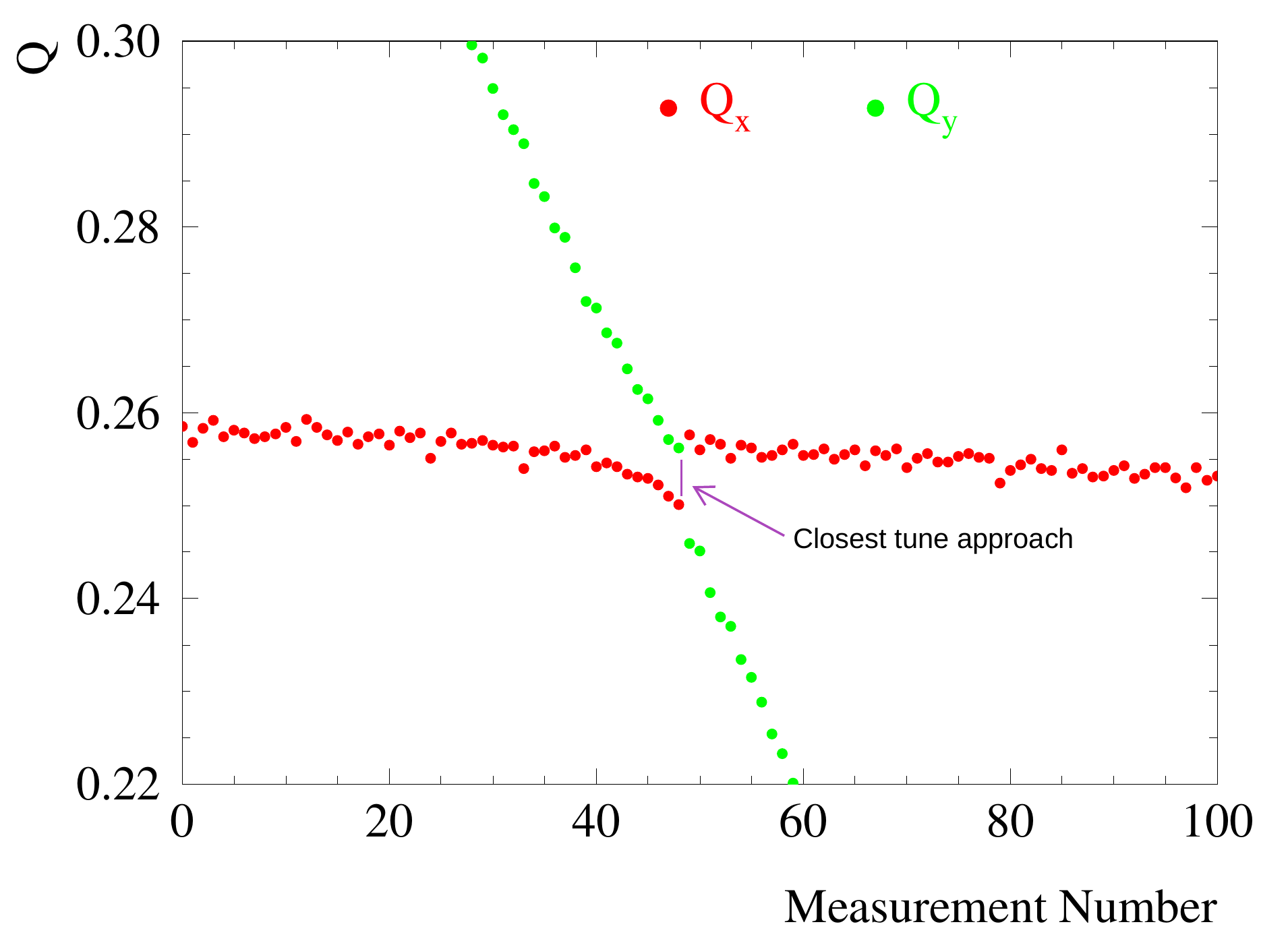}
   \end{center}
\caption{Example of a closest tune approach measurement at the LEP. The minimum approach distance is clearly visible.}
\label{fig:closest-q}
\end{figure}

The global machine coupling can also be determined directly using the closest tune approach~\cite{LEP-PERF}. This measurement technique requires  moving the tunes close to each other or even  moving one of the tunes across the other, as shown in Fig.~\ref{fig:closest-q}. It  provides information on the global coupling based on the tune measurement from a single location. The closest distance of approach of the tunes, $\Delta Q_{\min}$, corresponds to the coupling parameter~$C-$.

Coupling can also be determined along the machine from multi-turn beam position data (with, for example, AC dipole excitation), using pairs or groups of BPMs to reconstruct the coupling locally \cite{Q-COUP-LHC}. Such a technique provides detailed local coupling information, including the phase. The global coupling value can be obtained by integration of the local coupling. This method, although more invasive and operationally complex, provides excellent deterministic corrections.

The coupling correction scheme used depends on the machine design. Ideally experimental solenoids should be compensated by local anti-solenoids to correct the coupling source as close to the source as possible. At high-energy hadron colliders such as the LHC, the solenoids of the experiments contribute very little to the machine coupling because of the very high momentum. The global machine coupling is usually corrected with distributed skew quadrupoles, either using two orthogonal knobs (similar to a tune correction) or with more refined local corrections. For measurements that do not provide phase information (only global coupling), the two orthogonal knobs must be scanned by trial and error to determine a correction.
As an alternative, orbit bumps in sextupoles may be used for coupling corrections, but this can lead to problems with dispersion, requiring a careful combined correction of both parameters.

\section{Chromaticity}

The linear chromaticity defines the dependence of tune on momentum:
\begin{equation}\label{eq:chroma}
  Q' = \frac{\Delta Q}{\Delta p/p}\:.
\end{equation}
For a synchrotron lattice without any sextupole, $Q'$ is usually negative and  in general almost equals the tune, $Q' \approx -Q$. Sextupoles are used to adjust the chromaticity and are usually installed in regions with horizontal dispersion. As a consequence of the dispersion, the particles in the beam are sorted horizontally by momentum proportional to their relative energy offset $\Delta p/p$. For a position offset $x_0 = D_x \Delta p/p$, an additional quadrupole field with strength proportional to $K_2 x_0 = K_2 D_x \Delta p/p$ is generated according to Eq.~\eqref{eq:sext-offset-2}. The corrections are usually distributed over many sextupoles in the form of orthogonal knobs for the two planes to avoid degrading the dynamic aperture of the machine. Above transition energy, $Q'$ is normally set to be slightly positive ($Q' \sim 2$--$10$); below transition it is set  slightly negative. The choice of sign is determined by the need to control collective effects such as head--tail modes.

\begin{figure}[tbp]
  \begin{center}
\includegraphics[width=0.65\linewidth]{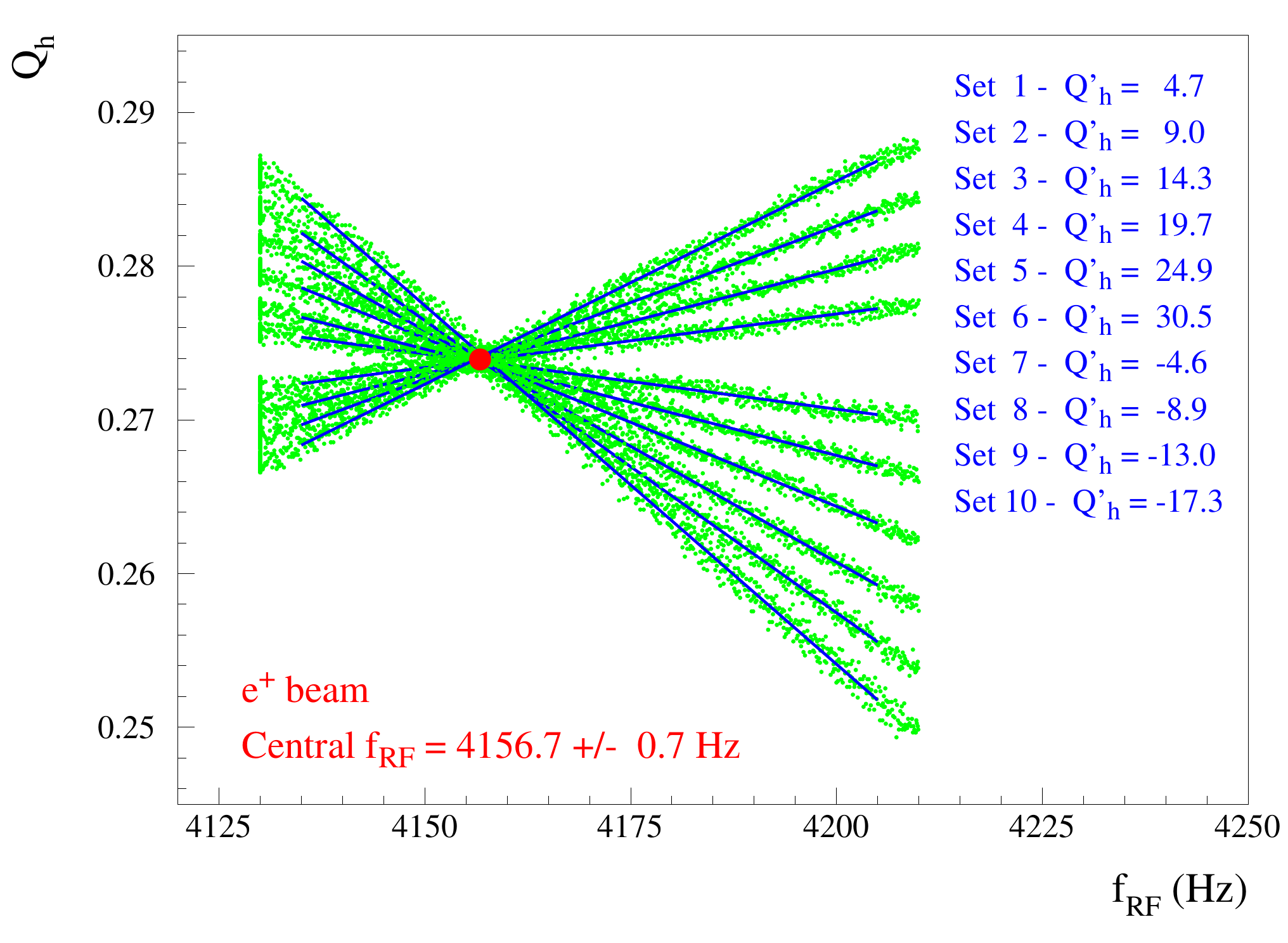}
   \end{center}
\caption{Measurement of the chromaticity at the LEP with a PLL technique. Each line corresponds to a $Q'$ measurement from the tune $Q$'s dependence on the beam momentum change (proportional to  the radio frequency change). The different lines correspond to different chromaticity settings, with slopes proportional to the chromaticity. At the crossing point of all the curves, the tune is independent of the chromaticity, which implies that the beam is centred on average in the machine sextupoles.}
\label{fig:chroma-pll-lep}
\end{figure}

The chromaticity is generally measured by changing or modulating the energy offset $\Delta p/p$ through the radio frequency while recording the tune change $\Delta Q_u$. An example is given in Fig.~\ref{fig:chroma-pll-lep} for the LEP.
A Schottky monitor can also be used to determine $Q'$ for hadron beams without the need for a radial modulation, providing a non-invasive measurement; $Q'$ is related to the difference in width of the upper and lower Schottky side-bands~\cite{Q-SHOT-LHC}.

\section{Linear optics}

Knowledge of the beam optics is essential for the good performance of an accelerator. Tools for measuring and correcting the optics towards a design model are essential at any modern facility.

\begin{figure}[tbp]
  \begin{center}
\includegraphics[width=0.85\linewidth]{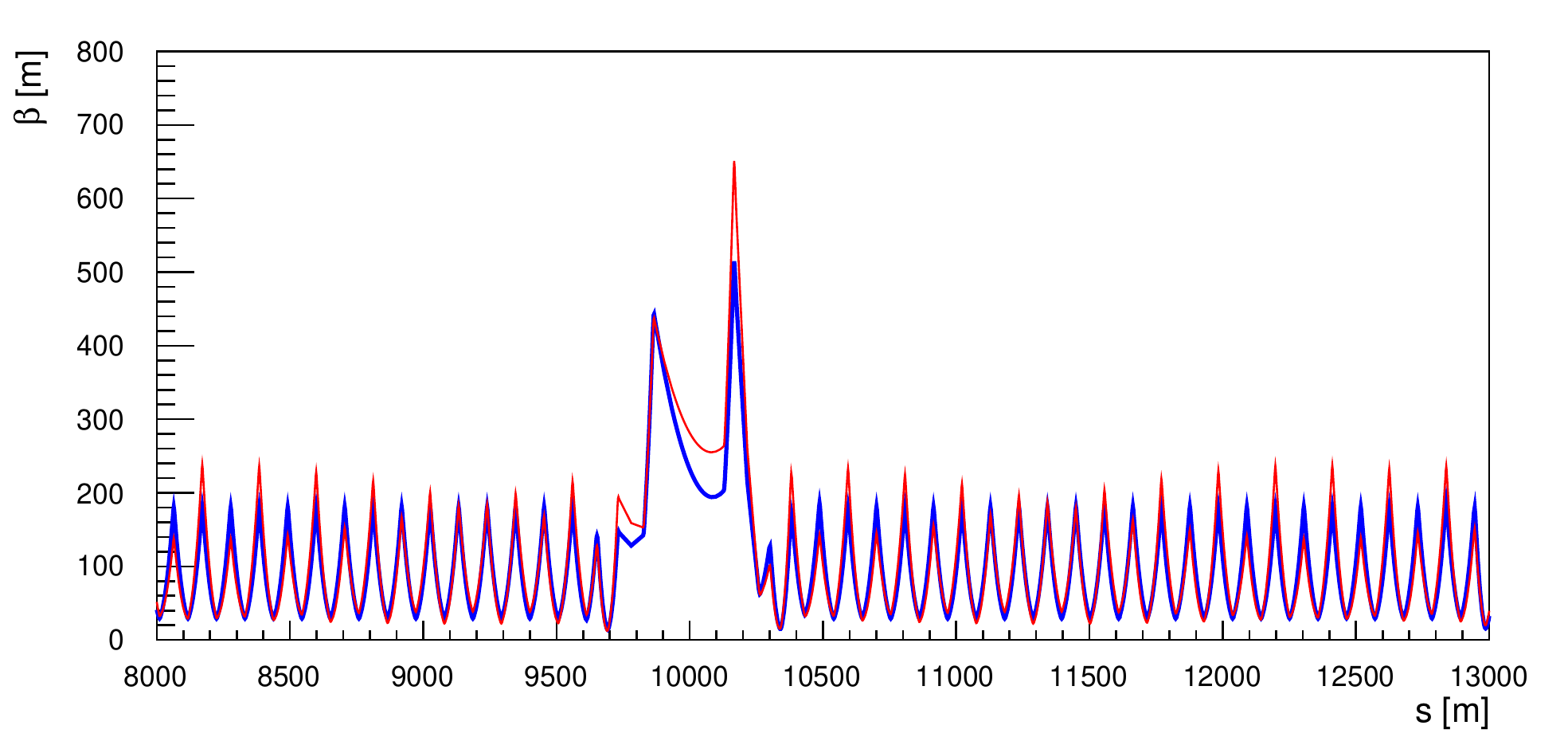}
\includegraphics[width=0.85\linewidth]{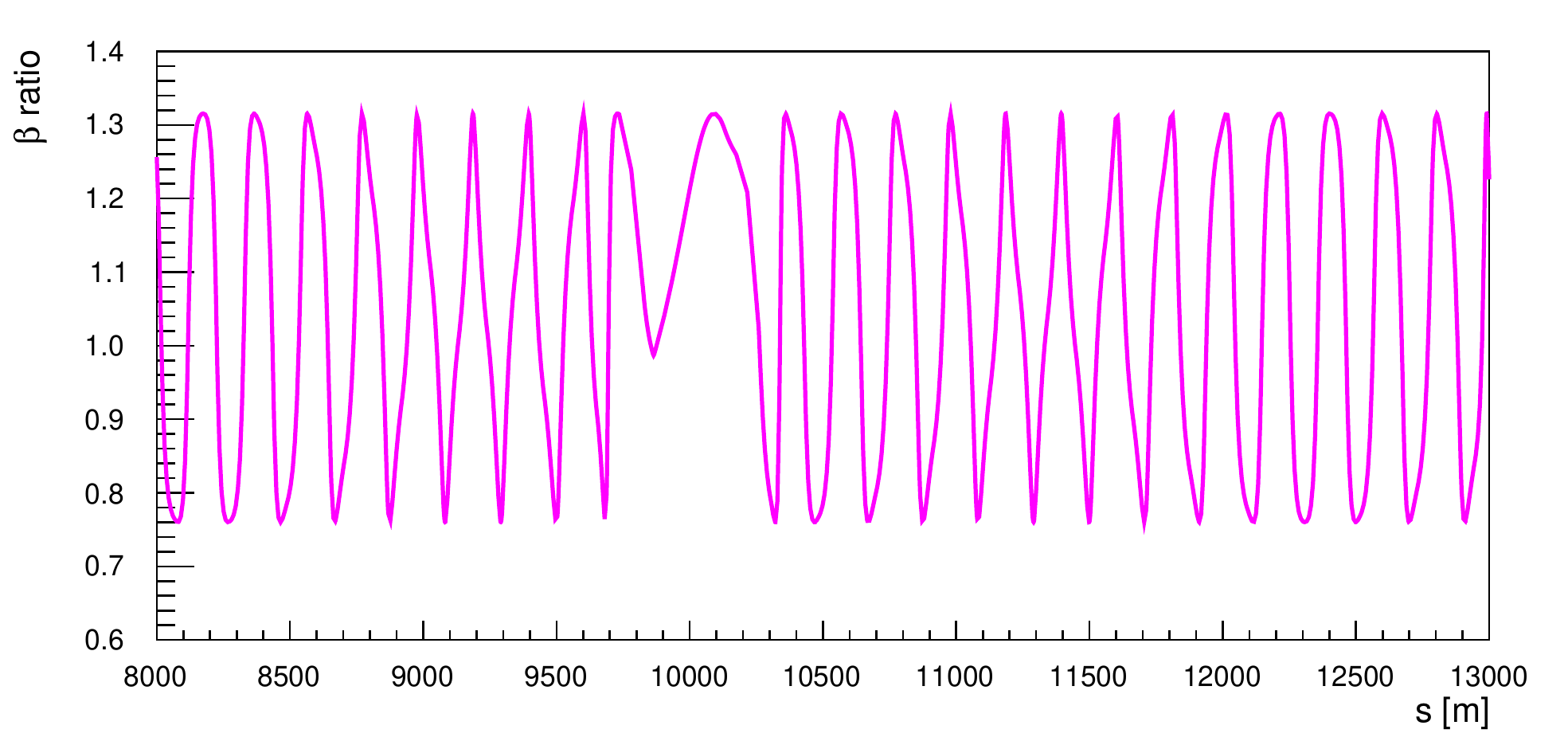}
\includegraphics[width=0.85\linewidth]{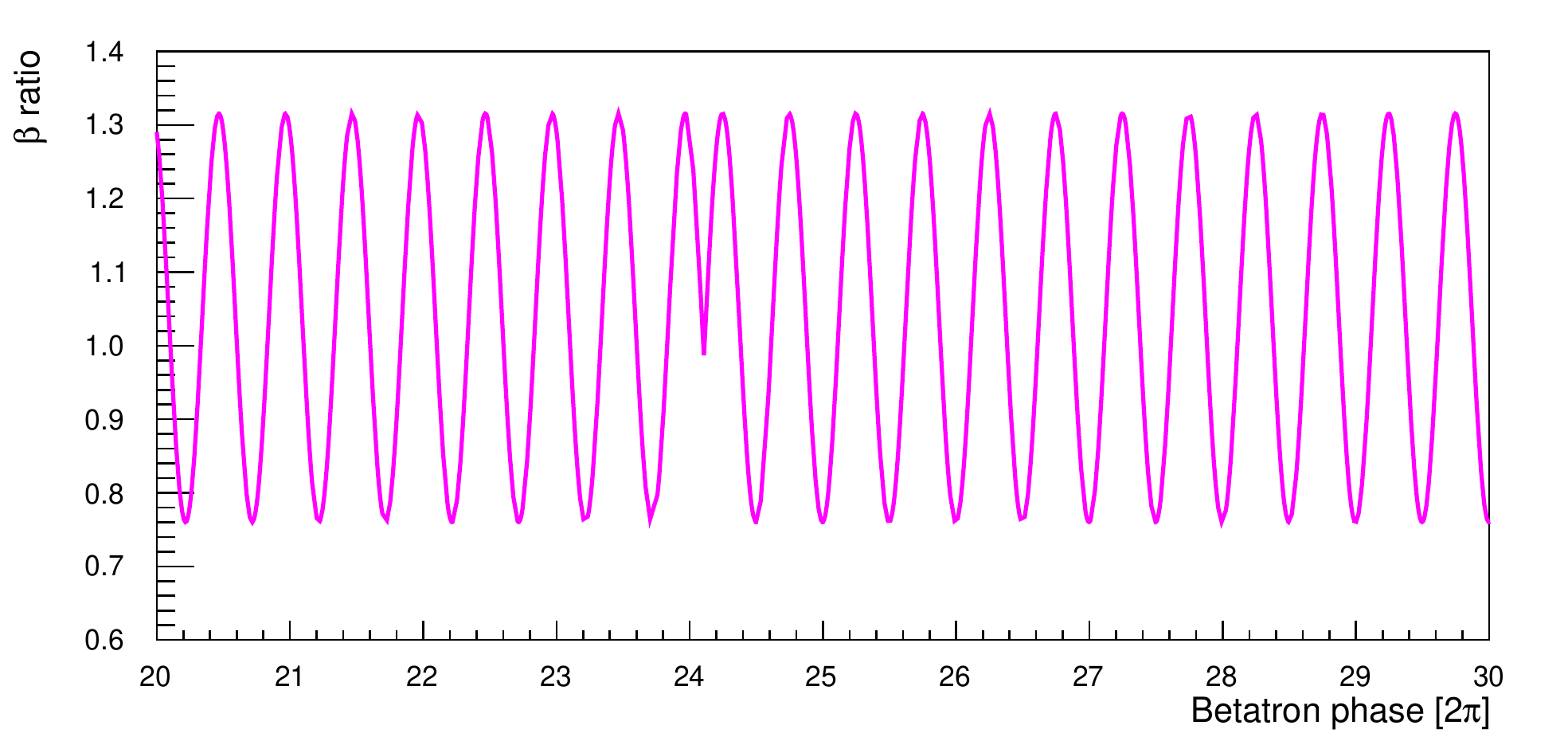}
   \end{center}
\caption{Example of beta-beating in a section of the  LHC. In the top graph the nominal (blue) and perturbed (red) betatron functions are plotted as functions of the longitudinal coordinate $s$. The middle plot shows the ratio of the perturbed to the nominal betatron function as a function of $s$. In the bottom plot the $s$ coordinate has been replaced by the nominal phase advance $\mu$ to reveal the oscillation and the kink at the location of the error more clearly.}
\label{fig:beta-beat-exam}
\end{figure}

A quadrupole gradient error, as shown in Fig.~\ref{fig:particle-in-lattice-deltak}, affects the machine tune and the optical functions (betatron function and phase advance). While for a linac the error only propagates to downstream sections, for a circular machine the error affects the entire ring  in a similar fashion to the closed-orbit error. Figure~\ref{fig:beta-beat-exam} displays the effect of a local error on the betatron function along the ring. The beam optics perturbation exhibits an oscillating pattern; the ratio of the perturbed to the nominal betatron function has an oscillating pattern called the betatron function beating, or ‘beta-beating’. The normalized amplitude of the perturbation is the same over the entire ring, and a kink appears on the oscillation at the location of the error in a similar fashion to the closed-orbit error. A careful inspection of Fig.~\ref{fig:particle-in-lattice-deltak} reveals, however, that there are two oscillation periods per $2\pi$ ($360^\circ$) phase advance: the beta-beating frequency is twice the tune frequency.

The betatron function error (beta-beating) at an observation point $j$ due to a number of strength errors $\Delta k_i$ is given to first order by
\begin{equation}\label{eq:beta-beat}
  \frac{\Delta \beta_j}{\beta_j} \simeq \sum_i \frac{\Delta k_i l_i \beta_i}{2 \sin(2 \pi Q)} \cos\bigl(2 \pi Q - 2|\mu_j - \mu_i|\bigr) = \sum_i B_{ij} \Delta k_i\:,
\end{equation}
where $l_i$ is the length of the element generating the error $\Delta k_i$. The denominator $\sin(2 \pi Q)$ generates a diverging correction when $Q = N$ and when $Q = N + 0.5$ (with $ N \in \mathbb{N}$); such errors drive integer and half-integer resonances.
In contrast to the case of orbit kicks, gradient errors have a non-linear effect on the betatron function. A correct treatment must be self-consistent; the equation above is only an approximation. The problem can, however, be linearized using the matrix elements $B_{ij}$ and solved iteratively with the SVD or MICADO algorithm based on measurements of $\Delta \beta_j$, since the mathematical structure is identical to that of the closed-orbit case. After each correction iteration the matrix elements $B_{ij}$  must, however, be re-evaluated.

\section{Optics measurement and correction}

There are three main techniques for measuring and reconstructing the machine optics~\cite{O-REV}.
\begin{itemize}
  \item K-modulation: the strength of individual quadrupoles is modulated to determine the local optics function.
  \item Orbit (trajectory) response: the orbit or trajectory response matrix is measured with orbit corrector kicks (see orbit correction), and a fit to the response is used to reconstruct and correct the machine model.
  \item Multi-turn beam position data: the beam is excited and multi-turn beam position data is recorded to determine the betatron phase advance between beam position monitors; the betatron function is reconstructed from the phase advance information.
\end{itemize}

\subsection{K-modulation technique}

K-modulation has already been described as a means to determine the BPM offset with respect to the quadrupole magnetic axis.
This technique can also be used to determine the average betatron function inside the modulated quadrupole, since the tune change $\Delta Q$ due to a gradient change $\Delta k$ is~\cite{O-KMOD-LHC, O-BSTAR-LHC, O-MT-LHC2}
\begin{equation}\label{eq:tune-kmod}
  \Delta Q = \frac{1}{4 \pi} \int_{s_0}^{s_0+l} \Delta k \beta(s) \,\rd s \; .
\end{equation}
The average betatron function in the quadrupole is then given by
\begin{equation}\label{eq:beta-kmod}
\beta_Q = \frac{2}{l \Delta k} \left[ \mathrm{cotg}(2 \pi Q) - \frac{\cos(2 \pi (Q + \Delta Q))}{\sin(2 \pi Q)} \right] \, .
\end{equation}
This technique is powerful and simple but requires the quadrupoles to be powered individually. Such an individual powering scheme is found frequently at  synchrotron light sources, but not at large machines in which only  a subset of quadrupoles may be changed individually due to the cost of the power converters.

\subsection{Orbit response technique}

The method of orbit (or trajectory) response (ORM) exploits the large amount of information that is encoded in the orbit response matrix $\matR$ given  in Eqs.~\eqref{eq:rij-co} and \eqref{eq:rij-st}. The principle behind the technique, available in the popular LOCO code~\cite{O-LOCO-BDYN}, is to excite steering elements and record the BPM response. This provides a measurement of the response matrix folded with BPM calibration factors $b_i$ and orbit corrector deflection calibration factors $\kappa_j$:
\begin{equation}
  \vd = {\bf R}^{\rm meas} \vk  \qquad \text{with } \;   R_{ij}^{\rm meas} = \frac{ b_i \kappa_j \sqrt{\beta_i \beta_j} \,\cos(| \mu_i - \mu_j |) }{2 \sin(\pi Q)}\:.
\end{equation}
All the elements of the response matrix $\matR$ are potential observables for a model fit.

\begin{figure}[tbp]
  \begin{center}
\includegraphics[width=0.7\linewidth]{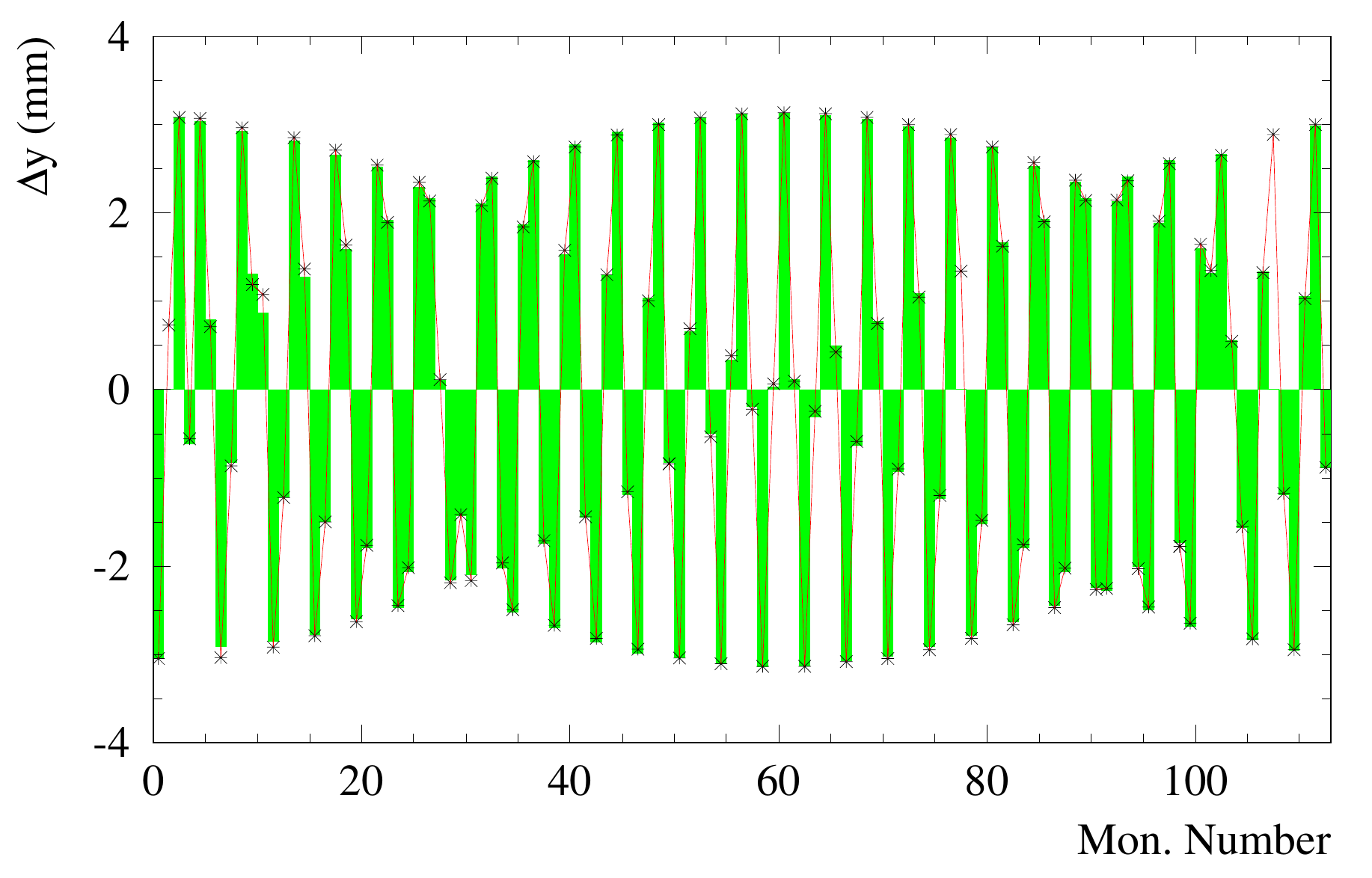}
\includegraphics[width=0.7\linewidth]{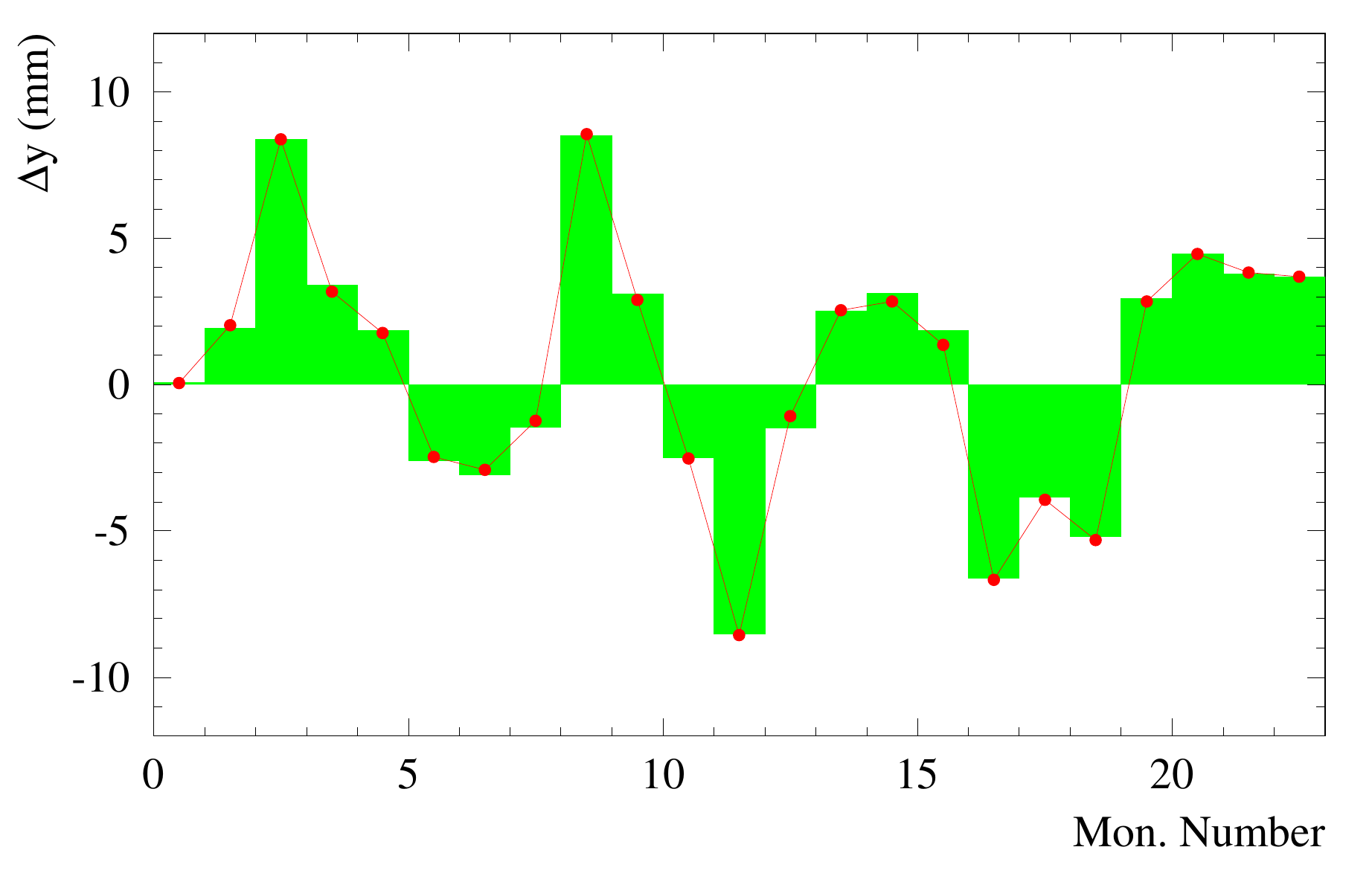}
   \end{center}
\caption{Example of orbit response measurements (green histogram) and fits (stars/dots) for the SPS and the CERN Neutrinos to Gran Sasso (CNGS) transfer lines~\cite{O-LOCO-SPS, O-LOCO-CNGS}.}
\label{fig:orm-examples}
\end{figure}

To fit the response data, the ORM matrix is linearized in the form of a vector $\vec{r}$, where each element $r_i$ corresponds to one element of the response matrix $\matR$~\cite{O-LOCO-SPS}:
\begin{equation}
  r_i = \frac{R_{ij}^{\rm meas} - R_{ij}^{\rm model}}{\sigma_{ij}}\:,
\end{equation}
normalized by the estimated measurement error $\sigma_{ij}$.
When all elements are measured, the size of the vector $\vec{r}$ corresponds to the $N \times M$ elements of $\matR$.
The fitted parameter vector $\vec{c}$ may be composed of:
\begin{itemize}
  \item BPM calibration factors and roll angles;
  \item steering element calibrations and roll angles;
  \item quadrupole gradients (skew and normal);
  \item any other model parameters.
\end{itemize}
A response matrix $\matG$ is constructed from the dependence of each $r_i$ on any $c_j$, $G_{ij} = \partial r_i / \partial c_j$, leading to the linear equation
\begin{equation}
  \vec{r} + \matG \Delta \vec{c} = 0 \: .
\end{equation}
The response analysis is coupled to an accelerator design tool such as MAD or PTC in order to determine the sensitivity to the quadrupole gradients and other model parameters. Once the system is cast in matrix form, it is solved by SVD inversion since the equation is structurally equivalent to an orbit correction. The tricks to filter noise by eliminating small eigenvalues, as discussed for beam steering, are employed here as well. A few iterations may be required for convergence. At each iteration $\matG$ must be re-evaluated. Figure~\ref{fig:orm-examples} shows an example of response data and fits  for the CERN Super Proton Synchrotron (SPS)~\cite{O-LOCO-SPS,O-LOCO-CNGS}.
ORM has been used with much success at synchrotron light sources, where it is a standard tool with
typically hundreds of BPMs, steerers, and quadrupoles~\cite{O-LOCO-SOLEIL}. The size of the matrix $\matG$ grows rapidly; for a machine with 100 BPMs and 100 steering elements, there are 10\,000 lines and over 200 columns. For very large machines such as the LHC and FCC-hh, the data volumes are immense and multi-turn methods are faster; therefore ORM techniques are mainly useful in calibrating BPMs and steerers.

\subsection{Multi-turn technique}

\begin{figure}[tbp]
  \begin{center}
\includegraphics[width=0.7\linewidth]{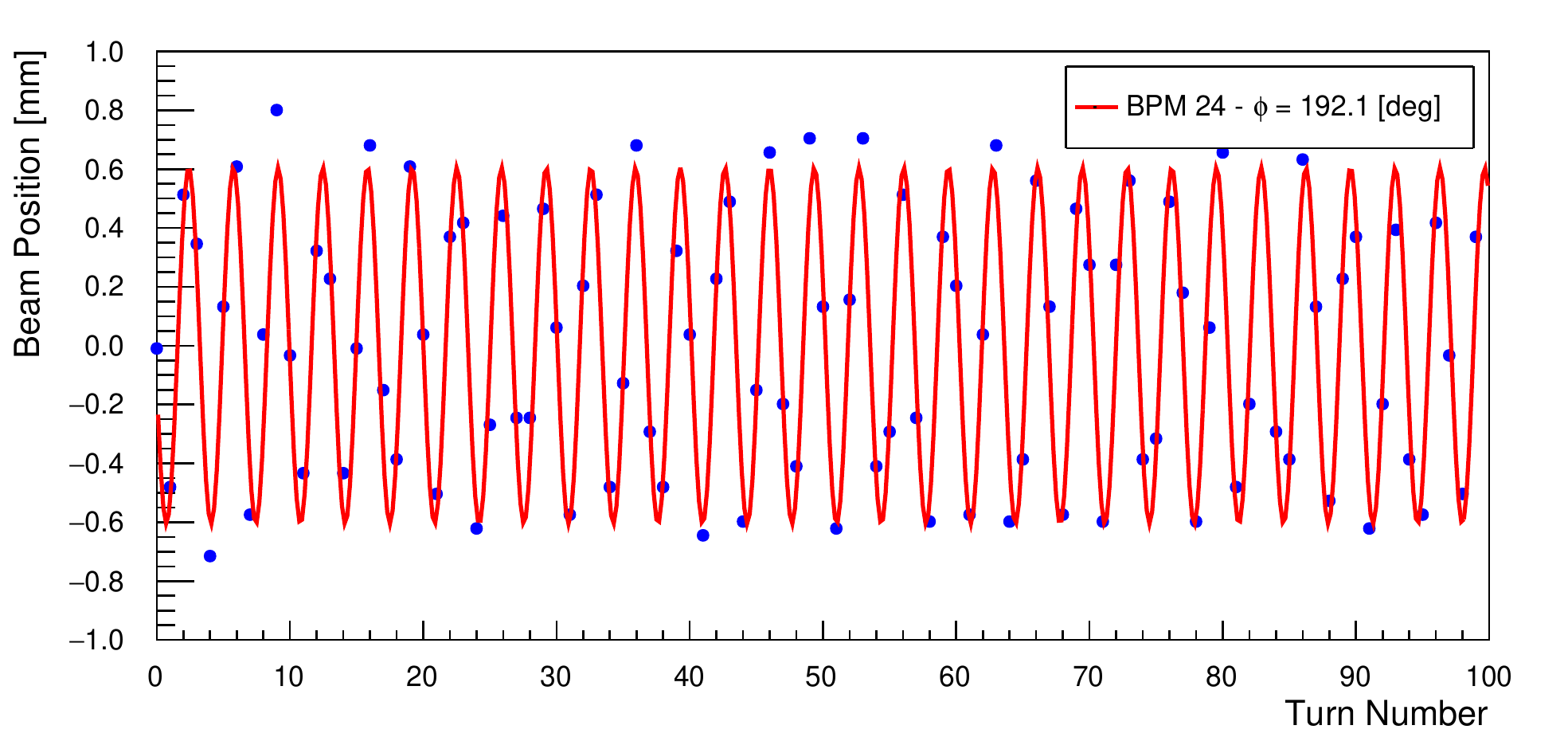}
\includegraphics[width=0.7\linewidth]{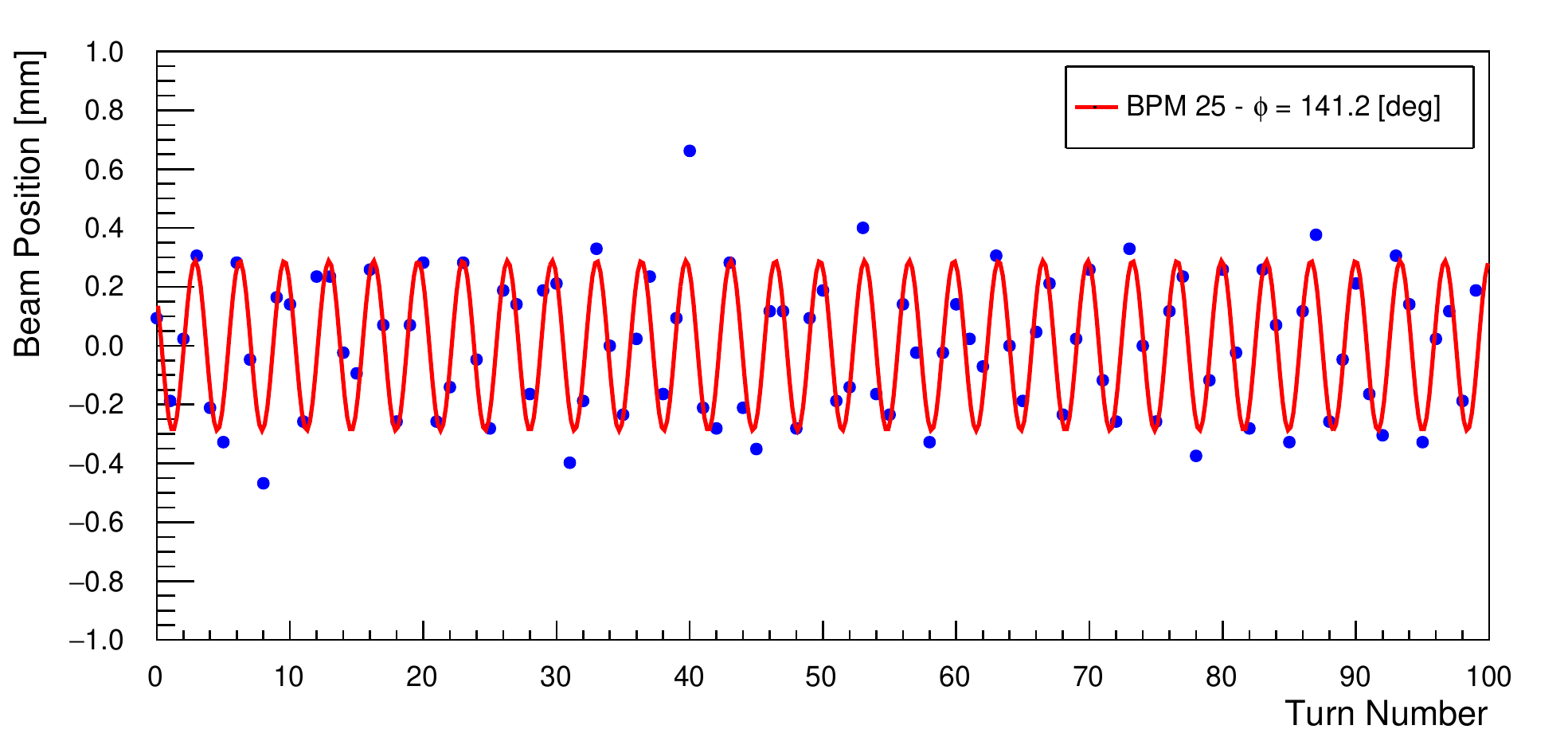}
\includegraphics[width=0.7\linewidth]{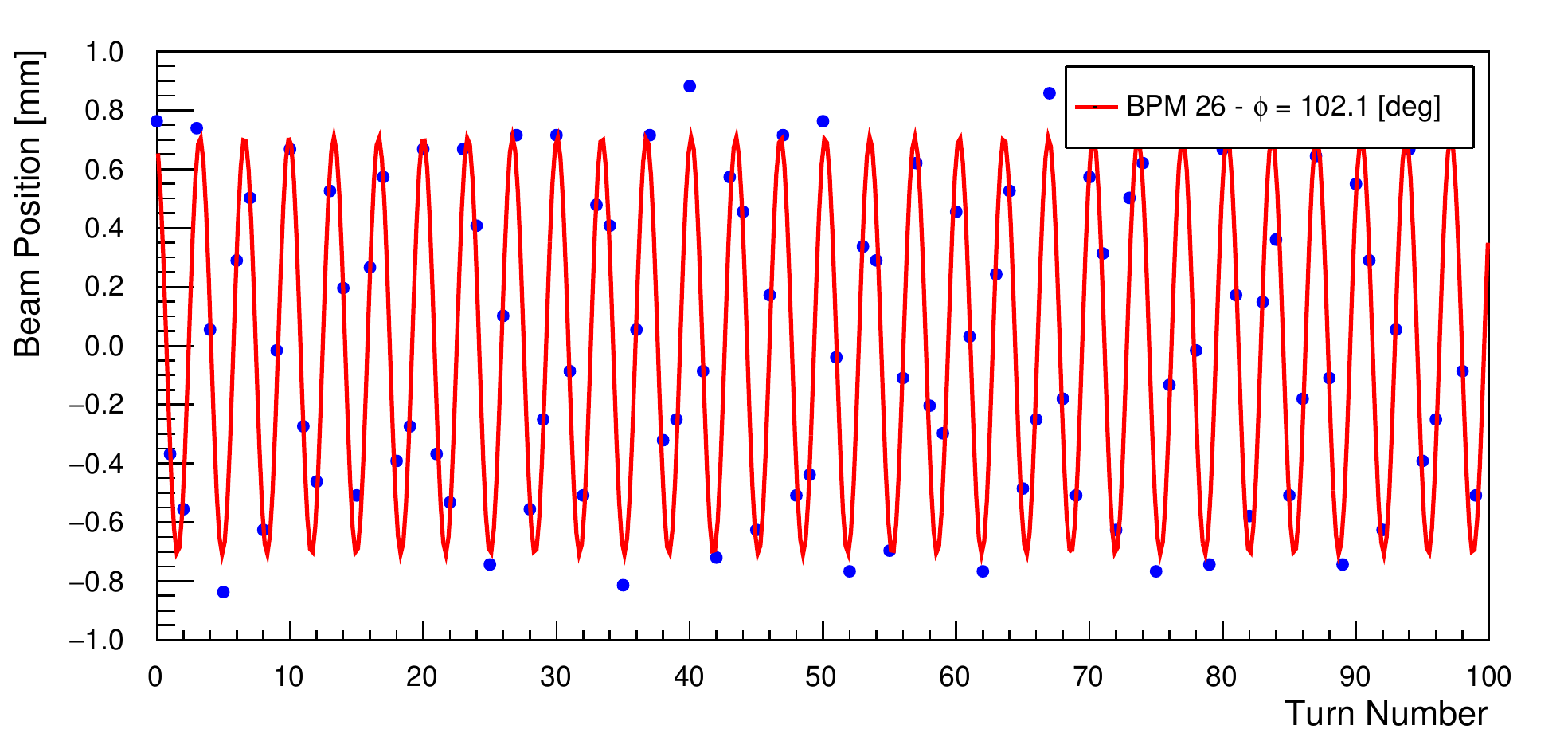}
   \end{center}
\caption{Example of multi-turn position data (blue points) and the associated oscillation fit (red curve) for an AC dipole excitation at the LHC. The three panels correspond to three consecutive BPMs labelled 24, 25, and 26. The phase advance is $50.9^{\circ}$ between BPM~24 and BPM~25, and is $39.1^{\circ}$ between BPM~25 and BPM~26.}
\label{fig:bpms-phase-advance}
\end{figure}

Multi-turn optics measurements rely on a beam excitation for a certain number of turns, typically a few thousand to obtain sufficient resolution~\cite{O-MT-LHC2, O-MT-LHC1}. The beam oscillation phase is extracted for each BPM; this phase corresponds to the betatron phase $\mu$ at each BPM. The betatron phase advance $\Delta \mu$ between two BPMs can then be extracted in a straightforward way. An important advantage is that the phase measurement does not depend on the BPM calibration, though it remains sensitive to BPM non-linearities that may bias the phase reconstruction. Exciting the beam with a single kick is often limited by the decoherence of the oscillation (or by radiation damping).
Excitation by an AC dipole (Fig.~\ref{fig:kicking-beam}) is more favourable, since the number of turns can be increased to obtain better measurement accuracy. Figure~\ref{fig:bpms-phase-advance} presents an example of multi-turn data for BPMs at the LHC: the betatron phase advance is directly obtained from the phase of the oscillation between adjacent BPMs. Obviously such a measurement implies that the data of all the BPMs must be correctly synchronized to the same turn.

The betatron function can be reconstructed from the phases obtained for three BPMs under the assumption that there are no sources of errors between those BPMs. For three consecutive BPMs labelled 1, 2, and 3, the measured betatron function at BPM~1, $\beta_1^{\rm meas}$, may be obtained with input from the model as
\begin{equation*}
  \beta_1^{\rm meas} = \beta_1^{\rm model} \frac{\coth(\Delta \mu_{12}^{\rm meas}) - \coth(\Delta \mu_{13}^{\rm meas})}{\coth(\Delta \mu_{12}^{\rm model}) - \coth(\Delta \mu_{13}^{\rm model})}\:.
\end{equation*}
The raw turn data is often filtered for noise by SVD before the phase is extracted, and multi-BPM interpolation techniques have been developed to improve the accuracy of this technique, in particular when the phase advances between neighbouring BPMs are unfavourable~\cite{O-MULTI-BPM, O-REV}.

An advantage of the multi-turn technique is that the optics for both planes is obtained from two measurements which can be very fast, and for large machines the data volumes are not as immense as for ORM techniques. In addition, this method is not sensitive to BPM calibrations.
A disadvantage of the multi-turn technique is that a fast kicker is required, and the beam must be excited to sufficiently large amplitude relative to the BPM turn-by-turn resolution, which can be an issue when the free aperture for kicking the beam is limited.
While for ORM measurements the BPM noise is in general not an issue (averaged over many turns), it can be an problem for single-turn measurements at low bunch intensity, as may be required for a superconducting machine.

Once the optics data is extracted from the multi-turn data, an optics modelling tool must to used to fit machine errors to the data, and to establish some corrections. Alternatively, SVD correction can be applied iteratively using a phase or betatron function response matrix.

In a machine with a low-beta section where the peak betatron function is much larger than the ring average, the local optics errors may  completely dominate the beta-beating. In such a configuration it is better to first correct the local errors before trying to correct the beta-beating in the rest of the machine.

While the k-modulation and multi-turn techniques provide direct measurements of the optical function (betatron phases and betatron functions), this is not the case for the ORM technique, where the same information requires first a fit to the data with a model.
In all cases, due to measurement uncertainties and  the non-linear response of optical functions to gradient changes, iterations may be required to converge to a satisfactory situation.

\section{Imperfection summary}

Table~\ref{tab:feed-down} presents a summary of the imperfections due to dipole, quadrupole, and sextupole magnets and their impact on the machine parameters. Note that as a consequence of a misalignment, a field error corresponding to a lower order is generated: a shifted quadrupole generates a dipole error, while a shifted sextupole generates a (skew) quadrupole error. This effect is commonly described as a feed-down.

\begin{table}[h]
\caption{Table of feed-down errors up to sextupoles}
\begin{center}
\begin{tabular}{llll} \hline \hline
  Field type & Imperfection & Error type & Impact \\ \hline
  Dipole & Field error & Dipole & Orbit, trajectory, energy \\
  Dipole & Roll & Dipole & Orbit, trajectory \\
  Quadrupole & Field error & Quadrupole & Tune, optics \\
  Quadrupole & Offset & Dipole & Orbit, trajectory \\
  Quadrupole & Roll & Skew quadrupole & Coupling \\
  Sextupole & Field error & Sextupole & Chromaticity \\
  Sextupole & Horizontal offset & Quadrupole & Tune, optics  \\
  Sextupole & Vertical offset & Skew quadrupole & Coupling  \\ \hline \hline
\end{tabular}
\label{tab:feed-down}
\end{center}
\end{table}

\end{document}